%% file: Turing_Foam.tex
\pdfoutput=1
\documentclass[
    aps,prx,twocolumn,
    reprint,
    superscriptaddress,
    nofootinbib,
    floatfix,
    amssymb,
    longbibliography 
]{revtex4-2}

\usepackage[english]{babel}

\usepackage{amsmath}
\usepackage{amsfonts}
\usepackage{amssymb}
\usepackage{graphicx}
\usepackage{siunitx}
\usepackage{bm}

\usepackage[utf8]{inputenc}

\usepackage{nicefrac}

\usepackage[normalem]{ulem}

\usepackage{xcolor}

\definecolor{linkColor}{rgb}{0,0.3,0.7}
\usepackage[colorlinks=true,
            allcolors=linkColor,
            pdfborder={0 0 0},
            pdfencoding = auto
            ]{hyperref}
            
\usepackage{url}

\definecolor{myGreen}{rgb}{0.1,0.5,0.1}

\definecolor{myGray}{rgb}{0.6,0.6,0.6}

\begin{document}
\title{Deciphering the Interface Laws of Turing Mixtures and Foams}

\author{Henrik Weyer}
\author{Tobias A. Roth}
\affiliation{Arnold Sommerfeld Center for Theoretical Physics and Center for NanoScience, Department of Physics, Ludwig-Maximilians-Universit\"at M\"unchen, Theresienstra\ss e 37, D-80333 M\"unchen, Germany}
\author{Erwin Frey}
\email{frey@lmu.de}
\affiliation{Arnold Sommerfeld Center for Theoretical Physics and Center for NanoScience, Department of Physics, Ludwig-Maximilians-Universit\"at M\"unchen, Theresienstra\ss e 37, D-80333 M\"unchen, Germany}
\affiliation{Max Planck School Matter to Life, Hofgartenstraße 8, D-80539 Munich, Germany}

\date{September 27, 2024}

\begin{abstract}
For cellular functions like division and polarization, protein pattern formation driven by NTPase cycles is a central spatial control strategy.
Operating far from equilibrium, no general theory links microscopic reaction networks and parameters to the pattern type and dynamics.
We discover a generic mechanism giving rise to an effective interfacial tension organizing the macroscopic structure of non-equilibrium steady-state patterns.
Namely, maintaining protein-density interfaces by cyclic protein attachment and detachment produces curvature-dependent protein redistribution which straightens the interface.
We develop a non-equilibrium Neumann angle law and Plateau vertex conditions for interface junctions and mesh patterns, thus introducing the concepts of ``Turing mixtures'' and ``Turing foams''.
In contrast to liquid foams and mixtures, these non-equilibrium patterns can select an intrinsic wavelength by interrupting an equilibrium-like coarsening process.
Data from \textit{in vitro} experiments with the \textit{E.\ coli} Min protein system verifies the vertex conditions and supports the wavelength dynamics.
Our study uncovers interface laws with correspondence to thermodynamic relations that arise from distinct physical processes in active systems.
It allows the design of specific pattern morphologies with potential applications as spatial control strategies in synthetic cells.
\end{abstract}
\maketitle

Heterogeneous systems are governed by interfaces that separate spatially differentiated regions such as grain boundaries in crystal growth \cite{Godreche1992}, membranes \cite{Safran2018}, liquid-liquid boundaries in multi-component mixtures \cite{Mao.etal2019}, and foams \cite{DeGennes.etal2004,Weaire.Hutzler2001}.
Interfaces are also central to diverse biological processes, such as segregation and spreading in developmental tissue \cite{Foty.Steinberg2005}, morphogenesis \cite{Noll.etal2017,Fernandez.etal2021}, and selection within dense microbial colonies \cite{Kayser.etal2018,Giometto.etal2021}.
Moreover, foundational models of active matter are governed by their interface properties \cite{Cates.Tailleur2015,Fausti.etal2021}.
In these systems, interface properties such as mechanical surface tensions entirely govern the macroscopic system structure and dynamics.
Can a similarly simple but powerful concept like surface tension arise far from equilibrium without any mechanical interactions?
How may the resulting laws governing the interface dynamics, like the Gibbs--Thomson and Young--Laplace relations in thermodynamics, differ and induce new behavior?

\begin{figure*}
	\includegraphics{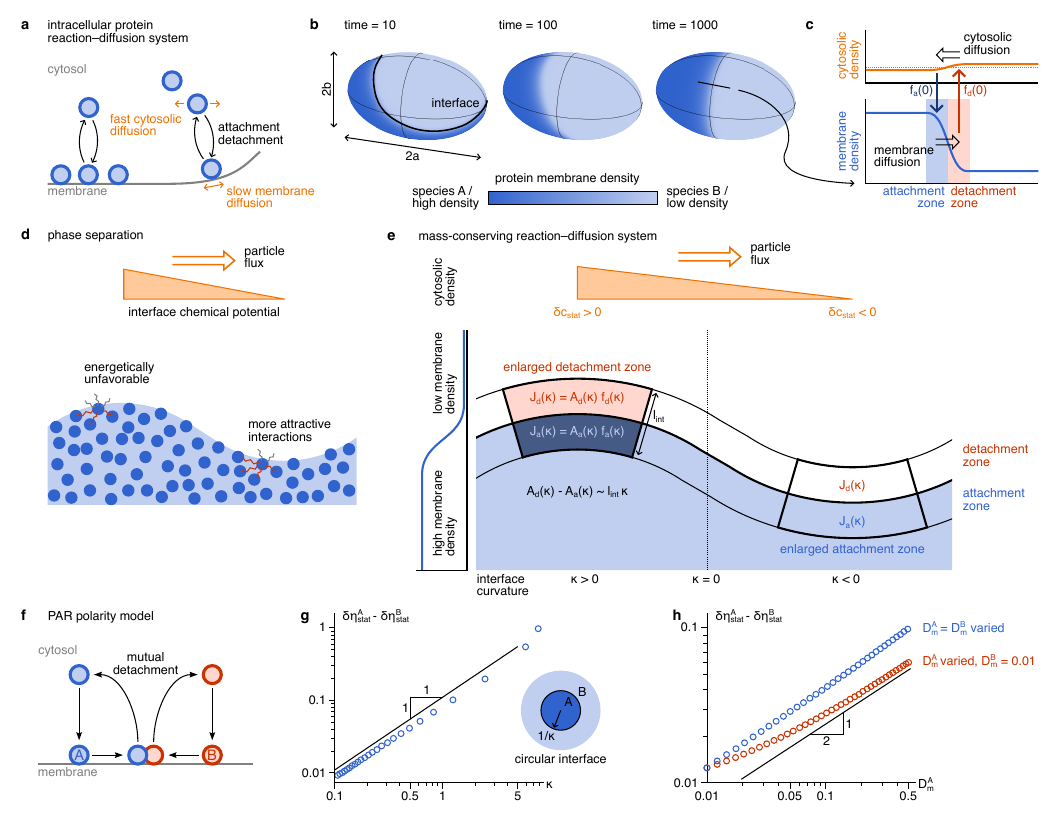}
	\caption{Effective interfacial tension results from cyclic protein attachment and detachment at pattern interfaces.
    \textbf{a}
    Coupling membrane attachment and detachment with fast cytosolic and slow membrane diffusion creates intracellular reaction--diffusion patterns.
    \textbf{b}
    Patterns of diverse McRD systems are described by the movement of the interface (black line) between domains of high- and low-densitiy or domains dominated by protein A and B.
    The membrane density of species A is shown in blue in the symmetric PAR system (\textbf{f}) on the surface of an ellipsoidal cell [$(a,b)=(1,0.6)$].
    \textbf{c}
    The pattern interface is sustained by cyclic fluxes of attachment and detachment $f_\mathrm{a,d}(0)$ in an attachment and detachment zone coupled to a shallow cytosolic gradient, which counteracts membrane diffusion.
    \textbf{d}
    In phase-separating liquids, the (exchange) chemical potential at the interface shows a gradient from outward- to inward-curved interfaces (orange).
    \textbf{e}
    In McRD systems, an area difference of the attachment and detachment zones arises at curved interfaces.
    The blue shading symbolizes the membrane-density gradient (left).
    The (local) reactive turnover balance of the total attachment--detachment fluxes ${J_\mathrm{a}(\kappa) = J_\mathrm{d}(\kappa)}$ can only be fulfilled if the local attachment--detachment fluxes $f_\mathrm{a,d}(\kappa)$ change compared to their values $f_\mathrm{a,d}(0)$ at a straight interface.
    The resulting curvature-induced cytosolic density gradient (Eq.~\eqref{eq:cytosolic-shift}) induces mass transport from outward- to inward-bend regions (orange).
    \textbf{f}
    In the symmetric PAR model, anterior (A, blue) and posterior (B, red) PAR proteins detach each other mutually (see Methods Sec.~\ref{methods:par}).
    \textbf{g,h}
    The non-equilibrium Gibbs-Thomson relation is verified in numerical simulations of a circular interface (inset) of the symmetric PAR system for varying interface curvature $\kappa$ and membrane diffusion coefficients $D_m^\mathrm{A,B}$ (${\kappa\approx 0.14}$).
    The model equations and simulation parameters for panels \textbf{b}, \textbf{g}, \textbf{h} are given in Methods Sec.~\ref{methods:par}.
    }
	\label{fig:1}
\end{figure*}

Reaction-diffusion systems describe ensembles of particles that diffuse independently of each other and undergo chemical reactions without mechanical interactions between them.
In biological cells, the spatial control of key cellular processes relies on NTPase-driven reaction--diffusion mechanisms forming heterogeneous non-equilibrium (steady-state) protein patterns (Fig.~\ref{fig:1}a)~\cite{Edelstein-Keshet.etal2013,Gross.etal2019,Loose.etal2011,Halatek.etal2018,Burkart.etal2022a}. 
The extensive diversity of macroscopic, highly nonlinear patterns in these systems---nicely exemplified by the paradigmatic \textit{E.\ coli} Min protein system \cite{Vecchiarelli.etal2016,Glock.etal2019,Brauns.etal2021b,Ren.etal2023a}---remains largely unexplained.
This limits our ability to control intracellular pattern formation and design specific reaction--diffusion patterns with targeted features.

Phenomenologically, the PAR polarity system in \textit{C. elegans} suggests the presence of a mechanism generating interface tension in intracellular protein patterns, as it robustly aligns the interface pattern between anterior and posterior PAR proteins along the shortest cell circumference \cite{Mittasch.etal2018}.
Although the actomyosin pseudocleavage furrow plays a significant role for this alignment in wild-type, these mechanical processes are known to be dispensable \cite{Bhatnagar.etal2023a} and simulations show that interface-length minimization occurs solely due to a reaction-diffusion mechanism (Fig.~\ref{fig:1}b) \cite{Gessele.etal2020}.
There is also evidence from conceptual models for the emergence of an effective interface tension in mass-conserving reaction-diffusion (McRD) systems, which describe intracellular protein pattern formation \cite{Frey.Brauns2022}.
For instance, elementary McRD systems are known to minimize the interface length by pattern reorientation \cite{Singh.etal2022,Maree.etal2012} and coarsening, i.e., a continuous growth of the characteristic pattern wavelength \cite{Otsuji.etal2007,Brauns.etal2021,Tateno.Ishihara2021}.
Interface-length minimization has been shown mathematically to occur generally in McRD models for a single protein species with two chemical states \cite{Miller.etal2023,Roth.etal}.
However, the mere introduction of a third component or the coupling to a reservoir can interrupt the coarsening process~\cite{Chiou.etal2021,Brauns.etal2021,Weyer.etal2023} and induce interface instabilities~\cite{Ohta.etal1989,Petrich.Goldstein1994,Muratov.Osipov1996,Carter.etal2023}.
Given the limited understanding of the underlying mechanism controlling length minimization, we are left with the central question: Can the self-organization of multi-species and multi-component reaction-diffusion systems be comprehensively understood through the concept of an effective interfacial tension?

Here, by exploring multi-species intracellular protein systems, we discover a generic mechanism giving rise to an effective interfacial tension in non-equilibrium steady-state patterns:
cyclic protein attachment and detachment at interfaces---driven by an NTPase cycle.
This effective interfacial tension organizes multi-species systems via a non-equilibrium Gibbs--Thomson relation and Neumann law, that show systematic differences to the corresponding thermodynamic relations.
Based on our discovery that these systems exhibit a non-equilibrium Neumann law and generalized Plateau vertex conditions, we introduce a new class of non-equilibrium systems, which we term \textit{Turing mixtures} and \textit{Turing foams}.
These patterns resemble liquid mixtures undergoing phase separation and liquid foams but arise entirely from the reaction--diffusion mechanism.
We demonstrate the Turing foam in experiments with the \textit{in vitro} Min protein system.
In contrast to liquid foams, we showcase that equilibrium-like coarsening can be interrupted and an intrinsic pattern wavelength selected.
The resulting coarsening (collapse) of small, and the splitting of large ``foam bubbles'' is observed in experimental data of the Min system.

\section{Effective interfacial tension}
\label{sec:eff-interfacial-tension}
The dynamics of interfaces in liquid mixtures exhibiting phase separation are driven by surface tension $\sigma$: 
Since interactions between particles of the same type are energetically favorable, a particle position at an outwardly curved interface (curvature ${\kappa>0}$) is associated with a higher (exchange) chemical potential ${\delta\mu \sim \sigma \kappa}$ than particles located in an inwardly curved interface region (\textit{Gibbs--Thomson relation}; Fig.~\ref{fig:1}d).
The resulting gradient in the chemical potential induces a particle flux that straightens the interface and reduces its length~\cite{Bray2002}.
Such an energetic argument cannot apply to pattern-forming reaction--diffusion systems in which ideal particles interact through chemical reactions but not through mechanical (pairwise) interaction potentials.

Intracellular protein patterns form by the regional accumulation and depletion of proteins on the cell membrane due to their attachment and detachment from and to the cytosol with rates depending on the protein membrane densities~\cite{Halatek.Frey2018,Brauns.etal2020,Halatek.etal2018} (Fig.~\ref{fig:1}a,b).
These patterns show sharp interfaces between regions of high and low protein-density on the membrane, while they are shallow in the cytosol because diffusion is significantly faster in the cytosol.
The basic mechanism of maintaining sharp interface gradients in the membrane densities in the presence of membrane diffusion lies in the formation of attachment and detachment zones coupled with (reverse) cytosolic protein fluxes \cite{Halatek.Frey2018, Halatek.etal2018,Brauns.etal2020}; see Fig.~\ref{fig:1}c.
Thus, driving the attachment--diffusion--detachment cycle by NTP hydrolysis, a fixed pattern interface is sustained as a non-equilibrium steady state.
In reverse, this implies that the membrane-density dependence of the reaction rates must allow dominant attachment (detachment) at high (low) membrane densities \cite{Brauns.etal2020}.
The resulting width of the attachment and detachment zones, that is, of the interfacial region can be estimated by the membrane-diffusion length ${\ell_\mathrm{int} \sim \sqrt{D_\mathrm{m} \tau_\mathrm{r}}}$, given a timescale $\tau_\mathrm{r}$ of reactive attachment and detachment and the membrane diffusion coefficient $D_\mathrm{m}$~\cite{Brauns.etal2020}.

What is driving the dynamics of interfaces in McRD systems?
For a curved interface, there is an area difference between the attachment zone and the detachment zone, ${A_\mathrm{d}(\kappa) - A_\mathrm{a}(\kappa) \sim \ell_\mathrm{int}\kappa}$, which depends on the magnitude and sign of its curvature $\kappa$ (Fig.~\ref{fig:1}e).
These imbalances change the cyclic protein fluxes. 
At outward-bent interfaces, proteins would accumulate indefinitely in the cytosol due to the area imbalance if the average local attachment and detachment fluxes, denoted as $f_\mathrm{a,d}(\kappa)$, remained unchanged compared to their values at a straight interface $f_\mathrm{a,d}(0)$ (cf.\ Fig.~\ref{fig:1}c).
However, we assume that the cytosolic density increase ${\delta c(\kappa) = c(\kappa)-c(0)}$ raises the local attachment flux because more proteins become available for attachment.
Therefore, the cytosolic accumulation saturates when the increased local attachment flux ${f_\mathrm{a}(\kappa) \approx f_\mathrm{a}(c(\kappa)) \approx f_\mathrm{a}(0)+ \partial_c f_\mathrm{a}|_{\kappa=0} \delta c(\kappa)}$ compensates for the area imbalance (${\delta c<0}$ for inward-bent interfaces).
We neglect curvature-induced changes in the membrane densities compared to their high-amplitude jump at the interface and approximate the detachment flux by ${f_\mathrm{d}(\kappa) \approx f_\mathrm{d}(0)}$ (see Methods Sec.~\ref{methods:eff-interfacial-tension} and SI Sec.~2).
Then, attachment and detachment balance each other at the curved interface for ${A_\mathrm{a}(\kappa)f_\mathrm{a}(\kappa)=A_\mathrm{d}(\kappa)f_\mathrm{d}(\kappa)}$ (reactive turnover balance \cite{Brauns.etal2020}), causing a steady-state cytosolic shift
\begin{equation}
\label{eq:cytosolic-shift}
    \delta c_\mathrm{stat}(\kappa) \approx \frac{f_\mathrm{d}}{\partial_c f_\mathrm{a}} \, \ell_\mathrm{int}
    \kappa \, .
\end{equation}
The resulting cytosolic protein gradients lead to fluxes that redistribute proteins laterally between differently curved interface regions (within a quasi steady state approximation). 
The analogy to the Gibbs--Thomson relation suggests that cyclic attachment and detachment of proteins at the pattern interface induces large-scale dynamics governed by an effective interfacial tension scaling with the interface width ${\sigma \sim \ell_\mathrm{int} \sim \sqrt{D_\mathrm{m}}}$.

Membrane-density changes and membrane diffusion between differently curved interface regions, neglected above, can be accounted for by analyzing the gradients of the sum of all protein states $u_i$ weighted by their diffusion coefficients, i.e., the gradients of the ``mass-redistribution potential'' ${ \eta = \sum_i D_i u_i/D_\mathrm{c} }$ (normalizing by one cytosolic diffusion constant $D_\mathrm{c}$); see Methods Sec.~\ref{methods:model-systems} and SI Sec.~2) \cite{Otsuji.etal2007,Halatek.Frey2018,Brauns.etal2021}.
As for the cytosolic density, the imbalances of attachment and detachment at curved interfaces induce a shift $\delta\eta_\mathrm{stat}$ in the mass-redistribution potential (see Methods~\ref{methods:eff-interfacial-tension}). 
For weakly curved interfaces, one finds the interfacial dynamics in McRD systems with two components in closed form \cite{Miller.etal2023,Roth.etal} and $\delta\eta_\mathrm{stat}$ scales with $\ell_\mathrm{int}\kappa$ \cite{Brauns.etal2021,Tateno.Ishihara2021}.

In multi-species phase separation, the thermodynamic Gibbs--Thomson relation reads ${\Delta\bm{\rho}\cdot \delta\bm{\mu}=-\sigma\kappa}$, where $\Delta \rho_i$ and $\delta \mu_i$ represent the (total) density jump across the interface and the chemical potential of species $i$, respectively (SI Sec.~3.1 and Ref.~\cite{Bronsard.etal1998}).
This relation follows from the condition that the sum of the osmotic pressures of all species balances the Laplace pressure.
Let us compare this with the effective interfacial tension in a conceptual multi-species model of the PAR polarity system in \textit{C. elegans} (Methods Sec.~\ref{methods:par}; cf.\ Fig.~\ref{fig:1}b) \cite{Trong.etal2014}.
More generally, for systems with multiple protein species and symmetric, second-order mutual detachment (Fig.~\ref{fig:1}f), we find that the shifts $\delta\bm{\eta}_\mathrm{stat}$ of the mass-redistribution potentials for all species fulfill ${ \Delta\bm{\rho}\cdot \mathbf{F}_\mathrm{int} \delta\bm{\eta}_\mathrm{stat}=-\sigma\kappa \sim - \ell_\mathrm{int}\kappa}$. 
The derivation of this relation and the general relation for systems with multiple protein species, where each species has a membrane and a cytosolic component, are given in SI Sec.~3.
The matrix $\mathbf{F}_\mathrm{int}$ of (scaled) attachment rates is diagonal with positive entries (SI Sec.~3).
Thus, except for the rescaling of the shifts $\delta\bm{\eta}_\mathrm{stat}$ by the attachment rates in McRD systems, these shifts behave analogously to the chemical-potential shifts $\delta\bm{\mu}$ in multi-component liquid mixtures.

For the PAR polarity model with symmetric reaction rates (Methods Sec.~\ref{methods:par}), the non-equilibrium analogue of the Gibbs-Thomson relation simplifies to ${\delta \eta_\mathrm{A}^\mathrm{stat}-\delta \eta_\mathrm{B}^\mathrm{stat} \approx -\sigma \kappa/(\Delta\rho_\mathrm{A} k_\mathrm{c}^\mathrm{A})}$, where $k_\mathrm{c}^\mathrm{A}$ denotes the attachment rate, if ${D_c\gg D_m}$.
We numerically verified (Methods Sec.~\ref{methods:simulation}) that the difference in the mass-redistribution potentials ${\delta \eta_\mathrm{A}^\mathrm{stat}-\delta \eta_\mathrm{B}^\mathrm{stat}}$ is proportional to $\kappa$ and ${\sigma\sim\sqrt{\max(D_m^{\mathrm{A}},D_m^{\mathrm{B}})}}$ (Figures~\ref{fig:1}g,h).
This demonstrates an interfacial-tension-like behavior and that the larger membrane diffusion coefficient determines the interface width and tension asymptotically when ${D_c\gg D_m}$ (SI Sec.~3.5.5).

Taken together, an effective interfacial tension $\sigma\sim\ell_\mathrm{int}$ arises due to the curvature-induced (local) increase or decrease of the detachment zone compared to the attachment zone.
Proteins detaching at positively curved interface regions are redistributed towards regions of negative interface curvature, straightening the interface (Fig.~\ref{fig:1}e).

\begin{figure*}
    \includegraphics{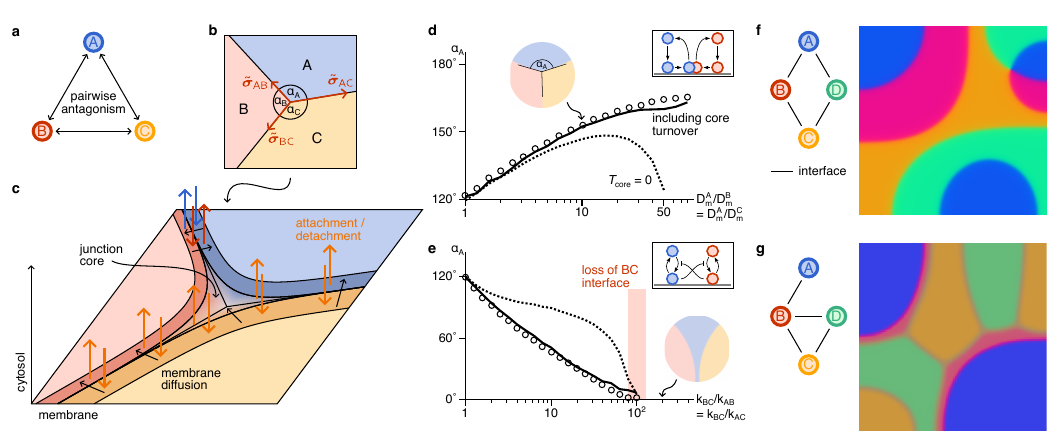}
\caption{
    Non-equilibrium Neumann law arising from the balance of attachment and detachment.
    \textbf{a}
    McRD systems with three or more species can be constructed by introducing pairwise antagonistic interactions resembling the PAR system (\textbf{d}) or inhibited attachment (\textbf{e}).
    \textbf{b}
    Junctions angles $\alpha_\mathrm{A,B,C}$ are constrained by the non-equilibrium Neumann law [Eq.~\eqref{eq:steady-state-angles}] for the effective interfacial tensions $\tilde{\bm\sigma}_\mathrm{AB,AC,BC}$.
    \textbf{c}
    In steady state, protein attachment (dark shadings) and detachment (in the attachment regions of the opposing species) must balance.
    In the junction core, all three domains meet.
    \textbf{d} 
    The junction angle $\alpha_A$ increases as a function of the membrane diffusion coefficient $D_\mathrm{m}^\mathrm{A}$ (keeping ${D_\mathrm{m}^\mathrm{B}=D_\mathrm{m}^\mathrm{C}}$ constant) for a symmetric three-species system with mutual detachment (inset) where all rates are equal; based on the solution of Eqs.~\eqref{eq:multiSpecies-twoComps},~\eqref{eq:PAR-multi} with parameters in Tab.~\ref{tab:PAR-rates} (circles; see Methods Sec.~\ref{methods:pattern-analysis}).
    The non-equilibrium Neumann law including the core turnover $\mathbf{T}_\mathrm{core}$ (black line) and for setting ${\mathbf{T}_\mathrm{core}=0}$ (dashed line) are shown. 
    The circular inset shows the membrane densities of the steady-state pattern.
    \textbf{e}
    The junction angle $\alpha_A$ decreases as a function of the attachment inhibition $k_\mathrm{BC}$ between species B and C (circles; see Methods Sec.~\ref{methods:pattern-analysis}) until the BC interface vanishes (red-shaded area) in a three-species system with inhibited attachment (blunt arrows in the inset).
    The model equations and parameters are given in Methods Sec.~\ref{methods:inhibited-attachment}).
    \textbf{f}, \textbf{g}
    Two four-species systems based on attachment inhibition are designed [panel \textbf{e}; Methods Sec.~\ref{methods:inhibited-attachment}] in which the non-equilibrium analogue of full wetting only allows interfaces to form between the species connected in the contact graphs.
    The membrane-density patterns are shown at time ${t=10\,000}$.
    The simulation parameters are given in Tab.~\ref{tab:PAR-rates} and we choose $k_{\alpha\beta}=k_{\beta\alpha}$ with ${k_{12}=k_{14}=k_{23}=k_{34}=3}$, ${k_{13}=k_{24}=600}$ in \textbf{f} and ${k_{12}=k_{23}=k_{24}=k_{34}=0.6}$, ${k_{13}=k_{14}=2\,400}$, ${\mu_{\alpha\beta}=4}$ in \textbf{g}.
    The system \textbf{f} forms stable four-fold junctions as in phase-separating liquids \cite{Mao.etal2020}.
	}
	\label{fig:2}
\end{figure*}

\section{Non-equilibrium Neumann law}
\label{sec:neumann_law}
In multi-species McRD systems, mutually antagonistic relationships can be established between all pairs of species (Fig.~\ref{fig:2}a), analogously to pairwise interactions in phase-separating systems with multiple components.
This entails, for example, mutual detachment from the membrane and inhibited attachment to the membrane (Figs.~\ref{fig:2}d,e).
We find a pattern morphology with interfaces and their junctions that resembles multi-component liquid mixtures  (Fig.~\ref{fig:2}b), and we refer to these patterns as \textit{Turing mixtures} since they arise exclusively by a reaction-diffusion mechanism.

In liquid mixtures, the meeting angles at (quasi-) stationary interface junctions are determined by force balance:
The vectorial sum of the three surface tensions must vanish, known as \textit{Neumann law}~\cite{DeGennes.etal2004}.
In contrast, in the reaction-diffusion systems considered here, these interface junctions are stationary if the overall \textit{attachment--detachment turnover} of each protein species is balanced (Fig.~\ref{fig:2}c).
For McRD systems with two components per protein species, we derive the non-equilibrium Neumann law (SI Sec.~4):
\begin{equation}
\label{eq:steady-state-angles}
    \Tilde{\bm{\sigma}}_\mathrm{AB}
    +
    \Tilde{\bm{\sigma}}_\mathrm{AC}
    +
    \Tilde{\bm{\sigma}}_\mathrm{BC} 
    = 
    \mathbf{T}_\mathrm{core}
    \, .
\end{equation}
Here, the mathematical expressions for the effective interfacial tensions $\Tilde{\sigma}_{ij}$ between domains ${i,j = \mathrm{A, B, C}}$ differ from those in the non-equilibrium Gibbs--Thomson relation but similarly scale with the interface widths $\ell_\mathrm{int}^{ij}\sim \sqrt{\max(D_m^{i},D_m^{j})}$ (SI Sec.~4).
The turnover balance leading to Eq.~\eqref{eq:steady-state-angles} replaces the osmotic-pressure balance that must be fulfilled in liquid mixtures.
Because liquid mixtures are governed by a free energy functional, the osmotic-pressure balance condition at the junction agrees with the mechanical equilibrium of the surface tensions (SI Sec.~4.2.1) \cite{Bronsard.etal1998}.
This is not the case in the non-equilibrium reaction-diffusion system.
The reactive turnover in the junction core---where all three species meet---is qualitatively different from the reactive flows at the binary interfaces.
The source term $\mathbf{T}_\mathrm{core}$ is the contribution of the junction core to the reactive turnover (SI Sec.~4).
Similarly, non-equilibrium currents modify the Young--Dupr\'e equation for partial wetting in systems undergoing motility-induced phase separation \cite{Zhao.etal2024}.

Figures~\ref{fig:2}d,e show how the non-equilibrium Neumann law prescribes the junction angles in two different three-species systems.
The junction angle $\alpha_\mathrm{A}$ of species A increases as the effective interfacial tensions $\Tilde{\sigma}_\mathrm{AB,AC}$ rise with $D_m^\mathrm{A}$.
In contrast, increasing the rate of mutual antagonism between species B and C (Methods Sec.~\ref{methods:inhibited-attachment}) enlarges $\Tilde{\sigma}_\mathrm{BC}$, reducing the angle $\alpha_\mathrm{A}$ until the BC interface vanishes when ${\alpha_\mathrm{A} = 0}$.
This represents a non-equilibrium version of the transition to full wetting in phase separation.
Full wetting occurs, akin to thermal-equilibrium systems \cite{DeGennes.etal2004,Mao.etal2020}, if ${\tilde{\sigma}_\mathrm{BC}>\tilde{\sigma}_\mathrm{AB}+\tilde{\sigma}_\mathrm{AC}}$ because the core turnover $\mathbf{T}_\mathrm{core}$ vanishes at the transition, as does the junction core itself (Fig.~\ref{fig:2}e; SI Sec.~4.2.3).
Choosing the effective interfacial tensions relative to each other to fulfill or not the bound ${\tilde{\sigma}_\mathrm{BC}<\tilde{\sigma}_\mathrm{AB}+\tilde{\sigma}_\mathrm{AC}}$ determines which species contact each other~\cite{Mao.etal2020}.
This enables the design of specific patterns by linking the microscopic parameters, reaction rates, and diffusion coefficients to the macroscopic pattern morphology (Fig.~\ref{fig:2}f,g).

\begin{figure*}
	\includegraphics{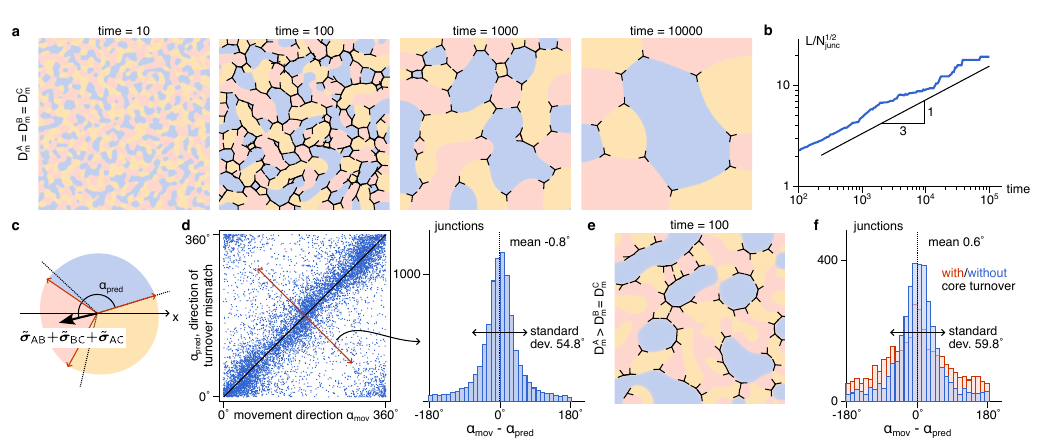}
	
 \caption{Coarsening and junction dynamics.
    \textbf{a}
    The pattern formed by the symmetric, three-species mutual-detachment model (cf.\ Fig.~\ref{fig:2}d; Methods Sec.~\ref{methods:par}) out of small random perturbations around uniform densities coarsens over time.
    The snapshots show the membrane densities.
    The triple junctions are marked in black.
    \textbf{b}
    The characteristic pattern wavelength ${\langle\Lambda\rangle =  L/N_\mathrm{junc}^{1/2}}$, determined from the junction density $N_\mathrm{junc}/L^2$, increases approximately as $\langle\Lambda\rangle\sim t^{1/3}$.
    The data is an average over three independent simulations.
    \textbf{c}
    We find the imbalance ${ \tilde{\bm\sigma}_\mathrm{AB} + \tilde{\bm\sigma}_\mathrm{BC} + \tilde{\bm\sigma}_\mathrm{AC} }$ in the effective interfacial tensions describes the junction movement direction.
    \textbf{d}
    The direction of junction movement $\alpha_\mathrm{mov}$ and the junction angles are determined from consecutive time points throughout the simulation (Methods Sec.~\ref{methods:pattern-analysis}).
    The average direction $\alpha_\mathrm{pred}$ of the imbalance ${ \tilde{\bm\sigma}_\mathrm{AB} + \tilde{\bm\sigma}_\mathrm{BC} + \tilde{\bm\sigma}_\mathrm{AC} }$ is determined from the junction angles using the effective interfacial tensions measured in Fig.~\ref{fig:2}d.
    Each measurement for each junction is shown as a point (three independent simulations).
    From these measurements, the histogram of the deviations ${\alpha_\mathrm{mov}-\alpha_\mathrm{pred}}$ is derived (accounting for the $\ang{360}$ periodicity of the angles).
    \textbf{e}
    A snapshot of a simulation of the asymmetric three-species mutual-detachment model with ${D_\mathrm{m}^\mathrm{A}=0.2>D_\mathrm{m}^\mathrm{B}=D_\mathrm{m}^\mathrm{C}=0.01}$ (same reaction terms as in \textbf{a}) is shown.
    \textbf{f}
    The histograms of the angular deviations ${\alpha_\mathrm{mov}-\alpha_\mathrm{pred}}$ for the two cases that $\alpha_\mathrm{pred}$ is chosen as the direction of the imbalance ${ \tilde{\bm\sigma}_\mathrm{AB} + \tilde{\bm\sigma}_\mathrm{BC} + \tilde{\bm\sigma}_\mathrm{AC} }$ (blue) or the imbalance ${ \tilde{\bm\sigma}_\mathrm{AB} + \tilde{\bm\sigma}_\mathrm{BC} + \tilde{\bm\sigma}_\mathrm{AC} -\mathbf{T}_\mathrm{core}^\mathrm{stat}}$ including the core turnover $\mathbf{T}_\mathrm{core}^\mathrm{stat}$ of the junction in steady state (red).
    The simulations are performed on square domains of length ${L=30}$.
    The parameters not specified in this caption are listed in Tab.~\ref{tab:PAR-rates}. 
    }
\label{fig:3}
\end{figure*}

\section{Dynamics of interface junctions}
Phase-separated liquid mixtures undergo Ostwald ripening:
Larger droplets expand over time while smaller droplets shrink concurrently, such that the typical droplet size grows as ${\langle\Lambda\rangle\sim t^{1/3}}$ with time $t$, known as LSW--scaling~\cite{Lifshitz.Slyozov1961, Wagner1961, Bray2002, Kohn.Yan2004}.
Here, we find the same scaling in large-scale simulations of Turing mixtures with symmetric three-species mutual-detachment (Fig.~\ref{fig:3}a and Movie~1; cf.~Fig.~\ref{fig:2}d and Methods Sec.~\ref{methods:par}); this scaling law has also been found analytically in two-component systems~\cite{Brauns.etal2021, Weyer.etal2023, Tateno.Ishihara2021}.

What mechanism drives the movement and annihilation of the junctions during the dynamics of Turing mixtures?
In equilibrium phase separation, a mismatch in the surface tensions, ${\sum_i \bm{\sigma}_i\neq 0}$, generates a mechanical force that moves the junction in the direction of this mismatch (Fig.~\ref{fig:3}c).
Interestingly, both our simulations of the symmetric model (Fig.~\ref{fig:3}d) and of the model with distinct membrane diffusion coefficients (Fig.~\ref{fig:3}e,f) reveal a strong correlation between the direction of junction movement and the turnover mismatch $\sum_i \tilde{\bm{\sigma}}_i$.
Although this mismatch does not generate any mechanical force, any deviation of the contact angles from the steady-state condition (Eq.~\ref{eq:steady-state-angles}) leads to an imbalance in the net attachment (or detachment) fluxes.
If, for example, the angle $\alpha_\mathrm{A}$ is reduced, the A-protein domain will retract due to the increased detachment relative to attachment induced by the increased interface curvature (cf.\ Sec.~\ref{sec:eff-interfacial-tension}).
The reduced angle $\alpha_\mathrm{A}$ leads to a mismatch $\sum_i \Tilde{\bm{\sigma}}_i$ that points more towards the A-domain hinting at a correlation. 
However, it remains an exciting open challenge to identify a mathematical reasoning of why a strong correlation is obtained without the core turnover $\mathbf{T}_\mathrm{core}$ (cf.~Fig.~\ref{fig:3}f), even though it is essential for the stationary angles (Fig.~\ref{fig:2}d,e).

\begin{figure*}
	\includegraphics{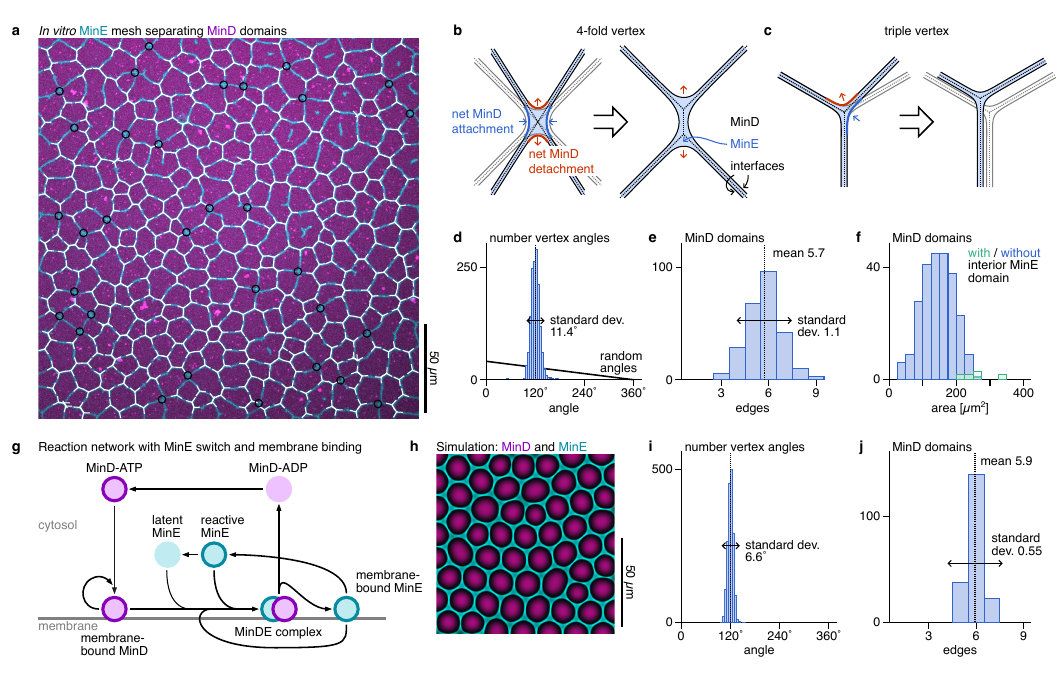}
	\caption{
    Turing foam in the \textit{in vitro} Min system.
	\textbf{a} \textit{In vitro}, the \textit{E.\ coli} Min system can form mesh patterns made up of thin MinE branches (cyan) separating MinD domains (magenta).
    Total concentrations: ${[\text{MinD}] = \SI{1.5}{\micro M}}$ and ${[\text{MinE{-}His}] = \SI{3}{\micro M}}$ (cf.~Fig.~\ref{fig:e1}; experimental data obtained and described in Ref.~\cite{Glock.etal2019}).
    Branches of triple vertices are marked in white, and
    4-fold vertices are labeled with black circles (Methods Sec.~\ref{methods:pattern-analysis}).
    \textbf{b}, \textbf{c} The instability of 4-fold and higher-order vertices (\textbf{b}) and branch rearrangement at triple vertices towards $\ang{120}$ junction angles (\textbf{c}) can be explained based on the curvature dependence of net attachment (blue) and detachment (red). 
    \textbf{d}, \textbf{e}, \textbf{f} Histograms of the angles at the triple vertices, the number of edges of each MinD domain, and the MinD domain areas are determined from the experimental data in \textbf{a} (see also Fig.~\ref{fig:e1}a); see Methods Sec.~\ref{methods:pattern-analysis}.
    For comparison, the vertex-angle histogram of vertices with random angles is shown (black line, SI Sec.~5).
    The largest MinD domains have an internal MinE domain disconnected from the mesh (green).
    \textbf{g} Reaction network of a minimal model of MinD (magenta) and MinE (cyan) membrane attachment and detachment supplemented by a conformational MinE switch and persistent membrane binding of MinE.
    \textbf{h} This model reproduces stationary mesh patterns of MinD (magenta) and MinE (cyan) in two-dimensional simulations (model equations and parameters given in Methods Sec.~\ref{methods:min}).
    The detected triple junctions are marked in white.
    \textbf{i}, \textbf{j} Histograms of the vertex angles and the edge number of the MinD domains are determined from six independent simulations (Methods Sec.~\ref{methods:pattern-analysis}).
	}
	\label{fig:4}
\end{figure*}

\begin{figure*}
	\includegraphics{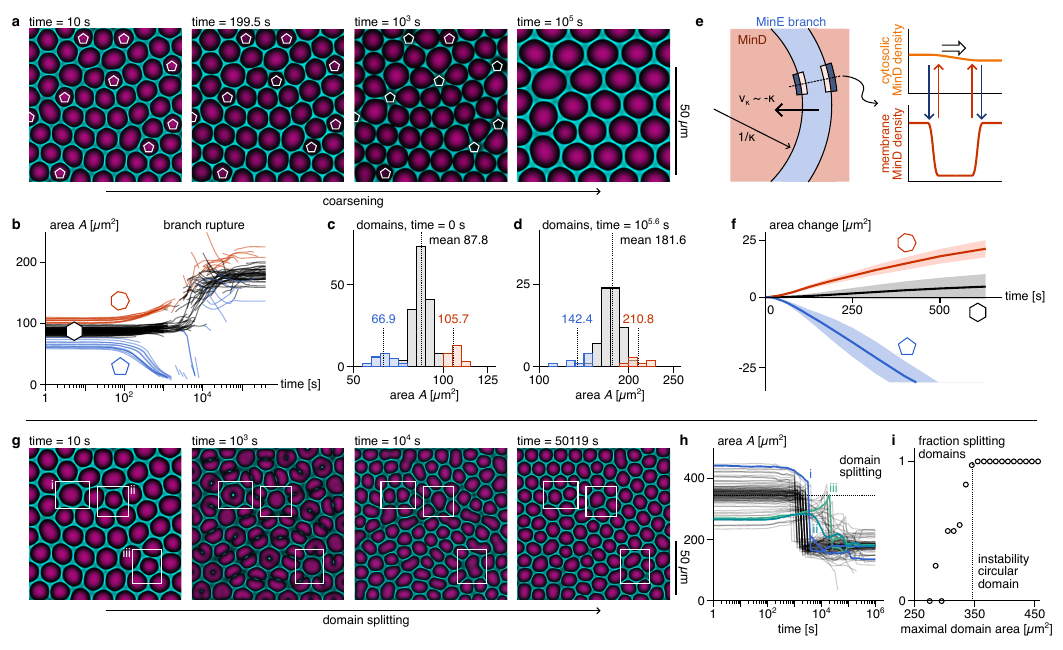}
	\caption{Interrupted coarsening and domain splitting of Turing foams.
    \textbf{a}
    The dynamics of a stationary mesh pattern is shown that is reduced in size by rescaling the simulation-domain length $L$ to $L/\sqrt{2}$ (simulation parameters given in Tab.~\ref{tab:Min-rates}; MinE membrane density shown in cyan and MinD in magenta).
    White pentagons label pattern domains with 5 edges (reflecting the pattern at the boundary).
    \textbf{b}
    The trajectories of the areas $A$ of one-half of the domains with $n=5,6,7$ (blue, black, red) in six independent simulations are shown.
    \textbf{c}, \textbf{d}, \textbf{f}
    The statistics from these six independent simulations are aggregated to calculate the distribution of initial (\textbf{c}) and final domain areas (\textbf{d}), as well as the area change ${A(t)-A(0)}$ compared to the initial domain area $A(0)$ independently averaged over the domains with ${n=5,6,7}$ edges (\textbf{f}, shaded regions depict the standard deviation). 
    \textbf{e}
    The curvature-induced branch movement with a normal velocity $v_\kappa$ is caused by the curvature-induced transport of MinD (red shaded) through the MinE branch (blue shaded) from the inside to the outside (black double arrow; cf.\ Fig.~\ref{fig:1}e).
    Sections of the MinD attachment and detachment zones are sketched as in Fig.~\ref{fig:1}e. 
    \textbf{g}
    The dynamics of a stationary mesh pattern on an enlarged domain ${L \to \sqrt{2}L}$ is shown. 
    Domains split by the growth of an internal MinE branch (domain i) or by the growth of a branch out of an existing branch (domain iii).
    Small domains do not split (domain ii).
    \textbf{h}
    One-half of the domain-area trajectories from six independent simulations are shown.
    The trajectories of the domains i, ii, and iii are labeled.
    \textbf{i}
    For domains with a given maximal domain area over time, the fraction that splits is shown (black circles).
    It is calculated from the histogram of the maximal areas of all domains and those that eventually split in the six simulations.
    The threshold of domain splitting is approximately given by the area at which a single circular domain forms an internal MinE domain (dotted line in \textbf{h}, \textbf{i}; Methods Sec.~\ref{methods:simulation}).
	}
	\label{fig:5}
\end{figure*}

\section{Turing foam in the \textit{in vitro} Min protein system}
\label{sec:min_patterns}

Can these generalized interface laws be applied to understand other biochemical systems?
While the PAR system has not been reconstituted yet, the \textit{E.\ coli} Min protein system is well-studied \textit{in vitro} \cite{Loose.etal2008,Vecchiarelli.etal2016, Wu.etal2016,Glock.etal2019}.
Patterns emerge from the interaction of two proteins: MinD, an ATPase that attaches to the lipid membrane in its ATP-bound state; and MinE, the ATPase-activating protein recruited by MinD onto the membrane \cite{Lutkenhaus2007,Halatek.etal2018,Ramm.etal2019}.
Using MinE proteins with a C-terminal His tag or wild-type MinE, the Min system forms (quasi-) stationary mesh (Fig.~\ref{fig:4}a) and labyrinthine patterns for most protein concentrations \cite{Glock.etal2019}. 
Currently, we lack both a model that robustly reproduces these network patterns and a comprehensive theoretical framework that elucidates their macroscopic dynamics and morphology.
Our interface theory predicts a distinct foam-like morphology for the mesh patterns, which we refer to as \textit{Turing foams}.

The mesh pattern is formed by thin MinE branches separating extended MinD domains and meeting at vertices (Fig.~\ref{fig:4}a; further data analyzed in Fig.~\ref{fig:e1}). 
Each branch has two MinD--MinE interfaces.
Our theory predicts that only triple vertices with three branches are stable (Fig.~\ref{fig:4}b,c).
Deviations in the vertex angles in higher-order vertices change the interface curvatures, prompting attachment and detachment fluxes that let the MinD domains retract where the vertex angles are reduced (Sec.~\ref{sec:eff-interfacial-tension}).
As a result, the vertex core elongates, and the vertex splits.
Similarly, attachment and detachment fluxes steer an initially asymmetric triple junction towards a symmetric configuration with $\ang{120}$ vertex angles.
These vertex conditions, which are referred to as Plateau vertex laws in the field of two-dimensional liquid foams, determine the structure of these systems~\cite{Weaire.Hutzler2001}.

Reanalyzing the experimental data in Ref.~\cite{Glock.etal2019} we find that the Min mesh pattern shows about 94\% triple junctions (Fig.~\ref{fig:4}a) whose vertex angles are distributed with a remarkably small standard deviation of about $\ang{11.4}$ around $\ang{120}$ (Fig.~\ref{fig:4}d; further data analyzed in Fig.~\ref{fig:e1}).
The mean number of domain edges is $5.7\pm 1.1$ (Fig.~\ref{fig:4}e).
In contrast to equilibrium foams, the largest MinD domains show disconnected, internal MinE branches (Fig.~\ref{fig:4}f), which we discuss in the next section.

Using an extended model based on biochemically suggested interactions of the Min proteins (Fig.~\ref{fig:4}g, Methods Sec.~\ref{methods:min}), we are able to reproduce the statistics of Turing foams in numerical simulations (Fig.~\ref{fig:4}h--j).
Moreover, during the development of the pattern, 4-fold vertices decay into pairs of triple vertices (Movie~2) as predicted.

\section{Von Neumann's law and domain splitting}

The dynamics of two-dimensional liquid foams follow \textit{von Neumann's law}, which states that the area of bubbles with less than six edges shrink and eventually collapse while bubbles with more than six edges grow~\cite{vonNeumann1952,Weaire.Hutzler2001}.
Although the observed Min patterns have a foam-like morphology (Fig.~\ref{fig:4}a), they violate this fundamental property of continuous coarsening.
Specifically, the larger MinD domains contain internal MinE branches that are disconnected from the mesh structure (Fig.~\ref{fig:4}a,f).
This implies a mechanism that limits the maximum size of stationary domains beyond which nucleation of internal MinE domains leads to the division of these MinD domains.
In the following, we demonstrate that the dynamics of the Turing foam is governed by interrupted coarsening and domain splitting identified previously in two-component reaction--diffusion systems \cite{Brauns.etal2021,Weyer.etal2023}.

What determines the final domain size and pattern wavelength?
For small domain sizes the branch curvature is large, suggesting a dominant role of the effective interfacial tension and liquid-foam like behavior.
Indeed starting the dynamics with a stationary pattern rescaled in length by the factor $1/\sqrt{2}$ (Fig.~\ref{fig:4}h), the pattern coarsens until the domain sizes become comparable to the initial (unscaled) pattern and then comes to a halt (Fig.~\ref{fig:5}a--d and Movie~3).
Notably, coarsening follows (qualitatively) von Neumann's law and proceeds by the collapse of domains with five edges and the growth of domains with seven edges before branches also rupture later in the dynamics (Fig.~\ref{fig:5}b,f).

How does von Neumann's law come about in the context of reaction--diffusion dynamics? 
In liquid foams, it is a consequence of the Laplace pressure, which drives gas diffusion between the bubbles \cite{Weaire.Hutzler2001}.
In Turing foams, in contrast, we derive von Neumann's law for the area change of an $n$-sided domain
\begin{equation}\label{eq:von-Neumann}
    \partial_t A_n \sim (n-6)
    \, ,
\end{equation}
based on the effective interfacial tension, which leads to a redistribution of (MinD) proteins due to cytosolic gradients across curved branches (Fig.~\ref{fig:5}e and Methods Sec.~\ref{methods:von-Neumann}).
The simulations might not fulfill the symmetry ${|\partial_t A_5|=|\partial_t A_7|}$ (Fig.~\ref{fig:5}f) because the interface widths are not sufficiently small compared to the domain sizes.

Do the simulations also recapitulate the experimentally observed splitting of large domains? 
Enlarging the stationary mesh pattern by a factor of $\sqrt{2}$, we observe the formation of additional MinE branches (Fig.~\ref{fig:5}g,h and Movie~4).
This can occur either by the formation and expansion of internal MinE branches or through branches that grow out of the MinE network in elongated domains, akin to a fingering instability of the branches (Fig.~\ref{fig:5}g).
We estimate the critical size for domain splitting as the size at which a single circular MinD domain surrounded by a MinE branch splits (Fig.~\ref{fig:5}i; Methods Sec.~\ref{methods:simulation}). 
After the splitting processes, the stationary domain areas become comparable again to those in the original (unscaled) steady-state pattern.

Both of the predicted behaviors of domain collapse and splitting are observed in the time course of the \textit{in vitro} Min mesh pattern (Fig.~\ref{fig:e2} and Movie~5).

\section{Discussion}
We have shown that a central concept of equilibrium physics---surface tension---emerges from an entirely distinct physical process in (mass-conserving) reaction--diffusion systems: orchestrated cyclic attachment and detachment at the pattern interface, which is a hallmark of the non-equilibrium nature of these steady states.
Deriving a non-equilibrium Neumann law and Plateau vertex conditions, we introduce Turing mixtures and foams.
Importantly, the non-equilibrium nature of the reaction--diffusion systems allows the selection of an intrinsic pattern length scale, interrupting an equilibrium-like coarsening process.

\begin{figure}[htb]
	\includegraphics{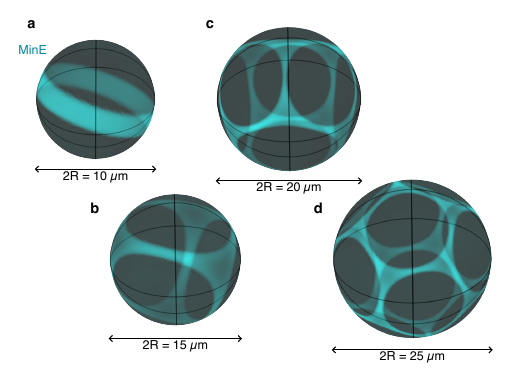}
	\caption{Interrupted coarsening stabilizes Turing foam on spherical surfaces.
    Polyhedral MinE meshes (translucent cyan on a translucent black background; MinD not shown) with an increasing number of faces (foam domains) form in simulations as the sphere radius $R$ is increased.
    An equatorial branch forms for $2R=\SI{10}{\micro m}$ (\textbf{a}), a tetrahedral foam for ${2R=\SI{15}{\micro m}}$ (\textbf{b}), a polyhedron with 4- and 5-sided domains at ${2R=\SI{20}{\micro m}}$ (\textbf{c}), and a dodecahedral foam for ${2R=\SI{25}{\micro m}}$ (\textbf{d}).
    The simulations are not drawn to scale.
    The model equations and parameters are given in Methods Sec.~\ref{methods:min}.
	}
	\label{fig:6}
\end{figure}

Both the instability of four-fold vertices and the emergence of stable $\ang{120}$ vertex angles have been observed in experiments and simulations based on (non-mass-conserving) pH-feedback reactions \cite{Goldstein.etal1996, Szalai.etal2012} related to the classical ferrocyanide-iodate-sulfite (FIS) reaction \cite{Lee.etal1993,Lee.Swinney1995}.
The FIS reaction also shows the Turing-foam dynamics governed by interrupted coarsening and domain splitting in a simulation (Movie~6, Methods~\ref{methods:fis}).
Notably, these systems are not mass-conserving suggesting a broader relevance of Turing foams beyond strict mass conservation.
As first steps, wavelength selection by interrupted coarsening has been derived in two-component reaction--diffusion systems with weakly broken mass conservation~\cite{Brauns.etal2021,Weyer.etal2023}.
Additionally, one-component reaction--diffusion models describing bistable media are mathematically equivalent to Model A dynamics, allowing the definition of an interfacial tension, for example, for competing microbial populations \cite{Schlogl1972,Lavrentovich.Nelson2019,Giometto.etal2021}.

Further experimental investigations should probe the statistics of coarsening and domain splitting as well as patterns on complex geometries, e.g., on structured membranes or vesicles.
While liquid foams are always unstable on surfaces with positive Gaussian curvature \cite{Avron.Levine1992}, simulations suggest that polyhedral domain arrangements on spheres are stabilized by interrupted coarsening in Turing foams (Fig.~\ref{fig:6}).

Foam-like structures are observed as well in scalar \cite{Fausti.etal2021}, polar \cite{DeLuca.etal2024}, and nematic active matter \cite{Maryshev.etal2020}, and in experimental active matter systems \cite{Lemma.etal2022}.
By revealing how far-from-equilibrium microscopic processes can yield equilibrium-like interface dynamics, our findings offer an approach to mechanistically understand the pattern dynamics, macroscopic morphologies, and wavelengths in active systems more broadly.

\begin{figure*}
	\includegraphics{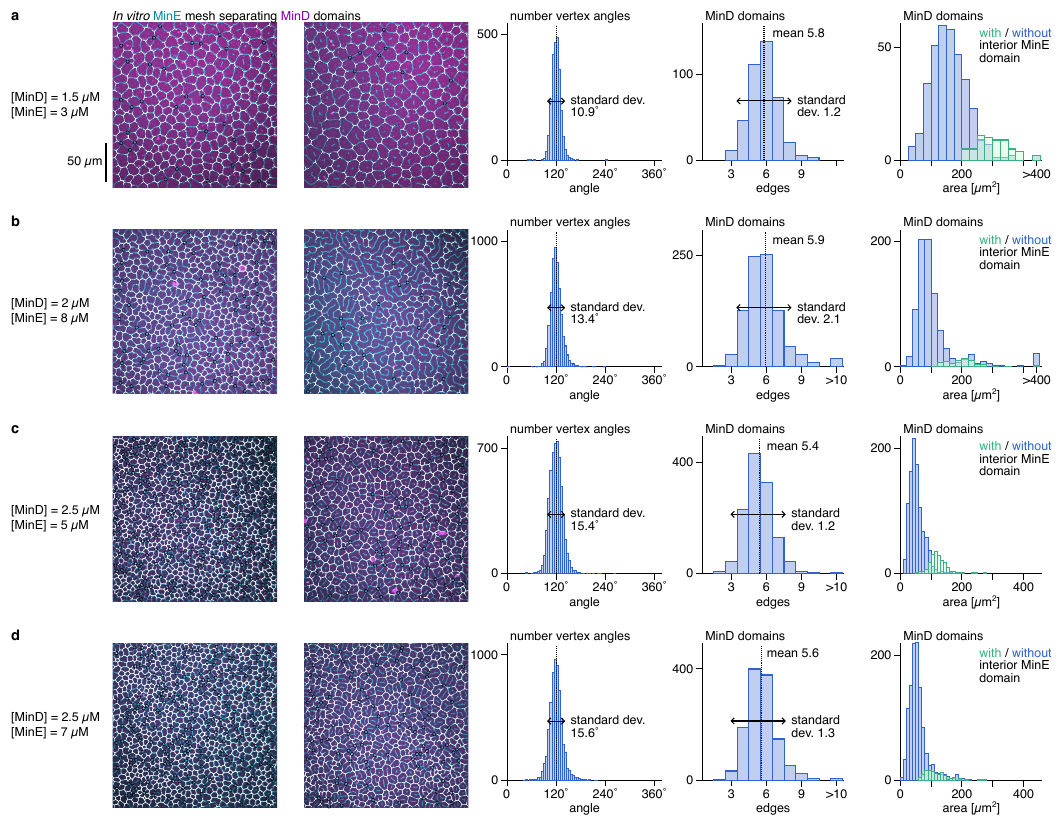}
	\caption{Vertex angles, edge numbers, and domain areas for different experimental conditions.
    The statistics of 3-fold vertices are analyzed as in Fig.~\ref{fig:4} for several snapshots under different total concentrations of MinD and MinE indicated in the figure.
    The data was recorded in the experiments for Ref.~\cite{Glock.etal2019}.
    The snapshots for each concentration combination stem from different fields of view of the same sample, which were prepared separately for the different concentration combinations.
    The fitted triple vertices are overlayed over the pattern (MinE fluorescence intensity in blue, MinD fluorescence intensity in red) in white.
    The 4-fold vertices are marked by black circles.
    \textbf{a} 
    The aggregated statistics for the snapshot analyzed in Fig.~\ref{fig:4} (first panel in this figure) and a second one are shown.
    \textbf{b} 
    The three histograms are obtained from three snapshots, two of which are shown. 
    At these total concentrations of MinD and MinE, the pattern transitions to a labyrinthine structure (second panel) which is seen in the histograms by the large domains with many edges.
    \textbf{c}, \textbf{d} 
    Two further concentration combinations (two analyzed images each) show the same statistics of Turing foams as the condition \textbf{a}.
    }
	\label{fig:e1}
\end{figure*}

\begin{figure*}
	\includegraphics{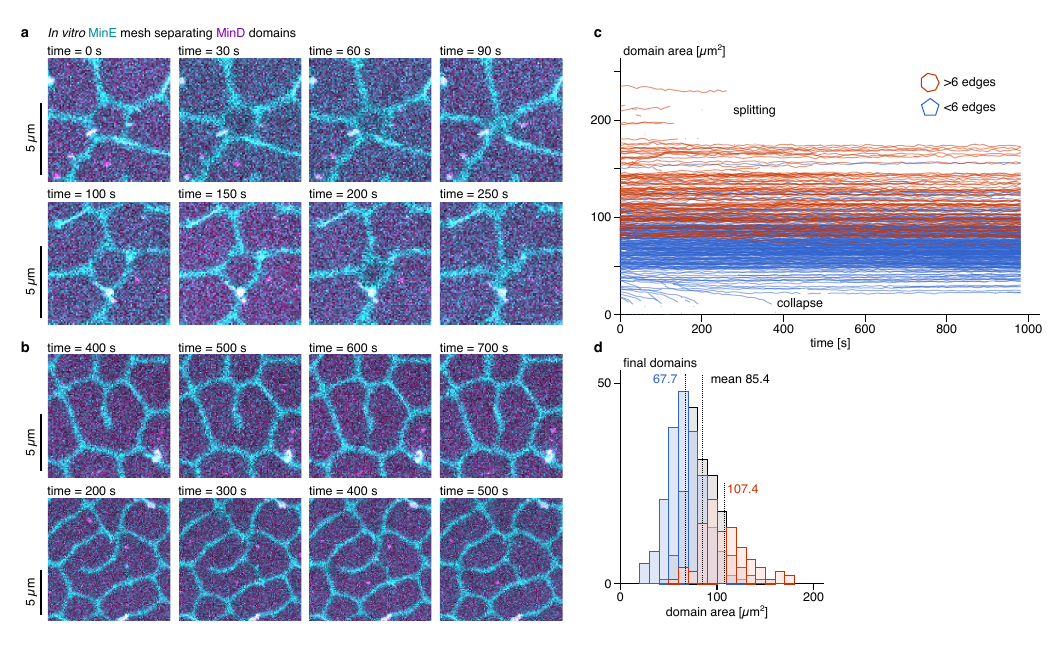}
	\caption{Domain collapse and splitting in the \textit{in vitro} Min system.
    \textbf{a} 
    The time series of single domains in the \textit{in vitro} Min protein Turing foam at the total concentrations of MinD ${[\text{MinD}] = \SI{2}{\micro M}}$ and MinE-His ${[\text{MinE{-}His}] = \SI{8}{\micro M}}$ are shown.
    The fluorescence intensity of labeled MinD (red) and MinE (blue) from the experiments for Ref.~\cite{Glock.etal2019} are shown for a domain with five (top row) and four (lower row) edges.
    Movie~5 shows the full pattern evolution recorded every 10 seconds.
    \textbf{b} 
    Two domains are shown that split by the growth of new MinE branches (cf.\ Movie~5).
    \textbf{c} 
    The domain-size trajectories for domains with less (blue) or more (red) than six edges are given (see Methods Sec.~\ref{methods:pattern-analysis}).
    \textbf{d}
    The histograms of the final domain sizes of domains with less (blue), equal to (black), and more (red) than six edges are shown in \textbf{d} (see Methods Sec.~\ref{methods:pattern-analysis}).
	}
	\label{fig:e2}
\end{figure*}

\cleardoublepage\newpage

\appendix
\renewcommand{\appendixname}{Methods}

\centerline{\textbf{Methods}}

\section{Reaction--diffusion dynamics}
\label{methods:model-systems}
In reaction--diffusion systems, $n$ different components $u_i$ spread diffusively with diffusion constants $D_i$ (assuming the absence of density dependence and cross diffusion) and undergo (nonlinear) reactions $f_i(\mathbf{u})$.
The components' evolution with time $t$ is described by coupled, nonlinear partial differential equations
\begin{equation*}
    \partial_t \mathbf{u}(\mathbf{x},t) = \mathbf{D}\nabla^2 \mathbf{u} + \mathbf{f}(\mathbf{u}) \, .
\end{equation*}
Here, the diffusion matrix is $\mathbf{D} = \mathrm{diag}(D_1,\dots,D_n)$, and we define $\mathbf{u} = (u_i)_{i=1,\dots,n}$ and $\mathbf{f} = (f_i)_{i=1,\dots,n}$.

In \emph{mass-conserving} reaction--diffusion (McRD) systems, the components constitute different (conformational) states or complexes of $N$ different types of molecules (species) whose total number is conserved.
Thus, the reactions solely describe the conversion of the molecules between the different components, e.g., membrane attachment and detachment (examples in the following subsections).
We refer to $N$ as the number of species while $n$ is the number of components of the system.
The mass-conservation constraint is formalized by introducing the stoichiometric vectors $\mathbf{s}_\alpha$ for the molecular species ${\alpha=1,\dots, N}$ (or $\mathrm{A,B},\dots$) which count how many of those molecules are contained in the different components \cite{Brauns.etal2021a}.
Consequently, the total density $\rho_\alpha(\mathbf{x},t)$ of the molecular species $\alpha$ reads
\begin{equation*}
    \rho_\alpha(\mathbf{x},t) = \mathbf{s}_\alpha\cdot\mathbf{u}(\mathbf{x},t).
\end{equation*}
For example, a system with components A, AB, and $\mathrm{B}_2$ would have the stoichiometric vectors ${s_\mathrm{A}=(1,1,0)}$ and ${s_\mathrm{B}=(0,1,2)}$.
Conservation of the total molecule numbers (``total mass'') implies that the evolution of the total densities $\rho_\alpha(\mathbf{x},t)$ follow the continuity equations \cite{Otsuji.etal2007,Ishihara.etal2007,Brauns.etal2020}
\begin{align}
    \partial_t \rho_\alpha(\mathbf{x},t) &= \mathbf{s}_\alpha\cdot\mathbf{D}\nabla^2 \mathbf{u} + \mathbf{s}_\alpha\cdot\mathbf{f}\nonumber\\
    &= \nabla^2 \left( \mathbf{s}_\alpha\cdot\mathbf{D} \mathbf{u} \right)\nonumber\\
    &\equiv D_c^\alpha \nabla^2 \eta_\alpha \, .
    \label{eq:cont-eq-general}
\end{align}
Note, that the greek index is not summed over.
The constraint
\begin{equation*}
    0 = \mathbf{s}_\alpha\cdot\mathbf{f}
\end{equation*}
describes that molecules leaving one component are added in a different component such that the individual terms $s_{\alpha i} f_i$ cancel.
As a result, the average total densities
\begin{equation*}
    \bar{\rho}_\alpha = \frac{1}{|\Omega|}\int_\Omega \mathrm{d}^d x\, \rho_\alpha(\mathbf{x},t)
\end{equation*}
are set by the initial conditions on domains $\Omega\subset \mathbb{R}^d$ (volume $|\Omega|$) with reflective or periodic boundary conditions.

Moreover, Eq.~\eqref{eq:cont-eq-general} defines the \emph{mass-redistribution potentials} $\eta_\alpha(\mathbf{x},t)$ ($\alpha=1,\dots, N$) as a sum of the different molecular states weighted by their diffusion coefficients \cite{Otsuji.etal2007,Halatek.Frey2018,Brauns.etal2021}.
The gradients of the mass-redistribution potential(s) determine the redistribution of the total densities analogously to the chemical potential(s) in Model B dynamics although they are not given as derivatives of a free energy \cite{Bray2002, Elliott.Luckhaus1991}.
The mass-redistribution potentials are normalized by one diffusion coefficient $D_c^\alpha$, e.g., the largest one whose component contributes most strongly to the redistribution of the total density.
For intracellular protein patterns, this is the cytosolic diffusion coefficient (assuming different cytosolic states of the same molecule diffuse similarly quickly).

\subsection{Multi-species McRD systems with two components per species}
\label{methods:multi-species-two-comp}
Consider a McRD systems where each molecular species $\alpha$ has only two states $m_\alpha$ and $c_\alpha$ (e.g., cytosolic and membrane-bound states) with diffusion coefficients $D_m^\alpha$ and ${D_c^\alpha>D_m^\alpha}$, respectively.
With reaction terms $f_\alpha(\mathbf{m},\mathbf{c})$ for each species, the reaction--diffusion equations for species $\alpha$ read
\begin{subequations}\label{eq:multiSpecies-twoComps}
\begin{align}
    \partial_t m_\alpha(\mathbf{x},t) &= D_m^\alpha \nabla^2 m_\alpha + f_\alpha(\mathbf{m},\mathbf{c}) \, ,\\
    \partial_t c_\alpha(\mathbf{x},t) &= D_c^\alpha \nabla^2 c_\alpha - f_\alpha(\mathbf{m},\mathbf{c}) \, .
\end{align}
\end{subequations}
The total densities ${\rho_\alpha=m_\alpha+c_\alpha}$ follow the continuity equations 
\begin{equation*}
    \partial_t \rho_\alpha 
    = D_c^\alpha \nabla^2 \eta_\alpha \, ,
\end{equation*}
with the mass-redistribution potential \begin{equation*}
 \eta_\alpha 
 = 
 c_\alpha + d_\alpha \, m_\alpha 
 \, ,
\end{equation*} 
where ${d_\alpha = D_m^\alpha/D_c^\alpha<1}$ defines the ratio of the diffusion constants.
In the intracellular context, one has ${d_\alpha\ll 1}$ because diffusion on the membrane is much slower than diffusion in the cytosol.

The mass-redistribution potential satisfies
\begin{equation}\label{eq:eta-dynamics-twoComps}
    \partial_t \eta_\alpha = (D_m^\alpha+D_c^\alpha) \nabla^2 \eta_\alpha - D_m^\alpha \nabla^2 \rho_\alpha - \Tilde{f}_\alpha
    \, ,
\end{equation}
with
\begin{equation*}
    \Tilde{f}_\alpha(\bm{\rho},\bm{\eta}) 
    = 
    (1-d_\alpha) \, f_\alpha\big(\mathbf{m}(\bm{\rho},\bm{\eta}),\mathbf{c}(\bm{\rho},\bm{\eta})\big)  \, . 
\end{equation*}

\subsubsection{PAR system}
\label{methods:par}
The \textit{C.\ elegans} PAR polarity system is modeled by McRD systems describing a membrane-bound and a cytosolic component $(m_\alpha,c_\alpha)$ for the aPAR and pPAR species labeled by $\alpha=\mathrm{A,B}$, respectively~\cite{Goehring.etal2011,Trong.etal2014}.
The reaction term includes linear attachment and detachment at the cell membrane as well as mutual detachment:
    \begin{align*}
        f_\mathrm{A} &= k_\mathrm{c}^\mathrm{A} c_\mathrm{A} - k_\mathrm{m}^\mathrm{A} m_\mathrm{A} - k_\mathrm{AB} m_\mathrm{B}^\mu m_\mathrm{A},\\
        f_\mathrm{B} &= k_\mathrm{c}^\mathrm{B} c_\mathrm{B} - k_\mathrm{m}^\mathrm{B} m_\mathrm{B} - k_\mathrm{BA} m_\mathrm{A}^\nu m_\mathrm{B}.
    \end{align*}
Commonly, the parameters are chosen symmetrically, i.e., ${k_\mathrm{c}^\mathrm{A} = k_\mathrm{c}^\mathrm{B}}$, ${k_\mathrm{m}^\mathrm{A} = k_\mathrm{m}^\mathrm{B}}$, and ${k_\mathrm{AB} = k_\mathrm{BA}}$, with the exponents for the mutual detachment term taken as ${\mu=\nu=2}$~\cite{Trong.etal2014}.
In what we call the ``symmetric PAR model", we maintain this symmetry by choosing identical diffusion coefficients on the membrane (${D_\mathrm{m}^\mathrm{A}=D_\mathrm{m}^\mathrm{B}}$)  and in the cytosol (${D_\mathrm{c}^\mathrm{A}=D_\mathrm{c}^\mathrm{B}}$).
The time and length scales are fixed by the attachment rate and cytosolic diffusion coefficient.
The standard parameter values are given in Table~\ref{tab:PAR-rates}.

We extend the PAR system to the interaction of ${N>2}$ different molecule species by introducing the mutual detachment between any pair of species.
In this case, the reaction terms read
\begin{equation}\label{eq:PAR-multi}
    f_\alpha = k_\mathrm{c}^\alpha c_\alpha - k_\mathrm{m}^\alpha m_\alpha - \sum_{\beta=1,\beta\neq\alpha}^N k_\mathrm{\alpha\beta} m_\beta^{\mu_{\alpha\beta}} m_\alpha.
\end{equation}
In the simulations we chose the exponents $\mu_{\alpha\beta}$ to be the same for all pairs of species.

\subsubsection{Inhibited attachment}
\label{methods:inhibited-attachment}
Another scenario of mutual antagonism arises when molecular species hinder each other's attachment to the membrane.
We consider the reaction terms
\begin{equation*}
    f_\alpha = \frac{k_\mathrm{c}^\alpha}{1 + \sum_{\beta=1,\beta\neq\alpha}^N k_\mathrm{\alpha\beta} m_\beta^{\mu_{\alpha\beta}} } c_\alpha - k_\mathrm{m}^\alpha m_\alpha,
\end{equation*}
which describe the attachment inhibition by a Hill term.
The strength of inhibition is set by the parameters $k_{\alpha\beta}$ and the Hill exponents are $\mu_{\alpha\beta}$.
In the simulations we employed the parameter values given in Table~\ref{tab:PAR-rates} for all pairs of species.
Parameters modified in individual simulations are specified in the respective figure captions.

\begin{table}[btp]
    \centering
    \begin{tabular}{cc|cc}
         Parameter & Value & Parameter & Value \\ \hline &&\\[-2ex]
         $D_\mathrm{m}^\mathrm{A,B}$ & $0.01$ & $k_\mathrm{AB,BA}$ & $60$\\
         $D_\mathrm{c}^\mathrm{A,B}$ & $1$ & $\mu,\nu$ & $2$\\
         $k_\mathrm{c}^\mathrm{A,B}$ & $1$ & $\Bar{\rho}_\mathrm{A,B}$ & $2$ \\
         $k_\mathrm{m}^\mathrm{A,B}$ & $1$ & &
    \end{tabular}
    \caption{Simulation parameters for the mutual detachment systems. The parameters that were changed in individual simulations are specified in the figure captions. 
    For the system with inhibited attachment, the simulations employed the same parameters, only choosing ${k_\mathrm{AB,BA}=10}$ instead.
    }
    \label{tab:PAR-rates}
\end{table}

\subsection{Min system}
\label{methods:min}
Pattern formation in the Min system has been studied based on reaction-diffusion models using various biochemical networks and rate constants~\cite{Howard.etal2001,Huang.etal2003,Loose.etal2008,Halatek.Frey2012,Denk.etal2018}.
For our analysis, we started with a core reaction network comprising biochemically validated or plausible reactions \cite{Huang.etal2003, Halatek.Frey2012}, complemented by a molecular MinE switch \cite{Park.etal2011}, which has been shown to be important in both \textit{in vivo} \cite{Ren.etal2023a} and \textit{in vitro} \cite{Denk.etal2018} settings.
However, our numerical analysis of this core reaction network revealed that achieving stationary Turing foams requires, in addition, persistent membrane binding of MinE, as previously hypothesized (see full reaction network in Fig.~\ref{fig:4}g) \cite{Loose.etal2011a, Vecchiarelli.etal2016, Denk.etal2018, Fu.etal2023}.
A parameter study and comparison of the experimental and simulated phase diagrams will be published elsewhere.
For simplicity, we simulated the patterns on a two-dimensional membrane coupled to a two-dimensional cytosolic layer.
Thus, the model disregards any effects due to a (non-uniform) bulk height \cite{Halatek.Frey2018,Gessele.etal2020,Burkart.etal2022a}.

In the cytosol, the model includes ADP-bound MinD $c_\mathrm{DD}$ and ATP-bound MinD $c_\mathrm{DT}$ states, as well as the reactive and latent MinE states $c_\mathrm{E,r}, c_\mathrm{E,l}$.
On the membrane, the model includes membrane-bound MinD $m_\mathrm{d}$, MinD--MinE complexes $m_\mathrm{de}$, and membrane-bound MinE $m_\mathrm{e}$.
The stoichiometric vectors $s_\mathrm{D,E}$ for MinD and MinE and the components ${\mathbf{u} = (c_\mathrm{DD},c_\mathrm{DT},c_\mathrm{E,r},c_\mathrm{E,l},m_\mathrm{d},m_\mathrm{de},m_\mathrm{e})}$ are
\begin{equation*}
    s_\mathrm{D} = (1,1,0,0,1,1,0),\, s_\mathrm{E} = (0,0,1,1,0,1,1).
\end{equation*}
The average total densities $\bar{\rho}_\mathrm{D},\bar{\rho}_\mathrm{E}$ of MinD and MinE added to the system are control parameters in the simulation and experiment.
The reaction terms read
\begin{align*}
    \mathbf{f}
    &=
    \begin{pmatrix}
        k_\mathrm{de} m_\mathrm{de} -\lambda c_\mathrm{DD}\\
        \lambda c_\mathrm{DD} - (k_\mathrm{D}+k_\mathrm{dD} m_\mathrm{d}) c_\mathrm{DT}\\
        k_\mathrm{e} m_\mathrm{e} - \mu c_\mathrm{E,r} - k_\mathrm{dE,r} m_\mathrm{d} c_\mathrm{E,r}\\
        \mu c_\mathrm{E,r} - k_\mathrm{dE,l} m_\mathrm{d} c_\mathrm{E,l}\\
        (k_\mathrm{D}+k_\mathrm{dD} m_\mathrm{d}) c_\mathrm{DT} - (k_\mathrm{dE,r} c_\mathrm{E,r} + k_\mathrm{dE,l} c_\mathrm{E,l} + k_\mathrm{ed} m_\mathrm{e}) m_\mathrm{d}\\
        (k_\mathrm{dE,r} c_\mathrm{E,r} + k_\mathrm{dE,l} c_\mathrm{E,l} + k_\mathrm{ed} m_\mathrm{e}) m_\mathrm{d} - k_\mathrm{de} m_\mathrm{de}\\
        k_\mathrm{de} m_\mathrm{de} - k_\mathrm{e} m_\mathrm{e} - k_\mathrm{ed} m_\mathrm{d} m_\mathrm{e}
    \end{pmatrix}.
\end{align*}
The chosen rates, diffusion coefficients, and average densities are given in Tab.~\ref{tab:Min-rates}.

\begin{table*}[btp]
    \centering
    \begin{tabular}{cccc}
         Parameter & Value & Unit & Process \\\hline
         $D_\mathrm{D}$ & $60$ & $\si{\micro m^2 s^{-1}}$ & diffusion coefficient for $c_\mathrm{DD},c_\mathrm{DT}$\\
         $D_\mathrm{E}$ & $60$ & $\si{\micro m^2 s^{-1}}$ & diffusion coefficient for $c_\mathrm{E,r},c_\mathrm{E,l}$\\
         $D_\mathrm{d}$ & $0.1$ & $\si{\micro m^2 s^{-1}}$ & diffusion coefficient for $m_\mathrm{d}$\\
         $D_\mathrm{de}$ & $0.1$ & $\si{\micro m^2 s^{-1}}$ & diffusion coefficient for $m_\mathrm{de}$\\
         $D_\mathrm{e}$ & $0.001$ & $\si{\micro m^2 s^{-1}}$ & diffusion coefficient for $m_\mathrm{e}$\\
         $k_\mathrm{D}$ & $0.15$ & $\si{s^{-1}}$ & MinD membrane-attachment rate\\
         $k_\mathrm{dD}$ & $0.00075$ & $\si{\micro m^2 s^{-1}}$ & MinD self-recruitment rate\\
         $k_\mathrm{dE,r}$ & $0.75$ & $\si{\micro m^2 s^{-1}}$ & recruitment rate of reactive MinE by MinD\\
         $k_\mathrm{dE,l}$ & $0.5\cdot 10^{-5}$ & $\si{\micro m^2 s^{-1}}$ & recruitment rate of latent MinE\\
         $k_\mathrm{ed}$ & $0.1$ & $\si{\micro m^2 s^{-1}}$ & recruitment rate of membrane-bound MinE\\
         $k_\mathrm{de}$ & $1$ & $\si{s^{-1}}$ & ATP-hydrolysis rate\\
         $k_\mathrm{e}$ & $0.001$ & $\si{s^{-1}}$ & MinE membrane-detachment rate\\
         $\lambda$ & $5$ & $\si{s^{-1}}$ & nucleotide-exchange rate\\
         $\mu$ & $20$ & $\si{s^{-1}}$ & rate of MinE's conformational change\\
         $\bar{\rho}_\mathrm{D}$ & 8000 & $\si{\micro m^{-2}}$ & average total MinD density\\
         $\bar{\rho}_\mathrm{E}$ & 4000 & $\si{\micro m^{-2}}$ & average total MinE density\\
         $L$ & 100 & $\si{\micro m}$ & edge length of the quadratic simulation domain\\
    \end{tabular}
    \caption{The parameters are given for the Min model chosen to simulate the mesh pattern.
    The rate choice is based on the values found to describe the density phase diagram of the Min system \textit{in vivo} \cite{Ren.etal2023a}.
    The MinE membrane diffusion coefficient $D_\mathrm{e}$ is set to a small value following Ref.~\cite{Fu.etal2023}. 
    }
    \label{tab:Min-rates}
\end{table*}

\subsection{FIS system}
\label{methods:fis}
We show in simulations that Turing foams also arise in the classical ferrocyanide-iodate-sulfite (FIS) reaction \cite{Lee.etal1993,Lee.Swinney1995}.
In the simulation, the four-species G\'asp\'ar--Showalter model with the parameters employed in Ref.~\cite{Lee.Swinney1995}, Fig.~17, was used.
This model includes the four species $x,y,z,a$ with diffusion coefficients $D_x > D_y=D_z=D_a$.
The parameters are given in Tab.~\ref{tab:FIS-rates}.
This model is not mass-conserving because the reactants are supplied with a rate $k_0$, modeling the exchange between the gel matrix and a reservoir.
The reaction terms are \cite{Lee.Swinney1995}
\begin{align*}
    \mathbf{f} 
    &=
    \begin{pmatrix}
        - k_{-1} x + k_1 a y  - k_2 x - k_4 z x - k_0 x\\
        k_{-1} x -k_1 a y  + k_2 x - 2 k_3 y^2 + 3 k_4 z x + k_0 (y_0-y)\\
        k_3 y^2 - k_4 z x - k_5 z - k_0 z\\
        -k_1 a y + k_{-1} x + k_0 (a_0 -a)
    \end{pmatrix}.
\end{align*}

\begin{table}[btp]
    \centering
    \begin{tabular}{ccc}
         Parameter & Value & Unit \\\hline
         $D_\mathrm{x}$ & $2\cdot 10^{-3}$ & $\si{\milli m^2 s^{-1}}$\\
         $D_\mathrm{y}$ & $10^{-3}$ & $\si{\milli m^2 s^{-1}}$\\
         $D_\mathrm{z}$ & $10^{-3}$ & $\si{\milli m^2 s^{-1}}$\\
         $D_\mathrm{a}$ & $10^{-3}$ & $\si{\milli m^2 s^{-1}}$\\
         $k_1$ & $5\cdot 10^{10}$ & $\si{M^{-1} s^{-1}}$\\
         $k_{-1}$ & $8.1\cdot 10^3$ & $\si{s^{-1}}$\\
         $k_2$ & $6\cdot 10^{-2}$ & $\si{s^{-1}}$\\
         $k_3$ & $7.5\cdot 10^4$ & $\si{M^{-1} s^{-1}}$\\
         $k_4$ & $2.3\cdot 10^9$ & $\si{M^{-1} s^{-1}}$\\
         $k_5$ & $2.4$ & $\si{s^{-1}}$\\
         $k_0$ & $1.4\cdot 10^{-2}$ & $\si{s^{-1}}$\\
         $y_0$ & $9\cdot 10^{-3}$ & $\si{M}$\\
         $a_0$ & $8.93\cdot 10^{-3}$ & $\si{M}$\\
         $L$ & $\sqrt{2}$ & $\si{\milli m}$
    \end{tabular}
    \caption{The parameters are given for the FIS model chosen to simulate the mesh pattern on a quadratic domain of length $L$ with no-flux boundary conditions.
    The rates follow Ref.~\cite{Lee.Swinney1995}.
    In the simulation, the length unit is chosen as $1/(10\sqrt{2})\si{\milli m}$ such that the simulation domain has length $20$ and the diffusion constants are $D_\mathrm{x}=0.4$ and $D_\mathrm{y,z,a}=0.2$ in the simulation units.
    }
    \label{tab:FIS-rates}
\end{table}

\section{Numerical simulation}
\label{methods:simulation}

We performed simulations of the different models on a two-dimensional domain with no-flux boundary conditions using COMSOL Multiphysics 6.1.
For Fig.~\ref{fig:6}, the simulation was performed on a spherical, closed surface.
These simulations employ a finite-element discretization on a triangular mesh with linear Lagrange elements.

To estimate the threshold size of splitting in the simulated Min protein Turing foam (dashed line in Fig.~\ref{fig:5}h,i), a single, circular MinD domain surrounded by a MinE branch at the boundary of the simulation domain with a radius of ${R_0=\SI{5}{\micro m}}$ was simulated for ${10^5 \, \si{s}}$.
Instead of adiabatically increasing the radius of the simulation domain (and thereby the radius of the MinD domain), and observing the threshold size at which an internal MinE domain grows, all diffusion coefficients were adiabatically decreased with time $t$ by the common factor ${f = 1+t/10^5\si{s}}$.
Thereby, the length unit was rescaled, resulting in an increasing radius ${R(t)=R_0 f^{-1/2}}$.
The pattern was recorded every $\SI{500}{s}$, and the size was determined when an internal MinE domain of comparable density with the surrounding MinE branch had formed.

The parameter dependence of the non-equilibrium Gibbs--Thomson relation of a radially symmetric interface in the symmetric PAR system in Fig.~\ref{fig:1}g,h was determined using numerical pseudo-arclength continuation implemented in Mathematica previously \cite{Brauns.etal2020,Brauns.etal2021}:
Assuming radial symmetry, the stationary reaction--diffusion equations were converted into an algebraic system of equations by a finite-differences discretization in the radial direction.
The algebraic steady state equations are solved by Newton iteration.

The configuration files for the COMSOL simulations and the Mathematica notebooks are available at \url{https://github.com/henrikweyer/Turing-foams}.

\section{Pattern analysis}
\label{methods:pattern-analysis}
The statistical analysis of the simulation and experimental data was performed using Mathematica 13.1 and is explained in detail in the SI, Secs.~6,~7.
To measure the junctions and vertices in simulation and experimental images, the skeleton network of the interfaces or branches was determined using Mathematica's morphological algorithms.
Its branch points give rough junction or vertex positions.
In a second step, a junction or vertex model was fitted to the density fields and their gradients to determine precise junction and vertex angles.
The domain areas were determined from the skeleton network.
The Mathematica notebooks are available at \url{https://github.com/henrikweyer/Turing-foams}.

\section{Effective interfacial tension}
\label{methods:eff-interfacial-tension}

Here, we briefly elaborate on the arguments and assumptions leading to the curvature-induced cytosolic shift,  Eq.~\eqref{eq:cytosolic-shift}, and discuss its extension to the shift $\delta\eta_\mathrm{stat}$ of the mass-redistribution potential.

Let us consider one protein species that can be described by a cytosolic protein density $c(\mathbf{x},t)$ and its density $m(\mathbf{x},t)$ on the membrane.
Multi-component and multi-species systems are discussed in the SI, Secs.~2 and 3.
Given the attachment rate $a(m)$ and detachment rate $d(m)$, the reaction kinetics is described by $f(m,c) = a(m) \, c- d(m) \, m$.
Here, we allow the rates to depend on the membrane density to describe, for example, recruitment onto the membrane and enzymatic detachment.
No dependence of the rates on the cytosolic density is considered because the cytosolic concentration is typically small compared to the protein density in membrane regions where proteins accumulate, rendering the dependence on the cytosolic density weak in comparison.
One has ${\partial_c f = a(m) > 0}$ because the attachment flux increases with the number of cytosolic proteins that can attach.
Therefore, in the main text we assume ${\partial_c f_\mathrm{a}>0}$.
In general multi-component, multi-species systems, this derivative may become negative, which may give rise to interface instabilities.
It is an interesting future research direction for which types of interaction networks the sign changes and these instabilities arise.
In contrast, the detachment term $d(m) \, m$ is independent of the cytosolic concentration.
This detachment term dominates in the detachment region such that we assume ${\partial_c f_\mathrm{d} \approx 0}$.
Therefore, the change in the cytosolic concentration only affects $f_\mathrm{a}$.

At a stationary interface, the attachment and detachment fluxes must balance; otherwise, proteins are lost from one species.
With the attachment and detachment fluxes $f_\mathrm{a,d}$ in the attachment and detachment zone, respectively, this reactive turnover balance reads ${A_\mathrm{a}(\kappa) f_\mathrm{a}(\kappa) = A_\mathrm{d}(\kappa) f_\mathrm{d}(\kappa)}$.
At a straight interface, ${\kappa = 0}$, the areas $A_\mathrm{a,d}$ are equal such that the condition simplifies to ${f_\mathrm{a}(0) = f_\mathrm{d}(0)}$.
Taking the difference of the balance equations for the weakly curved and the straight interface, one has to first order in $\kappa\,\ell_\mathrm{int}$
\begin{equation*}
    [A_\mathrm{a}(\kappa)-A_\mathrm{d}(0)] f_\mathrm{d}(0) + A_\mathrm{a}(0) \delta c \partial_c f_\mathrm{a} \approx [A_\mathrm{d}(\kappa)-A_\mathrm{d}(0)] f_\mathrm{d}(0).
\end{equation*}
Here, we used the assumption ${f_\mathrm{d}(\kappa) \approx f_\mathrm{d}(0)}$.
Equation~\eqref{eq:cytosolic-shift} then follows from the geometric construction ${[A_\mathrm{d}(\kappa)-A_\mathrm{a}(\kappa)]/A_\mathrm{d}(0)\approx \ell_\mathrm{int} \kappa}$ (Fig.~\ref{fig:1}e).
The approximation ${f_\mathrm{d}(\kappa) \approx f_\mathrm{d}(0)}$ not only assumes that the detachment flux is independent of the cytosolic density but also that changes in the interface (membrane-density) profile are negligible for the non-equilibrium Gibbs--Thomson relation at weakly-curved interfaces.
Moreover, only considering protein fluxes induced by the cytosolic shifts neglects membrane diffusion due to curvature-induced gradients in the membrane densities.

To include both effects, we turn to an explicit calculation for the shift in the mass-redistribution potential ${\eta = c + D_m/D_c m \equiv c + d m}$ following the Supplemental Material of Ref.~\cite{Brauns.etal2021} and Ref.~\cite{Roth.etal}.
Note that, in the biologically relevant limit ${D_m \ll D_c}$, one has ${\eta \approx c}$.
First, we observe that the continuity equation Eq.~\eqref{eq:cont-eq-general} implies that the mass-redistribution potential is exactly constant for (flux-free) stationary patterns.
Different shifts of the stationary redistribution potential in different regions of the pattern due to different interfacial curvatures give (within a quasi steady state approximation) rise to gradients in $\eta$ that drive the redistribution of proteins via the continuity equation Eq.~\eqref{eq:cont-eq-general}.

To derive the local steady-state shift $\delta \eta$, we focus on one region of the interface with (approximately) constant curvature and describe the curved interface in polar coordinates.
With the radial coordinate $r$, the stationary interface $\rho_\mathrm{stat}(r)$ has to fulfill [cf.\ Eq.~\eqref{eq:eta-dynamics-twoComps}]
\begin{equation}\label{eq:profile-eq-2c-2D}
    0 = \frac{D_m}{r} \partial_r \rho_\mathrm{stat} + D_m \partial_r^2 \rho_\mathrm{stat} + \Tilde{f}(\rho_\mathrm{stat},\eta_\mathrm{stat}).
\end{equation}
We denote the stationary pattern for vanishing interface curvature by $\rho_\mathrm{stat}^0(r)$ with the stationary mass-redistribution potential $\eta_\mathrm{stat}^0$.
By multiplying by $\partial_r \rho_\mathrm{stat}$, integrating over the interface region, and neglecting small remaining gradients in the high- and low-density plateaus far from the (isolated) interface at the integration boundaries ${R\pm b}$ (${R\gg b\gg \ell_\mathrm{int}}$), one has
\begin{equation*}
    0 \approx \int_{R-b}^{R+b} \mathrm{d}r\frac{(\partial_r \rho_\mathrm{stat})^2}{r} + \int_{\rho_\mathrm{stat}(R-b)}^{\rho_\mathrm{stat}(R+b)}\mathrm{d}\rho\, \Tilde{f}(\rho,\eta_\mathrm{stat})
\end{equation*}
with the curvature radius ${R=\kappa^{-1}}$.
This integral equation represents the reactive turnover balance that must be fulfilled at stationary interfaces.
Interface curvature leads to a finite first integral, while the second integral over the reaction term (attachment minus detachment terms) must vanish exactly at straight interfaces.

Because the term $(\partial_r \rho_\mathrm{stat})^2$ is localized to the interface region, and the width of a weakly curved interface fulfills $\ell_\mathrm{int}\ll R$, the first term gives $\kappa \int_{R-b}^{R+b} \mathrm{d}r\, (\partial_r \rho_\mathrm{stat}^0)^2 + \mathcal{O}((\kappa\ell_\mathrm{int})^2)$.
Moreover, one has ${\Tilde{f}(\rho_\mathrm{stat}(R\pm b),\eta_\mathrm{stat})\approx 0}$ because all gradients approximately vanish far from the interface region in the pattern plateaus [cf.\ Eq.~\eqref{eq:profile-eq-2c-2D}].
Thus, the integration boundaries in the second integral can be replaced by the plateau densities $\rho_\mathrm{stat}^0(R\pm b) \equiv \rho_\mp$ up to corrections of the order ${\mathcal{O}((\kappa\ell_\mathrm{int})^2)}$.
Linearizing in the shift $\delta\eta = \eta_\mathrm{stat}-\eta_\mathrm{stat}^0$, one has
\begin{equation*}
    \delta\eta \int_{\rho_-}^{\rho_+}\mathrm{d}\rho\, \partial_\eta\Tilde{f}(\rho,\eta_\mathrm{stat}^0) \approx \kappa \int_{R-b}^{R+b} \mathrm{d}r\, (\partial_r \rho_\mathrm{stat}^0)^2.
\end{equation*}
All changes in the interface profile ${\rho_\mathrm{stat}-\rho_\mathrm{stat}^0}$ are negligible because they only give rise to higher-order corrections.
The integral on the left-hand side represents an average reaction rate that can be understood as an attachment rate given that ${\eta\approx c}$ for the biologically relevant limit ${D_m \ll D_c}$.
For the stability of (sharp) interfaces, it is required that this averaged reaction rate is positive \cite{Brauns.etal2021,Weyer.etal2023} such that ${\delta \eta \sim \kappa}$.
The dependence of the integrals on the interface width $\ell_\mathrm{int}$ is discussed in SI Sec.~3.5.5.

Concerning multi-component systems, in SI Sec.~2, we show that a shift ${\delta\eta_\mathrm{stat}\sim \ell_\mathrm{int}\kappa}$ arises under conditions on the monotonicity of the component densities at the interface if the slow-diffusing components induce the pattern-forming feedback, i.e., they increase their production (attachment) at high density and lead to net detachment at low densities while the rates depend linearly (monotonously increasing) on the fast-diffusing components.
In SI Sec.~3, we give explicit expressions for the effective interfacial tension for systems with two components per protein species.
In SI Sec.~4, the non-equilibrium Neumann law is derived using methods developed for multi-component Cahn--Hilliard systems \cite{Cahn.Hilliard1958,Eyre1993,Elliott.Luckhaus1991,Bronsard.Reitich1993,Bronsard.etal1998,Garcke.Novick-Cohen2000}.

\section{Von Neumann law in Turing foams}
\label{methods:von-Neumann}
To develop the von Neumann law for Turing foams, consider a single branch of the mesh (Fig.~\ref{fig:5}e).
In the following, we focus on MinD redistribution because its interface curvature dominates the dynamics due to MinD's larger interface width compared to MinE in the simulation [cf.~Eq.~\eqref{eq:cytosolic-shift}].
Given the branch is narrow compared to its radius of curvature ${R = 1/\kappa}$, its two interfaces (inner and outer interface with respect to the branch curvature) have curvature $\kappa$.
The steady-state cytosolic MinD concentration is altered by an amount proportional to ${\delta c_\mathrm{stat}\sim\pm\kappa}$ at the inner and outer interfaces, respectively.
Those shifts create a cytosolic density gradient between the two branch interfaces redistributing MinD proteins from the inner towards the outer interface (inner to outer MinD domain with respect to the branch curvature) and resulting in a curvature-induced normal velocity of the branch ${v_\kappa \sim -\kappa}$.
Thus, with the $\ang{120}$ angles at the vertices, the Gauss-Bonnet theorem gives for the temporal change of the domain area ${A_n=|\Omega_n|}$ of a single domain $\Omega_n$ with $n$ edges (corners)
\begin{equation*}
    \partial_t A_n = \int_{\partial\Omega_n}\mathrm{d}s\, v_\kappa \sim -\int_{\partial\Omega_n}\mathrm{d}s\, \kappa =\frac{\pi}{3} (n-6)
    \, .
\end{equation*}
Here, we have assumed that the width of the vertices is negligible compared to the branch lengths in the boundary integral.


\medskip

\centerline{\textbf{Data Availability}}

Data generated in this study is available from the corresponding author upon reasonable request.
The experimental data analyzed in this study was obtained from the authors of Ref.~\cite{Glock.etal2019}.
\medskip

\centerline{\textbf{Code Availability}}

The analysis code and exemplary setup codes to perform the numeric simulations of this study are available under \url{https://github.com/henrikweyer/Turing-foams}.
\medskip

\centerline{\textbf{Acknowledgements}}

We thank Philipp Glock and Petra Schwille for providing us with previously unpublished data on the mesh patterns obtained in the experiments for Ref.~\cite{Glock.etal2019}, and Fridtjof Brauns and Marius Predel for inspiring discussions.
Also, we thank Jona Kayser, Beatrice Nettuno, Steffen Rulands, and Davide Toffenetti for valuable feedback on the manuscript.
This work was funded by the Deutsche Forschungsgemeinschaft (DFG, German Research Foundation) through the Excellence Cluster ORIGINS under Germany’s Excellence Strategy – EXC-2094 – 390783311, 
the European Union (ERC, CellGeom, project number 101097810), and the Chan-Zuckerberg Initiative (CZI).
\medskip

\centerline{\textbf{Author contributions}}
H.W., T.A.R., and E.F. conceived the study.
H.W. developed the theory and performed the analysis.
T.A.R. assisted in the analysis.
H.W., T.A.R., and E.F. discussed and interpreted the results and wrote the manuscript.
\medskip

\centerline{\textbf{Competing interests}}
The authors declare no competing interests.
\medskip

\cleardoublepage\newpage

\input{Turing_Foam.bbl}

\end{document}


\maketitle

\tableofcontents

\clearpage

\newpage

\section{Model systems}
For the convinience of the reader, we repeat the general definition of the McRD dynamics here.
Reaction--diffusion systems with $n$ different components $u_i$ with diffusion constants $D_i$, which undergo (nonlinear) reactions described by the reaction terms $f_i(\mathbf{u})$, read
\begin{equation}\label{eq:RD-general}
    \partial_t \mathbf{u}(\mathbf{x},t) = \mathbf{D}\nabla^2 \mathbf{u} + \mathbf{f}(\mathbf{u}).
\end{equation}
Here, $t$ denotes time, the diffusion matrix is ${\mathbf{D} = \mathrm{diag}(D_1,\dots,D_n)}$, and we define ${\mathbf{u} = (u_i)_{i=1,\dots,n}}$ and ${\mathbf{f} = (f_i)_{i=1,\dots,n}}$.
The system domain is chosen as $\Omega\subset \mathbb{R}^d$ with no-flux or periodic boundary conditions.
We focus on two-dimensional patterns ($d=2$) to model patterns on the cell membrane or a supported lipid bilayer.\footnote{
Here, we consider a two-dimensional cytosolic layer close to the membrane.
This simplification of the geometry excludes any effects due to a non-uniform bulk height induced by the varying bulk-boundary ratio \cite{Halatek.etal2018,Gessele.etal2020,Burkart.etal2022a}, an important direction for the further analysis of intracellular protein systems.}

We consider \emph{mass-conserving} reaction--diffusion (McRD) systems.
In these, the components constitute states of $N$ different molecules whose total numbers are conserved.
One can therefore introduce the stoichiometric vectors $\mathrm{s}_\alpha$ for the molecular species $\alpha = 1,\dots, N$ which count how many of those molecules are contained in the different components \cite{Brauns.etal2021a}.
Consequently, the total density $\rho_\alpha(\mathbf{x},t)$ of the molecular species $\alpha$ within the system reads
\begin{equation}
    \rho_\alpha(\mathbf{x},t) = \mathbf{s}_\alpha\cdot\mathbf{u}(\mathbf{x},t) \, .
\end{equation}
Here and in the following, we always denote the vector scalar product by a dot while the matrix and matrix-vector products are written without a multiplication symbol.
Local mass conservation, i.e., that no new proteins are produced and no proteins are degraded anywhere in the system, plays out as the constraint 
\begin{equation}\label{eq:mass-cons-constraint}
    0 = \mathbf{s}_\alpha\cdot\mathbf{f} \, .
\end{equation}
The minimal example is the two-component McRD system describing one protein species ($N=1$) that has the two states $m$ and $c$ ($n=2$), e.g., a membrane-bound and a cytosolic state \cite{Edelstein-Keshet.etal2013,Otsuji.etal2007,Halatek.Frey2018,Brauns.etal2020,Frey.Brauns2022}. It reads 
\begin{subequations}
\label{eq:2cMcRD}
\begin{align}
    \partial_t m(\mathbf{x},t) &= D_m \nabla^2 m + f(m,c) \, ,\\
    \partial_t c(\mathbf{x},t) &= D_c \nabla^2 c - f(m,c) \, .
\end{align}
\end{subequations}
The single stoichiometric vector is $\mathbf{s}_1 = (1,1)^\mathsf{T}$.

With this local mass-conservation constraint, the total densities $\rho_\alpha(\mathbf{x},t)$ follow the continuity equations \cite{Otsuji.etal2007,Ishihara.etal2007,Halatek.Frey2018,Brauns.etal2020}
\begin{align}
    \partial_t \rho_\alpha(\mathbf{x},t) &= \mathbf{s}_\alpha\cdot\mathbf{D}\nabla^2 \mathbf{u} + \mathbf{s}_\alpha\cdot\mathbf{f}\nonumber\\
    &= \nabla^2 \mathbf{s}_\alpha\cdot\mathbf{D} \mathbf{u}\nonumber\\
    &\equiv D_c^\alpha \nabla^2 \eta_\alpha.\label{eq:cont-eq-general}
\end{align}
Note, that one does not sum over the Greek index $\alpha$.
The last equation defines the mass-redistribution potentials ($\alpha=1,\dots,N$)~\cite{Otsuji.etal2007,Ishihara.etal2007,Brauns.etal2020}
\begin{equation}
    \eta_\alpha 
    = 
    \frac{1}{D_c^\alpha} \, 
    \mathbf{s}_\alpha \cdot 
    \mathbf{D}\mathbf{u},
\end{equation}
where one diffusion coefficient $D_c^\alpha$ each was chosen as normalization.
The gradients of the mass-redistribution potentials drive the redistribution of the protein mass, i.e., the currents in the total densities $\rho_\alpha$.

Importantly, on domains $\Omega\subset \mathbb{R}^d$ (volume $|\Omega|$) with reflective or periodic boundary conditions the average total densities 
\begin{equation}
    \bar{\rho}_\alpha = \frac{1}{|\Omega|}\int_\Omega \mathrm{d}^d x\, \rho_\alpha(\mathbf{x},t)
\end{equation}
are set by the initial conditions.

\subsection{Multi-species McRD systems with two components per species}

In McRD systems where each molecular species $\alpha$ is modeled by two states $m_\alpha$ and $c_\alpha$ (e.g., fast-diffusing cytosolic and slow-diffusing membrane-bound states in the context of intracellular pattern formation) with diffusion coefficients $D_m^\alpha$ and $D_c^\alpha>D_m^\alpha$, respectively, the reaction--diffusion equations for species $\alpha$ read
\begin{subequations}\label{eq:multiSpecies-twoComps}
    \begin{align}
        \partial_t m_\alpha(\mathbf{x},t) &= D_m^\alpha \nabla^2 m_\alpha + f_\alpha(\mathbf{m},\mathbf{c}) \, ,\\
        \partial_t c_\alpha(\mathbf{x},t) &= D_c^\alpha \nabla^2 c_\alpha - f_\alpha(\mathbf{m},\mathbf{c}) \, .
    \end{align}
\end{subequations}
Such systems represent $N$ coupled two-component McRD systems Eq.~\eqref{eq:2cMcRD}.
Conservation of mass is ensured by the fact that the same reaction term $f_\alpha(\mathbf{m},\mathbf{c})$ occurs in both equations with the opposite sign.
Therefore, the total densities ${\rho_\alpha=m_\alpha+c_\alpha}$ follow the continuity equations 
\begin{equation}
\label{eq:cont-eq-2c}    
    \partial_t \rho_\alpha 
    = D_c^\alpha \nabla^2 \eta_\alpha \, ,
\end{equation}
with the mass-redistribution potentials
\begin{equation}
 \eta_\alpha 
 = 
 c_\alpha + d_\alpha \, m_\alpha 
 \, ,
\label{eq:mass-redistribution-potential} 
\end{equation} 
where we choose ${d_\alpha = D_m^\alpha/D_c^\alpha}<1$.

The mass-redistribution potentials satisfy the equations
\begin{equation}\label{eq:eta-dynamics}
    \partial_t \eta_\alpha = (D_m^\alpha+D_c^\alpha) \nabla^2 \eta_\alpha - D_m^\alpha \nabla^2 \rho_\alpha - \Tilde{f}_\alpha
    \, ,
\end{equation}
which defines the modified reaction term 
\begin{equation}\label{eq:mod-reac-term}
    \Tilde{f}_\alpha(\bm{\rho},\bm{\eta}) 
    = 
    (1-d_\alpha) \, f_\alpha\big(\mathbf{m}(\bm{\rho},\bm{\eta}),\mathbf{c}(\bm{\rho},\bm{\eta})\big)  \, . 
\end{equation}

In the next section Sec.~\ref{sec:multi-comp}, we estimate the effective interfacial tension in multi-component McRD systems.
Afterwards, we derive non-equilibrium Gibbs--Thomson relations and explicit formulas for the effective interfacial tension in multi-species McRD systems with two component per species.
In Sec.~\ref{sec:gen-Neumann-law}, the non-equilibrium Neumann law is derived.
In the last sections Secs.~\ref{sec:random-angles},~\ref{sec:junction-analysis},~\ref{sec:foam-analysis}, the statistical analysis of junctions and vertices is discussed.

\section{Effective interfacial tension in multi-component McRD systems}
\label{sec:multi-comp}

In this section, we discuss that a positive effective interfacial tension ${\sigma \sim \ell_\mathrm{int}}$ for interfaces in a protein species with multiple components follows from assumptions about the reaction kinetics.
We show that the arrangement of attachment and detachment zones discussed in the main text Sec.~``Effective interfacial tension'' follows.
Their local area change at weakly curved interfaces is then shown to introduce a positive effective interfacial tension.

To define the effective interfacial tension, one assumes that the reaction--diffusion dynamics results in patterns with well-defined (sharp) interfaces separating high- and low-density regions.
We then consider weakly-curved stationary interfaces and calculate the shift in the mass-redistribution potential relative to the straight stationary interface, that is, a non-equilibrium Gibbs--Thomson relation.
From this relation, we define the effective interfacial tension.
If the movement of the interfaces is slow compared to all other dynamic processes in the system, these changes in the stationary mass-redistribution potential induce gradients between differently curved interface regions of the pattern that drive the redistribution of mass between those (quasi-steady-state approximation).

The argument for the positive effective interfacial tension given in this section rests on the following assumptions.
First, we assume that the slow-diffusing (membrane-bound) species induce the pattern-forming feedback that results in the formation of attachment and detachment zones which stabilize the interfaces.
These slow-diffusing species show steep density gradients at the pattern interface and determine the direction of the interface gradient of the total protein density.
Secondly, fast-diffusing (cytosolic) components are assumed to show weak reverse gradients (cf.\ main text Fig.~1c and Fig.~\ref{fig:attachment-detachment-balance}b).
We assume that the total density and the fast-diffusing components vary monotonously across the interface.
Moreover, we assume that these fast-diffusing components do not induce feedback.
These assumptions are motivated by the attachment--detachment kinetics of intracellular protein systems \cite{Halatek.etal2018}.
They must to be checked for the specific system under consideration.

\subsection{Reaction-diffusion equations and non-equilibrium steady states}

Consider a general McRD system, Eq.~\eqref{eq:RD-general}, with $N$ distinct protein species denoted by ${\alpha=1,\dots, N}$.
Each protein species $\alpha$ has ${i = 1, \dots,M_\alpha}$ different components, i.e., protein states, with the corresponding densities $u^\alpha_i(\mathbf{x},t)$ and diffusion coefficients $D^\alpha_i$.
For the component $i$ of species $\alpha$, the reaction term $f^\alpha_i(\mathbf{u})$ may depend on all components $u^\alpha_i$ of all species.
The reaction--diffusion equations read [cf.\ Eq.~\eqref{eq:RD-general}]
\begin{equation}\label{eq:McRD}
    \partial_t u^\alpha_i = D^\alpha_i \nabla^2 u^\alpha_i + f^\alpha_i(\mathbf{u}) \, .
\end{equation}
Mass conservation of each protein species implies ${\sum_{i=1}^{M_\alpha} f^\alpha_i = 0}$. As a consequence, the total protein densities ${\rho_\alpha=\sum_i u^\alpha_i}$ fulfill the continuity equations [cf.\ Eq.~\eqref{eq:cont-eq-general}]
\begin{equation}\label{eq:cont-eq}
    \partial_t \rho_\alpha = \nabla^2 \sum_i D^\alpha_i u^\alpha_i \equiv \nabla^2 \Tilde{\eta}_\alpha \, , 
\end{equation}
where we defined the (modified) mass-redistribution potentials $\Tilde{\eta}_\alpha$.

The continuity equations imply that the (modified) mass-redistribution potentials ${\Tilde{\bm\eta} = \Tilde{\bm\eta}^0}$ are constants for stationary patterns $\mathbf{u}_\mathrm{stat}$ on a domain with no-flux or periodic boundary conditions.
Moreover, the stationary pattern must fulfill for all components $i$ of each species $\alpha$
\begin{equation}\label{eq:reactive-turnover-balance}
    0 = \int_\Omega\mathrm{d}^dx\, f_i^\alpha(\mathbf{u}_\mathrm{stat}(\mathbf{x})) \, ,
\end{equation}
which follows from integrating Eq.~\eqref{eq:McRD} in steady state on a $d$-dimensional domain $\Omega$ with no-flux or periodic boundary conditions.
The solvability condition, Eq.~\eqref{eq:reactive-turnover-balance}, ensures that, in the steady state, the total reactive flux between the components integrated over the whole domain vanishes, and no component has a net gain or loss of molecules.
This condition, known as the \textit{reactive turnover balance}, determines the values $\Tilde{\eta}^0_\alpha$, serving as parameters of the stationary pattern profile \cite{Brauns.etal2020,Frey.Brauns2022}.
A more detailed analysis of this relation is given in the next section Sec.~\ref{sec:multi-species-two-components} for systems with two components per species.

\begin{figure}[btp]
    \centering
    \makebox[\textwidth][c]{\includegraphics{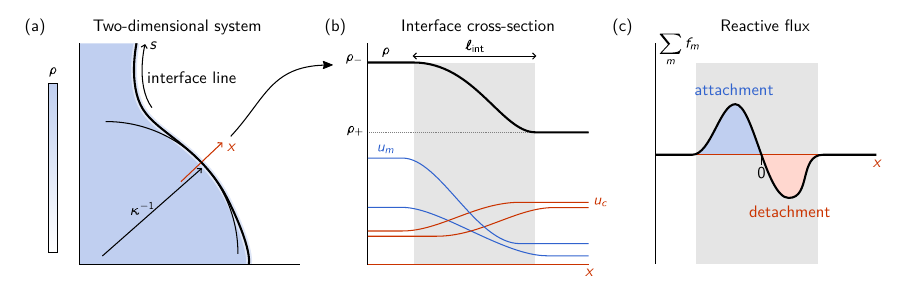}}
    \caption{Interfaces and density profiles in multi-component McRD systems.
    (a)
    We consider a pattern with a curved interface (with curvature $\kappa$) parametrized in terms of the arc-length $s$ on a two-dimensional plane.
    (b)
    Across the interface, i.e., along its normal direction with coordinate $x$ [red arrow in panel (a)], the total density $\rho$ (solid black line) of the considered protein species is assumed to decrease monotonously from $\rho_-$ at ${x\ll -\ell_\mathrm{int}}$ [blue shaded region in (a)] to $\rho_+$ at ${x\gg \ell_\mathrm{int}}$ within the interface width $\ell_\mathrm{int}$. 
    The counter-directed (cytosolic) components $u_c$ (red lines) increase monotonously in the interface region (shaded gray region).
    The other (membrane) components are denoted by $u_m$ (blue lines).
    These other components have to add up to a decreasing density profile in the interface region.
    (c)
    In the simplest case, the reactive flux $\sum_m f_m$ is positive (blue-shaded) in the attachment zone and negative (red-shaded) in the detachment zone.
    The reactive flux can, however, become negative (positive) in the attachment (detachment) zone.
    In that case, the integrated reactive flux in the attachment (${x<0}$) and detachment (${x>0}$) remains positive and negative, respectively.
    }
    \label{fig:attachment-detachment-balance}
\end{figure}

We now consider pattern-forming systems in two dimensions, e.g., protein patterns forming on a lipid membrane.
We further assume that the (nonlinear) dynamics result in patterns characterized by well-defined (sharp) interfaces between high- and low-density regimes (cf.\ Fig.~\ref{fig:attachment-detachment-balance}).
The width of the interface is denoted by $\ell_\mathrm{int}$.
Considering a straight and isolated interface, the reactive turnover balance simplifies to
\begin{equation}\label{eq:ttb-straight-int}
    0 
    \approx 
    \int_{-b}^b\mathrm{d}x\, f^\alpha_i(\mathbf{u}_\mathrm{stat}(x))
    \, ,
\end{equation}
where $x$ is a coordinate perpendicular to the interface and parametrizes the one-dimensional interface profile $\mathbf{u}_\mathrm{stat}(x)$ normal to the interface.
The interface is positioned around ${x=0}$.
Thus, choosing a distance ${b\gg \ell_\mathrm{int}}$ ensures that the integral runs over the whole interface region.
The exact choice of $b$ is irrelevant because the reaction--diffusion equations
Eqs.~\eqref{eq:McRD} in steady state imply that ${f_i^\alpha \approx 0}$ holds in the plateaus of the pattern far from the interface where the profile derivatives vanish approximately.

\subsection{Attachment--detachment balance}
\label{sec:attachment-detachment-balance}

In the above described general setting of multi-component McRDs many types of spatio-temporal patterns can emerge depending on the choice of the reaction terms.
As described in the introduction, here we focus on systems with well-defined and sharp interfaces that separate areas of high and low density.
We describe these interfaces as one-dimensional curves parameterized in terms of their arclength $s$.

We now focus on a single species $\alpha$ whose total density $\rho(\mathbf{x},t)$ we assume to vary monotonously across the interface, dropping the index $\alpha$ for simplicity.
The normal coordinate $x$ to the interface curves is chosen such that one has ${\partial_x \rho <0}$ (cf.\ Fig.~\ref{fig:attachment-detachment-balance}).

To determine the behavior of the interface when it is curved, we first define the attachment and the detachment processes.
Then, we show that attachment and detachment zones organize at the interface.
The balance of attachment and detachment at a stationary interface then leads to changes in the attachment and detachment fluxes at curved interfaces.

\textit{Defining ``membrane-bound'' and ``cytosolic'' species.\;---}
Starting from the reaction--diffusion equations, we must first define `cytosolic' and `membrane-bound' states to define `attachment' and `detachment'.
To this end, we consider those (``counter-directed'') components $u_{i_c}$ whose interface gradients are opposite to the total density profile, i.e., they fulfill ${\partial_x u_{i_c}>0}$ (Fig.~\ref{fig:attachment-detachment-balance}b).
In the next section, we show that the counter-directed components correspond (on average) to the fast-diffusing protein states (`cytosolic' states).
We denote by $i_c$, ${c=1,\dots,M_c}$ the indices of the counter-directed species and by $i_m$, ${m=1,\dots,M_m}$ the indices of all other species (`membrane-bound' states; cf.\ Fig.~\ref{fig:attachment-detachment-balance}b).
In the following, we abbreviate $i_c$ by the index $c$ and $i_m$ by $m$.

\medskip 

\textit{Deriving attachment and detachment zones.\;---}
Given the two classes of components, $c$ and $m$, and anticipating their roles as cytosolic and membrane-bound states, we define attachment to refer to molecules that transition from $c$-components to $m$-components, while detachment relates to molecules that transition from $m$-components to $c$-components.
As a result, the interface inherently displays the characteristic arrangement of attachment and detachment zones discussed in the main text due to the orientation and monotonicity of the $c$-components (Fig.~\ref{fig:attachment-detachment-balance}c).
To show this, consider a straight stationary interface.
Adding up and integrating the stationary equations for the $c$-components, Eqs.~\eqref{eq:McRD} with ${\partial_t u_c = 0}$, over the domain $[y,b]$ one obtains
\begin{align}
    0 
    = 
    \int_y^b\mathrm{d}x 
    \biggl[ \partial_x^2\sum_c D_c u_c +\sum_c f_c \biggr]
    \, ,
\end{align}
where ${b \gg \ell_\mathrm{int}}$ is again a point in the plateau region of the interface profile at a distance from the interface that is much greater than the interface width $\ell_\mathrm{int}$.
We also used that parallel to the interface line all gradients vanish.
Performing the integral in the first term and considering that the profile is flat in the plateau regions, one obtains
\begin{align}\label{eq:ttb-monotonicity}
    0 \approx -\partial_x \sum_c D_c u_c\Big|_{x=y} + F_c^+(y)
    \, ,   
\end{align}
where $F_c^+(y)$ is the total reaction term of the $c$-components to the ``right'' of $y$, i.e., in the domain $[y,b]$.
For ${y=-b}$, the term $F_c^+(-b)$ (approximately) agrees with the reactive turnover balance Eq.~\eqref{eq:ttb-straight-int} such that ${F_c^+(-b) \approx 0}$.
Moreover, since ${\partial_x u_c > 0}$, we conclude from Eq.~\eqref{eq:ttb-monotonicity} that the reactive turnover for the $c$-components must be positive over the whole interface, ${F_c^+(y) \geq 0}$.
Since ${F_c^+(-b) \approx 0}$, the reactive turnover function $F_c^+(y)$ must have at least one (local) maximum in the domain ${y \in [-b,b]}$ at which
\begin{equation}
    0 
    = 
    \partial_y F_c^+(y) 
    = 
    - \sum_c f_c = \sum_m f_m
    \, .
\end{equation}
The minus sign arises because $y$ is the lower bound of the integral $F_c^+(y)$.
The last equality follows from the mass conservation constraint on $\mathbf{f}$, Eq.~\eqref{eq:mass-cons-constraint}.

In summary, the above analysis shows that the reactive-turnover function $F_c^+(y)$ is zero in the plateau regions ${x = \pm b}$ and has a maximum in between.
Therefore, we can define the origin of the axis perpendicular to the interface such that ${x=0}$ corresponds to the global maximum of $F_c^+(y)$ at which ${0=\sum_c f_c = -\sum_m f_m}$.
If the turnover function $F_c^+(y)$ has only one (local) maximum, $F_c^+(y)$ is monotonically increasing for ${y<0}$, implying $\sum_m f_m > 0$ and defining the attachment zone, while analogously we have $\sum_m f_m < 0$ for ${y>0}$ defining the detachment zone; see Fig.~\ref{fig:attachment-detachment-balance}c for an illustration.
Even if the reaction turnover function $F_c^+(y)$ has additional local maxima, it is non-negative due to the monotonicity of the $c$-components and one has
\begin{equation}\label{eq:int-detachment-flux}
    \int_0^b\mathrm{d}x \sum_m f_m < 0,
\end{equation}
and the reactive turnover balance $F_c^+(-b)\approx 0$ implies
\begin{equation}
    \int_{-b}^0\mathrm{d}x \sum_m f_m = -\int_{-b}^0\mathrm{d}x \sum_c f_c \approx - 0 + \int_0^{b}\mathrm{d}x \sum_c f_c > 0 \, .
\end{equation}
Other than in the simple case shown in Fig.~\ref{fig:attachment-detachment-balance}c, local maxima imply that the local attachment--detachment flux $\sum_m f_m$ can also be negative (positive) in parts of the attachment (detachment) zone.
However, the integrated reactive flux in the attachment (detachment) zone is always positive (negative).
Choosing the global maximum as the boundary between the attachment and detachment zones ensures that these integrated reactive fluxes are largest, that is, the imbalance between attachment and detachment is largest in each single zone.

\medskip

\textit{Curved interfaces.\;---}
In the following, we determine how the attachment--detachment flux changes with interface curvature $\kappa$, assuming weakly curved interfaces ${\kappa\ll 1/\ell_\mathrm{int}}$.
We start from the reaction--diffusion equation, Eq.~\eqref{eq:McRD}, in the stationary limit where ${\partial_t u_i=0}$.
Considering a position $s$ on the interface with local curvature $\kappa$, we sum over all $m$-components and integrate over the entire interface region $[-b,b]$ along a small line element ${ [s,s+\mathrm{d}l]}$.
Since the interface width is small and the interface is weakly curved, ${\ell_\mathrm{int}\ll b\ll 1/\kappa}$, one obtains
\begin{equation}\label{eq:turnover-balance-multiComp-kappa}
    0 \approx 
    \int_{-b}^b\mathrm{d}x 
    \int_{s}^{s+\mathrm{d}l}\mathrm{d} \tilde s\, 
    \Bigl(\frac{1}{\kappa}+x\Bigr) 
    \sum_m 
    \bigl[ D_m \nabla^2 u_m + f_m \bigr].
\end{equation}
Here we have used that the curvilinear coordinates $(x,s)$ are local polar coordinates, and the Jacobian is proportional to the radial variable ${1/\kappa + x}$.

According to Gauss' theorem, the area integral over the term $\nabla^2 u_m$ can be written as a boundary integral for $ \nabla u_m$ projected onto the normal vector of the boundary.
This boundary integral vanishes for two reasons.
At the two boundaries at $s=s_0,s_0+\mathrm{d}l$, the derivatives $\partial_s u_m$ are integrated.
These vanish for weakly curved interfaces as the density profiles perpendicular to the interface line are assumed to not vary parallel to the interface curve.
Since the profile becomes flat far away from the interface (at $\pm b$), the boundary integral also vanishes at the remaining two sections of the boundary.
Consequently, in the limit ${\mathrm{d}l\to 0}$, Eq.~\eqref{eq:turnover-balance-multiComp-kappa} reduces to
\begin{equation}\label{eq:turnover-balance}
    0 
    = 
    \int_{-b}^b\mathrm{d}x \, 
    \bigl( 1+x\,\kappa \bigr) 
    \sum_m f_m \, .
\end{equation}
This equation represents the reactive turnover balance for a weakly curved interface, differing from Eq.~\eqref{eq:reactive-turnover-balance} by the inclusion of the dilation factor $1+x\,\kappa$, reflecting the altered area due to curvature: 
When $\kappa$ is positive, the outer region (${x>0}$) of the interface is stretched while the inner region (${x<0}$) is compressed.

We define the integrated attachment and detachment fluxes for the curved interface as
\begin{subequations}
    \begin{align}
        F^{\mathrm{att}}_\kappa 
        :=
        \int_{-b}^0\mathrm{d}x\, \sum_m f_m
        \, , \\
        F^{\mathrm{det}}_\kappa 
        :=
        \int_0^b\mathrm{d}x\, \sum_m f_m 
        \, .
    \end{align}
\end{subequations}
The curvature $\kappa$ enters via the curvature-induced changes in the stationary density profiles $u_i$ that determine the reactive fluxes $f_i(\mathbf{u})$.
Assuming that the interface profile is approximately symmetric with respect to ${x=0}$, we define the average attachment and detachment flux per unit area of the attachment and detachment zone by (cf.\ Sec.~``Effective interfacial tension'' in the main text)
\begin{equation}
    f_\mathrm{a,d}(\kappa) 
    \equiv
    \frac{F^{\mathrm{att,det}}_\kappa}{\ell_\mathrm{int}/2}
    \, .
\end{equation}

Because the reaction terms vary continuously with the densities, $\sum_m f_m$ goes to zero continuously at ${x=0}$ (Fig.~\ref{fig:attachment-detachment-balance}c).
Moreover, $\sum_m f_m$ approaches zero exponentially in the plateaus because all density gradients vanish in the plateaus and, due to Eq.~\eqref{eq:McRD} for ${\partial_t u_i=0}$, the plateau densities fulfill ${f_i^\alpha\approx 0}$.
Thus, $\sum_m f_m$ is localized in the attachment and detachment zones, and one may define averages over the attachment and detachment zones by\footnote{
Note that the term $\sum_m f_m$ may change sign within the attachment and detachment zones.
This is the case if the function $F_c^+$ has local maxima.
Thus, $\sum_m f_m/F_\kappa^\mathrm{att/det}$ may not be directly interpretable as a probability distribution.
}
\begin{subequations}
    \begin{align}
        \langle\bullet\rangle_{\mathrm{att},\kappa} = \frac{1}{F^{\mathrm{att}}_\kappa}\int_{-b}^0\mathrm{d}x\, \bullet\sum_m f_m,\\
        \langle\bullet\rangle_{\mathrm{det},\kappa} = \frac{1}{F^{\mathrm{det}}_\kappa}\int_0^b\mathrm{d}x\, \bullet\sum_m f_m.
    \end{align}
\end{subequations}
With this, the attachment--detachment balance, Eq.~\eqref{eq:turnover-balance}, may be written as
\begin{equation}
    0 = F^{\mathrm{att}}_\kappa 
    \bigl( 
    1 + \kappa \langle x\rangle_{\mathrm{att},\kappa} 
    \bigr)
    +
    F^{\mathrm{det}}_\kappa 
    \bigl(
    1 + \kappa \langle x\rangle_{\mathrm{det},\kappa}
    \bigr)
    \, ,
\end{equation}
where $1/\kappa +\langle x\rangle_{\mathrm{att/det},\kappa}$ are the average radial positions of the attachment and detachment zones with respect to the curvature of the interface.
Thus, the positions $\langle x\rangle_{\mathrm{att/det},\kappa}$ will scale with (half of) the interface width.
Approximating the interface as symmetric, we set ${\langle x\rangle_{\mathrm{att/det},\kappa} \approx \pm \ell_\mathrm{int}/4}$ given that the width of the attachment an detachment zones is approximately $\ell_\mathrm{int}/2$.
This yields
\begin{equation}
    0 \approx 
    F^{\mathrm{att}}_\kappa 
    \biggl( 1 - \frac{\ell_\mathrm{int}\kappa}{4} \biggr)
    +
    F^{\mathrm{det}}_\kappa 
    \biggl(1 + \frac{\ell_\mathrm{int}\kappa}{4} \biggr)
    \, .
\end{equation}
Denoting the curvature-induced change in the attachment and detachment fluxes by ${\delta F_\kappa^\mathrm{att/det}=F^{\mathrm{att/det}}_\kappa-F^{\mathrm{att/det}}_0}$, one arrives to leading order in $\kappa$ at
\begin{equation}\label{eq:turnover-imbalance-kappa}
    \delta F_\kappa^\mathrm{att}+\delta F_\kappa^\mathrm{det} 
    \sim 
    F^{\mathrm{att}}_0 \ell_\mathrm{int}\kappa.
\end{equation}
Note that $F_\kappa^\mathrm{att}>0$ (attachment) and $F_\kappa^\mathrm{det}<0$ (detachment). 
This holds not only for $\kappa = 0$ [cf.\ Eq.~\eqref{eq:int-detachment-flux}] but also for $\kappa\ell_\mathrm{int} \ll 1$ because the (sufficiently small) curvature-induced changes do change their signs.
Then, the same arguments as for the straight interface apply to the weakly curved interfaces.
Equation~\eqref{eq:turnover-imbalance-kappa} therefore shows that with increasing curvature $\kappa$ the attachment flux per interface length must increase relative to the detachment flux.
As explained in the main text, this is due to the decrease in the area of the attachment compared to the area of the detachment zone.

\subsection{Effective interfacial tension for systems with pattern-forming feedback due to slow-diffusing components}

What remains is to establish a connection between changes in the attachment and detachment fluxes with the densities, particularly the (modified) mass-redistribution potential $\Tilde{\eta}$ governing the redistribution of molecules in the system via the continuity equation Eq.~\eqref{eq:cont-eq}.
To this end, we first show that, on average, the counter-oriented components $c$ are the fast-diffusing species.
Afterward, we will relate the attachment--detachment balance to the (average) shift in the mass-redistribution potential by determining the shifts in the densities of fast- and slow-diffusing species.

\textit{Fast- and slow-diffusing components.\;---}
Recall that we consider the (quasi-) stationary, sharp interfaces of a single protein species on a domain with no-flux or periodic boundary conditions.
At the stationary interface, one has
\begin{equation}\label{eq:stat-int-eta}
    \sum_i D_i u_i(\mathbf{x}) = \Tilde{\eta} = \mathrm{const.}
\end{equation}
With the plateau densities $u_i^\pm$ at ${x \sim b}$ and ${x\sim -b}$, respectively, define the density differences across the interface as
\begin{align}
    \Delta u_i 
    = u_i^+-u_i^-
    \approx u_i|_{x=b}-u_i|_{x=-b} \, .
\end{align}
Equation~\eqref{eq:stat-int-eta} then yields the balance condition
\begin{equation}\label{eq:const-eta}
    - \sum_c D_c\Delta u_c = \sum_m D_m \Delta u_m 
    \, .
\end{equation}
Next, we define the averaged diffusion coefficients as
\begin{align}
    \langle D_c\rangle := \frac{\sum_c D_c\Delta u_c}{\Delta\rho_\mathrm{C}}
    \, , \qquad
    \langle D_m\rangle := \frac{\sum_m D_m\Delta u_m}{\Delta\rho_\mathrm{M}}
    \, ,
\end{align}
with the total-density field of counter-oriented components ${\rho_\mathrm{C} = \sum_c u_c}$, and ${\rho_\mathrm{M} = \rho-\rho_\mathrm{C} = \sum_m u_m}$.
With these definitions, the balance condition Eq.~\eqref{eq:const-eta} gives
\begin{equation}
    \bigl( \langle D_m \rangle - \langle D_c \rangle \bigr) \, \Delta\rho_\mathrm{C} 
    = 
    \langle D_m \rangle \, \Delta\rho
    \, .
\end{equation}
Consequently, as ${\Delta\rho<0}$, ${\Delta\rho_\mathrm{C}>0}$, and ${\langle D_c\rangle>0}$, we find
\begin{equation}\label{eq:diff-coeff-rel}
    0 < \langle D_m\rangle < \langle D_c\rangle.
\end{equation}
Intuitively, this relation arises because the amplitude of the density jumps of the different components at the interface is determined by their diffusion coefficients.
To have the total-density profile oriented with $\partial_x\rho<0$, the density jumps of the counter-directed components $c$ must be smaller than those of the other components $m$.
Therefore, at least on average, the diffusion coefficients of the counter-directed species must be larger.

\textit{Curvature-induced shift of the stationary mass-redistribution potential.\;---}
Motivated by the average relation Eq.~\eqref{eq:diff-coeff-rel}, we assume that the counter-directed components are the fast-diffusing components and the other components are the slow-diffusing ones.
To relate the curvature-induced relative increase of the attachment flux compared to the detachment flux with $\kappa$ described by Eq.~\eqref{eq:turnover-imbalance-kappa}, we must understand, how the fluxes depend on the densities of the slow- and fast-diffusing species.
We assume that the fast-diffusing species do not induce feedback, and the attachment flux increases with increasing cytosolic densities $u_c$.
Thus, defining reaction rates $F^\mathrm{att}_c>0$ and $F^\mathrm{att}_m$ that describe how the attachment flux depends on the (average) curvature-induced density changes $\delta u_{m,c}^\kappa$, one has 
\begin{equation}\label{eq:att-flux-densitity-rel}
    \delta F_\kappa^\mathrm{att} = \sum_c F^\mathrm{att}_c \delta u_c^\kappa + \sum_m F^\mathrm{att}_m \delta u_m^\kappa.
\end{equation}
In the simplest case, the attachment flux just depends linearly on the cytosolic density (densities) because the number of proteins available for attachment increases with the cytosolic density.

Concomitantly, we assume that the slow-diffusing components $u_m$ induce the pattern-forming feedback:
The arrangement of the attachment and detachment zone implies that the feedback must induce net attachment at high densities of the slow-diffusing components $m$ while net detachment prevails at low densities of the $m$ components.
As a result, we assume that (average) increases in the slow-diffusing components lead to increased attachment and decreased detachment fluxes.
Thus, one has $F^\mathrm{att}_m > 0$ and, with $F^\mathrm{det}_m > 0$, it holds (recall that the detachment flux $F_\kappa^\mathrm{det}$ is negative)
\begin{equation}\label{eq:det-flux-densitity-rel}
    \delta F_\kappa^\mathrm{det} = \sum_m F^\mathrm{att}_m \delta u_m^\kappa,
\end{equation}
where the cytosolic components do not contribute significantly because detachment is dominant in the detachment zone, and they are assumed not to induce feedback.
Note that all signs would turn around if the counter-propagating (fast-diffusing) components induced the pattern-forming feedback.

With Eqs.~\eqref{eq:att-flux-densitity-rel},~\eqref{eq:det-flux-densitity-rel}, the curvature-induced turnover imbalance Eq.~\eqref{eq:turnover-imbalance-kappa} suggests that
\begin{equation}
    \delta u_c^\kappa \sim \ell_\mathrm{int}\kappa,\qquad \delta u_m^\kappa \sim \ell_\mathrm{int}\kappa,
\end{equation}
holds at least for the majority of components.
As a result, also the shift of the stationary mass-redistribution potential ${\eta = \Tilde{\eta}/\langle D_c\rangle}$ fulfills
\begin{equation}
     \delta\eta_\kappa \approx \sum_c \frac{D_c}{\langle D_c\rangle} \delta u_c^\kappa + \sum_m \frac{D_m}{\langle D_c \rangle} \delta u_m^\kappa \sim \ell_\mathrm{int}\kappa.
\end{equation}
Thus, the prefactor in the non-equilibrium Gibbs--Thomson relation $\delta\eta_\kappa \sim \ell_\mathrm{int} \kappa$ can be expected to be positive and proportional to the interface width, which implies a positive effective interfacial tension $\sigma \sim \ell_\mathrm{int}$.

We arrive at a positive effective interfacial tension by assuming that the pattern-forming feedback is induced by the slow-diffusing (membrane) components whose interface gradients define the orientation of the total-density interface.
In contrast, the counter-oriented (cytosolic) components are assumed to enter the reaction terms without feedback.
These conditions are motivated by the attachment--detachment kinetics of intracellular pattern-forming systems but are also similar to the short-range activation necessary in activator--inhibitor systems.
The curvature-induced relative change of the attachment and detachment fluxes then suggest positive average density changes at the interface with curvature $\kappa$, and consequently also for the mass-redistribution potential $\eta$.
Exploring in what systems this behavior breaks down and how this might relate to interfacial instabilities observed in two-component reaction-diffusion systems without mass conservation~\cite{Petrich.Goldstein1994,Muratov.Osipov1996,Carter.etal2023} represents an interesting avenue for future research. 

\section{Effective interface tension in McRD systems with two components per species}
\label{sec:multi-species-two-components}

Having analyzed multi-component McRD systems in the previous section by dividing the components into slow- and fast-diffusing classes, we now derive exact nonequilibrium Gibbs-Thomson relations for multi-species systems in which each species has only two components.
For later comparison, we first derive the phase equilibria in multi-species phase-separating liquid mixtures.

\subsection{Phase equilibria in multi-component phase separation}
\label{sec:Gibbs-Thomson-phase-sep}
We compare the $N$-species McRD system with phase separation in an incompressible ${N+1}$-component mixture.
Assuming equal molecular volumes ${v=1}$ for the different components, the composition is specified by $N$ volume fractions ${\phi_i = N_i v/V}$ where $N_i$ is the number of particles of type $i$ and $V$ is the total volume of the system.
The last component ${N+1}$ is the solvent with volume fraction ${1-\sum_{i=1}^N \phi_i}$.
To arrive at the Gibbs-Thomson relation, we consider a circular droplet (considering two dimensions as in the RD system) with volume ${V_1=\pi r^2}$ of one phase with compositions ${\bm{\phi} = (\phi_1, \phi_2, \dots, \phi_N)}$ in a bath of a different phase with volume $V-V_1$ and compositions ${\bm{\psi} = (\psi_1, \psi_2, \dots, \psi_N)}$. The thermodynamics of this system is determined by the free energy~\cite{Doi2013}
\begin{equation}
    F(T, V_1,\bm{\phi}, V, \bm{\psi}) = V_1 f(\bm{\phi}) + (V-V_1) f(\bm{\psi}) + 2\sqrt{\pi V_1} \sigma
    \, ,
\end{equation}
where the first two terms are the bulk free energies of the two phases with volumes $V_1$ and $V-V_1$, respectively. 
The corresponding free energy densities $f$ depend on the volume fractions and the temperature, whereby the latter will be kept constant and has been omitted to simplify the notation.
The last term is a boundary term accounting for the interface energy, where $\sigma$ denotes the surface tension and  ${2\sqrt{\pi V_1} = 2 \pi r}$ is the circumference of the droplet.
Mass conservation implies that $\bar{\phi}_i V = \phi_i V_1 +  \psi_i (V-V_1)$, where the average volume fractions $\bar{\phi}_i$ have fixed values, as has the total volume $V$.

The thermodynamic equilibrium state is obtained by minimizing $F$ with respect to both the volume $V_1$ and the volume fractions $\bm{\phi}$, subject to the constraints of total volume and mass conservation.
This yields the (generalized) common-tangent construction
\begin{subequations}\label{eq:CTC}
\begin{align}
    0 &= \partial_{\phi_i} f(\bm{\phi}) - \partial_{\psi_i} f(\bm{\psi})
    \, , \\
    0 &= \kappa \, \sigma + f(\bm{\phi}) - f(\bm{\psi}) + \sum_{i=1}^N (\psi_i-\phi_i) \, \partial_{\phi_i}f(\bm{\phi}) \, ,
\end{align}
\end{subequations}
with the curvature ${\kappa=1/r}$.
The first equation states that the chemical potentials of all components have to be equal in both phases.
The second relation ensures that the whole system is in mechanical equilibrium, as we discuss in the following.

These conditions can be simplified using the exchange chemical potentials
\begin{equation}\label{eq:ex-chem-pot}
    \mu_i 
    \equiv
    \partial_{N_i} \bigl( V f(\bm{\phi}) \bigr) \Big|_V
    = \partial_{\phi_i}f(\bm{\phi})
    \, .
\end{equation}

Now, define the curvature-induced shifts in the exchange chemical potentials ${\delta\bm{\mu} = \bm{\mu}|_{\kappa} - \bm{\mu}|_{0}}$ and the volume fractions ${\delta\bm{\phi} = \bm{\phi}|_{\kappa} - \bm{\phi}|_{0}}$, ${\delta\bm{\psi} = \bm{\psi}|_{\kappa} - \bm{\psi}|_{0}}$.
Then, subtracting the conditions Eqs.~\eqref{eq:CTC} for a straight interface ${\kappa=0}$ and a system with a curved interface ${\kappa\neq 0}$, one finds to first order in the deviations $\delta\bm{\mu},\delta\bm{\phi},\delta\bm{\psi}$
\begin{equation}
	0 \approx \kappa \sigma + \left(\delta\bm{\phi}-\delta\bm{\psi}\right)\cdot \partial_{\bm{\phi}} f(\bm{\phi}|_{0}) + \sum_{i=1}^N 
    \big[ \left( \psi_i|_{0}-\phi_i|_{0}\right) \delta \mu_i + \left(\delta \psi_i -\delta \phi_i\right) \mu_i|_{0}
    \big] \, .
\end{equation}
Using the definition of the exchange chemical potentials Eq.~\eqref{eq:ex-chem-pot}, one arrives at the Gibbs-Thomson relation
\begin{equation}\label{eq:dMu-Gibbs-Thomson}
    \Delta\bm{\phi}\cdot \delta\bm{\mu} 
    = 
    - \sigma \, \kappa
    \, ,
\end{equation}
with the jumps ${\Delta\phi_i := \psi_i|_{0}- \phi_i|_{0}}$ in volume fraction between the two phases.
The left-hand side describes the osmotic-pressure difference between the two phases in contact, which is balanced by the Laplace pressure on the right-hand side. 
The relation Eq.~\eqref{eq:dMu-Gibbs-Thomson} has been derived mathematically for $N$-component Cahn-Hilliard equations describing ${N+1}$-component incompressible liquid mixtures by $N$ partial differential equations for the volume fractions $\phi_i$ \cite{Eyre1993,Elliott.Luckhaus1991,Bronsard.etal1998,Garcke.Novick-Cohen2000}.
We will use a similar approach to derive the non-equilibrium Gibbs--Thomson relations in the McRD systems below.

\subsection{Non-equilibrium steady states of McRD systems}

In this section, we analyze the non-equilibrium steady states of McRD systems with two components per species that describe stationary density interfaces.
We consider systems with two components for each molecular species as a model class for intracellular protein patterns that arise from the interaction of multiple protein species that attach onto the cell membrane and detach into the cytosol.
The result at the end of this section will be the derivation of a non-equilibrium Gibbs--Thomson relation for pattern interfaces in these systems.
This non-equilibrium Gibbs--Thomson relation then defines the effective interfacial tension of the interfaces.

In systems with two components per species, the patterns can be described by the total density and the mass-redistribution potential of each molecular species, cf.\ Eqs.~\eqref{eq:cont-eq-2c},~\eqref{eq:eta-dynamics}.
Any stationary pattern $[\bm{\rho}^\mathrm{stat}(\mathbf{x}),\bm{\eta}^\mathrm{stat}]$ for systems subject to no-flux boundary conditions must---due to the continuity equation Eq.~\eqref{eq:cont-eq-2c}---have spatially uniform mass-redistribution potentials   ${\eta_\alpha^\mathrm{stat}}$, and satisfy the profile equation [cf.\ Eq.~\eqref{eq:eta-dynamics}]
\begin{equation}\label{eq:profile-eq}
    0 = 
    D_m^\alpha \nabla^2 \rho_\alpha^\mathrm{stat}(\mathbf{x}) + \Tilde{f}_\alpha(\bm{\rho}^\mathrm{stat}(\mathbf{x}),\bm{\eta}^\mathrm{stat})
    \, .
\end{equation}
In this profile equation, the set of stationary mass-redistribution potentials $\bm{\eta}^\text{stat}$ act as parameters.
They must be chosen such that the profile equation has a solution that satisfies the no-flux boundary conditions.
By integrating the profile equation over the entire $d$-dimensional domain, the no-flux boundary conditions result in the reactive turnover balance condition [cf.\ Eq.~\eqref{eq:reactive-turnover-balance}]
\begin{equation}\label{eq:reactive-turnover-balance-2c}
    0 = \int\mathrm{d}^dx\, \Tilde{f}_\alpha(\bm{\rho}^\mathrm{stat}(\mathbf{x}),\bm{\eta}^\mathrm{stat})
    \, .
\end{equation}

\subsubsection{Straight interfaces}
\label{sec:straight-int}

Consider now a stationary solution of the McRD equations given by a well-separated straight interface (with curvature ${\kappa = 0}$ on a two-dimensional surface).
The density profiles $\rho_\alpha^\mathrm{stat,0}(\mathbf{x})$ for each species will then vary only in the normal direction, which we denote by $x$.
As above in Sec.~\ref{sec:multi-comp}, we position the interface around ${x=0}$.
The set of profile equations Eq.~\eqref{eq:profile-eq} for the straight interface then reads
\begin{equation}\label{eq:profile-eq-straight}
    0 = D_m^\alpha \partial_x^2 \rho_\alpha^\mathrm{stat,0}(x) + \Tilde{f}_\alpha(\bm{\rho}^\mathrm{stat,0}(x),\bm{\eta}^\mathrm{stat,0})
    \, ,
\end{equation}
and the reactive turnover balance Eq.~\eqref{eq:reactive-turnover-balance-2c} fixing $\bm{\eta}^\mathrm{stat,0}$ becomes
\begin{equation}\label{eq:reactive-turnover-balance-2c-straight}
    0 = \int\mathrm{d}x\, \Tilde{f}_\alpha(\bm{\rho}^\mathrm{stat,0}(x),\bm{\eta}^\mathrm{stat,0})
    \, ,
\end{equation}
where the integral runs over the whole domain or, as an approximation, over the whole interface region $[-b,b]$ with $b\gg\ell_\mathrm{int}$.

Mathematically, the profile equation Eq.~\eqref{eq:profile-eq-straight} can also be interpreted through a mechanical analogy known as the ``rolling ball analogy''~\cite{Mikhailov1990}.
Identifying $x$ with time, the profile equation describes the dynamics of a particle's position $\bm{\rho}$ (in a space with a dimension given by the total number of species) with (``anisotropic'') mass $\mathbf{D}_m = \mathrm{diag}(D_m^1,\dots,D_m^N)$ in the force field $-\Tilde{\mathbf{f}}$.
A pattern interface connects two plateaus at total densities $\bm{\rho}_\pm$ at ${x\gg \ell_\mathrm{int}/2}$ and ${x\ll -\ell_\mathrm{int}/2}$, respectively, in which the profiles become flat (${\partial_x\bm{\rho} \to 0}$).
This implies that the plateaus correspond to fixed points $(\bm{\rho}_\pm,0)$ in the phase space $(\bm{\rho},\partial_x\bm{\rho})$ of the ``dynamics'' Eq.~\eqref{eq:profile-eq-straight} at which ${0 = \Tilde{\mathbf{f}}}$.
Thus, in this phase space, the stationary interface profile $\bm{\rho}^\mathrm{stat,0}$ is a heteroclinic orbit connecting the two fixed points.
If a scalar potential $V$ exists for the ``force'' field such that ${\Tilde{\mathbf{f}} =\bm\nabla_\rho V}$ holds with the gradient ${\nabla_\rho = (\partial_{\rho_1},\dots, \partial_{\rho_{N}})}$, this corresponds to a particle starting out at a local maximum (or saddle point) of the potential, ``rolling'' through a potential valley, and ending up with zero velocity on top of a different local maximum (saddle point).
The reactive turnover balance conditions then correspond to fixing $\bm{\eta}$, which are parameters of the potential, at a specific value $\bm{\eta}^\mathrm{stat,0}$ such that the ball ends up with zero velocity on the second maximum (saddle point).

\begin{figure}[btp]
    \centering
    \includegraphics[width=0.9\textwidth]{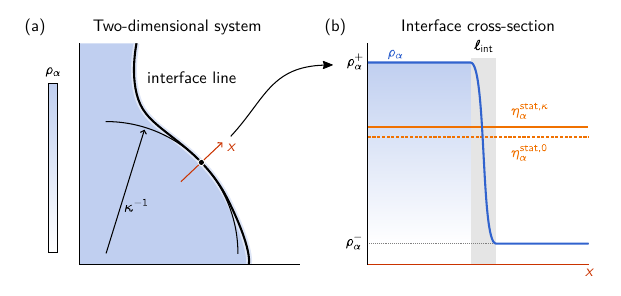}
    \caption{Analysis of a weakly curved interface.
    (a) 
    The density field $\rho_\alpha$ of species $\alpha$ is shown in the two-dimensional plane (blue).
    The interface corresponds to a line with a local curvature $\kappa$ (black line).
    The curvature radius $\kappa^{-1}$ is shown at one point of the interface (black arc and arrow) together with the normal coordinate $x$ (red arrow).
    (b) 
    Along this normal direction $x$, the density $\rho_\alpha$ (blue) varies strongly over the interface width ${\ell_\mathrm{int}\ll 1/\kappa}$, i.e., we consider a sharp interface (grey shading).
    The density approaches the plateau values $\rho_\alpha^\pm$ far from the interface in positive and negative $x$-direction, respectively.
    The stationary mass-redistribution potential $\eta_\alpha^{\mathrm{stat},\kappa}$ varies depending on the local interface curvature $\kappa$ (orange).
    The plateau densities $\rho_\alpha^\pm$ depend on the mass-redistribution potentials $\eta_\alpha^{\mathrm{stat},\kappa}$.}
    \label{fig:weakly-curved-int}
\end{figure}

\subsubsection{Weakly curved interfaces}
\label{sec:weakly-curved-int}
In the following, we assume that the interface is weakly curved so that the local curvature $\kappa$ is small compared to the reciprocal of the interface width $\ell_\text{int}$: ${\kappa \, \ell_\text{int} \ll 1}$; we will give a mathematical definition of the interface width in Sec.~\ref{sec:interface-width-tension}.
Polar coordinates can then be used to take the curvature of the interface into account locally, i.e., one can use curvilinear coordinates with a coordinate $x$ perpendicular and a coordinate $s$ parallel to the interface; see Fig.~\ref{fig:weakly-curved-int}.
Positioning the interface around $x=0$, the profile equation Eq.~\eqref{eq:profile-eq} reads in these polar coordinates
\begin{equation}
    0 = D_m^\alpha \left(\frac{1}{\kappa^{-1}+x} \, \partial_x + \partial_x^2\right) \rho_\alpha^{\mathrm{stat},\kappa}(x) + \Tilde{f}_\alpha(\bm{\rho}^{\mathrm{stat},\kappa}(x),\bm{\eta}^{\mathrm{stat},\kappa})
    \, ,
\end{equation}
where we have used the representation of the Laplacian in polar coordinates and assumed that the density profile remains constant along the interface.
Note also that ${\kappa = \kappa(s)}$ is the local interface curvature.

Since the densities vary significantly only in the interface region, the gradients $\partial_x\rho_\alpha^{\mathrm{stat},\kappa}(x)$ will be peaked around ${x=0}$ with a width of about $\ell_\mathrm{int}$.
Consequently, in the interface region of weakly curved interfaces, one may approximate the profile equation by (to first order in ${\kappa \, \ell_\text{int}}$, sharp interface approximation)
\begin{equation}\label{eq:profile-eq-weakly-bend}
    0 = D_m^\alpha (\kappa \, \partial_x + \partial_x^2) \rho_\alpha^{\mathrm{stat},\kappa}(x) + \Tilde{f}_\alpha(\bm{\rho}^{\mathrm{stat},\kappa}(x),\bm{\eta}^{\mathrm{stat},\kappa})
    \, .
\end{equation}
Because the reactive turnover is changed at curved interfaces (cf.\ Sec.~``Effective interfacial tension'' in the main text and Sec.~\ref{sec:multi-comp}), the mass-redistribution potentials will be shifted, i.e., $\bm{\eta}^{\mathrm{stat},\kappa} = \bm{\eta}^\mathrm{stat,0}+\delta\bm{\eta}$.
The goal of this section is to calculate these curvature-induced shifts.\footnote{
In the mechanical analogy, the curvature introduces ``friction'' proportional to the particle's ``velocity''.
Consequently, the ``force'' field $-\Tilde{\mathbf{f}}$, that is, the parameters $\bm{\eta}$ have to be changed to restore the heteroclinic orbit, giving modified stationary mass-redistribution potentials ${\bm{\eta}^{\mathrm{stat},\kappa} = \bm{\eta}^\mathrm{stat,0}+\delta\bm{\eta}}$.
}

In the following, in  Sec.~\ref{sec:stab-properties} we first discuss general stability properties that the reaction terms must fulfill, which are useful later to interpret the terms in the non-equilibrium Gibbs--Thomson relation.
In Sec.~\ref{sec:general-curved}, we give a general expression for the shifts $\delta\bm{\eta}$, i.e., the general form of the non-equilibrium Gibbs--Thomson relation for multi-species McRD systems with two components for each species.
A mathematical expression solely in terms of the stationary interface profile $\bm{\rho}^\mathrm{stat,0}$ can be given if the reaction terms can be derived from a single reaction potential.
This situation applies to the multi-species generalization of the minimal PAR model (Methods Sec.~A.1.a), and we will consider these systems in Sec.~\ref{sec:fully-solvable}.
At the end of the Sec.~\ref{sec:fully-solvable}, we discuss in detail the analogies between the derived non-equilibrium Gibbs--Thomson relations and the equilibrium thermodynamics of multi-component mixtures (derived in Sec.~\ref{sec:Gibbs-Thomson-phase-sep}).
While the reaction potential requires a specific choice of the reactive interaction terms between the different species, we consider the opposite case in Sec.~\ref{sec:binary-self-reac}.
In Sec.~\ref{sec:binary-self-reac}, we express the non-equilibrium Gibbs--Thomson relation in terms of $\bm{\rho}^\mathrm{stat,0}$ for binary systems with arbitrary reactive interaction terms but symmetric dependence of their reaction terms on their own densities (``self reactions'').

\subsection{General stability properties}
\label{sec:stab-properties}

Recall that the basic assumption of our theory is that the reaction--diffusion systems we consider form stable patterns characterized by interfaces that separate regions of different densities.
The goal of this section is to discuss some (simple) conditions that the reaction term $\tilde{\mathbf{f}}$ must satisfy to allow such solutions, specifically the formation of stable straight density interfaces separating density plateaus.
We first shortly discuss the pattern plateaus and afterward the interface region.
We will use these conditions later for analyzing the non-equilibrium Gibbs--Thomson relations.
Note that while these conditions are necessary for interface stability, they are not sufficient.

\subsubsection{The local reactive dynamics}
\label{sec:local-stability}

In a well-mixed system, all gradients vanish, and only the reactions induce dynamics.
Thus, Eqs.~\eqref{eq:cont-eq-2c},~\eqref{eq:eta-dynamics} reduce to
\begin{subequations}\label{eq:well-mixed}
\begin{align}
    \partial_t \rho_\alpha &=0
    \, ,\\
    \partial_t \eta_\alpha &= - \Tilde{f}_\alpha
    \, .
\end{align}
\end{subequations}
The total densities remain constant since the reactions conserve the total densities $\rho_\alpha$.
The dynamics in a spatially extended system, in which the densities vary spatially, can be understood as a well-mixed dynamics taking place locally within sufficiently small subregions (``boxes'') of the whole system \cite{Halatek.Frey2018}.
These ``boxes'' are then coupled through the diffusion of particles between adjacent boxes.
Therefore, we refer to the dynamics Eq.~\eqref{eq:well-mixed} as the \textit{local dynamics}.\footnote{
If a timescale separation exists between the local reactive dynamics and mass redistribution between the boxes, one can make an adiabatic assumption and approximate the component densities by the local reactive equilibria in each box and solve for the mass redistribution induced by diffusion given these densities (``local equilibria theory'' \cite{Brauns.etal2020}).}

Considering the nonlinear dynamics approach, much of the behavior of the nonlinear reaction term $\Tilde{\mathbf{f}}$ is captured by its nullcline (high-dimensional nullcline surface) $\bm{\eta}^*(\bm{\rho})$ \cite{Halatek.Frey2018,Brauns.etal2020} that is defined as the \emph{reactive equilibrium} (we assume there is no multistability in the local dynamics; see \cite{Brauns.etal2020} for a discussion of multistability in two-component systems) of the well-mixed system at given total densities $\bm{\rho}$:
\begin{equation}
    0 = \Tilde{\mathbf{f}}(\bm{\rho},\bm{\eta}^*(\bm{\rho}))
    \, .
\end{equation}

We are interested in patterns that consist of interfaces separating plateau regions with nearly uniform densities.
Since all the density gradients (approximately) vanish in the plateau regions, the stationary plateau values $(\bm{\rho}_\mathrm{p},\bm{\eta}_\mathrm{p})$ are given by reactive equilibria [cf.~Eq.~\eqref{eq:profile-eq}], i.e., they lie on the nullcline surface
\begin{equation}
    [\bm\rho_\mathrm{p},\bm{\eta}_\mathrm{p}] 
    = 
    [\bm\rho_\mathrm{p}, \bm{\eta}^*(\bm{\rho}_\mathrm{p})]
    \, .
\end{equation}
The plateaus are only stable if these reactive equilibria are stable in the local (well-mixed) dynamics Eq.~\eqref{eq:well-mixed}.
Linearization of the well-mixed equations, Eqs.~\eqref{eq:well-mixed}, implies that the reactive equilibrium $[\bm{\rho},\bm{\eta}^*(\bm{\rho})]$ is stable if the eigenvalues of ${(\partial_{\bm{\eta}}\Tilde{\mathbf{f}})_{\alpha\beta}=\partial_{\eta_\beta}\Tilde{f}_\alpha}$ have a positive real part.

\subsubsection{Stability in the interface region}
\label{sec:interface-region-stability}

Similar to the local stability of the plateaus, we demand stable dynamics for the mass-redistribution potential in the interface region.
Deviations $\delta\bm{\eta}$ from the stationary mass-redistribution potential $\bm{\eta}^\mathrm{stat,0}$ must relax to zero.
On the scale of the interface width, we assume that the deviations $\delta\bm{\eta}$ are approximately uniform (as is $\bm{\eta}^\mathrm{stat,0}$).
Then, the dynamics of the deviations in the interface region is obtained by linearizing Eq.~\eqref{eq:eta-dynamics} around the stationary pattern $[\bm{\rho}^\mathrm{stat,0}(x),\bm{\eta}^\mathrm{stat,0}]$, which yields
\begin{align}
    \partial_t \delta \eta_\alpha &\approx \sum_\beta 
    \Big[
    \left(- \delta_{\alpha\beta} D_m^\alpha \partial_x^2 - \partial_{\rho_\beta} \Tilde{f}_\alpha|_{\bm{\rho}^\mathrm{stat,0},\bm{\eta}^{\mathrm{stat},0}}
    \right) \delta\rho_\beta - \delta\eta_\beta \, \partial_{\eta_\beta}\Tilde{f}_\alpha|_{\bm{\rho}^\mathrm{stat,0},\bm{\eta}^{\mathrm{stat},0}}
    \Big]
    \nonumber\\
    &\equiv\sum_\beta 
    \big[ 
    \mathcal{L}_{\alpha\beta} \, \delta\rho_\beta - \partial_{\eta_\beta}\Tilde{f}_\alpha 
    \, \delta\eta_\beta
    \big]
    \, .
    \label{eq:eta-stability-dyn}
\end{align}
Here, we introduced the deviations $\delta\bm{\rho}$ from the stationary profiles for the total densities in the interface region, and defined the linear operator $\bm{\mathcal{L}}$.

Thus, to analyze the stability of the interface region, we again have to solve for the eigenvalues $\sigma$ and eigenfunctions $\delta\mathbf{u}=(\delta\rho_1,\dots,\delta\rho_N,\delta\eta_1,\dots,\delta\eta_N)^\mathsf{T}$ of the linear dynamics.
In short, the linear dynamics may be written as ${\partial_t \delta\mathbf{u} = \bm{\mathcal{M}} \, \delta\mathbf{u}}$ with a matrix operator $\bm{\mathcal{M}}$ that contains the diffusion operators $\sim\nabla^2$ and the linearized reaction terms $\partial_{\rho_\beta,\eta_\beta}\tilde{f}_\alpha|_{\bm{\rho}^\mathrm{stat,0},\bm{\eta}^{\mathrm{stat},0}}$.
In contrast to the flat plateau regions, here, the stationary state $\bm{\rho}^\mathrm{stat,0}$, around which the dynamics is linearized, varies spatially.
Therefore, the linear operator $\bm{\mathcal{M}}$, which must be diagonalized, explicitly depends on $x$ because $\partial_{\rho_\beta,\eta_\beta}\tilde{f}_\alpha|_{\bm{\rho}^\mathrm{stat,0},\bm{\eta}^{\mathrm{stat},0}}$ depend on $x$.
The resulting eigenvalue problem is similar to a Sturm--Liouville eigenvalue problem.
However, the linear operator is not necessarily self-adjoint.

Because the linear operator is explicitly space-dependent, the eigenfunctions are not Fourier modes but must be determined alongside the eigenvalues.
Here, the problem simplifies because we assume that the components $\delta\bm\eta$ of the eigenfunction $\delta\mathbf{u}$ are approximately constant in the narrow interface region.
We do not consider other eigenfunctions for which $\delta\bm\eta$ varies strongly within the interface region, i.e., that induce gradients in the mass-redistribution potential leading to mass redistribution within the interface region.

To proceed in determining the dynamics of $\delta\bm{\eta}$, one defines the scalar product (not to be confused with averages that we also denote by angular brackets)
\begin{equation}\label{eq:scalar-product}
    \langle \mathbf{u}, \mathbf{w}\rangle 
    = \int_{-b}^b\mathrm{d}x\, \mathbf{u}\cdot \mathbf{w}
    \, ,
\end{equation}
where the integral spans the interface region $[-b,b]$ with $b\gg\ell_\mathrm{int}$.
Then, one finds the adjoint operator $\bm{\mathcal{L}}^\dagger$ of $\bm{\mathcal{L}}$ with respect to this scalar product using integration by parts and neglecting gradients in the plateaus at ${x \sim\pm b}$.
This yields
\begin{equation}\label{eq:adjoint-L}
(\bm{\mathcal{L}}^\dagger)_{\alpha\beta} \approx - D_m^\alpha \partial_x^2 - \partial_{\rho_\alpha} \Tilde{f}_\beta\, ,
\end{equation}
which differs from $\bm{\mathcal{L}}$ by the interchanged indices of the linearized reaction terms $\partial_{\rho_\alpha} \Tilde{f}_\beta$. 
We then define the zero mode $\mathbf{v}(x)$ of the adjoint operator by
\begin{equation}
\label{eq:adjoint-zero-mode}
    0 = \bm{\mathcal{L}}^\dagger \mathbf{v}\, .
\end{equation}
We will use this zero mode below to solve the eigenvalue problem.
Does the zero mode always exist?
To answer this question, note that differentiation of the profile equation Eq.~\eqref{eq:profile-eq-straight} with respect to $x$ implies
\begin{align}\label{eq:transl-invar}
     0 &= \sum_\beta\left(D_m^\alpha\partial_x^2 + \partial_{\rho_\beta}\Tilde{f}_\alpha\right)\partial_x \rho^\mathrm{stat,0}_\beta =- \bm{\mathcal{L}} \, \partial_x \bm{\rho}^\mathrm{stat,0},
\end{align}
because the stationary mass-redistribution potentials are spatially uniform.
Thus, the translation mode $\partial_x\bm{\rho}^\mathrm{stat,0}$ is a right eigenmode with eigenvalue $0$ of the McRD equations on the infinite line $(-\infty,\infty)$ or if the boundaries are neglected.\footnote{
The derivative $\partial_x\bm{\rho}^\mathrm{stat,0}$ of the stationary pattern is the translation mode because, for ${a\ll \ell_\mathrm{int}}$, it holds that ${\bm{\rho}^\mathrm{stat,0}(x) + a \, \partial_x\bm{\rho}^\mathrm{stat,0}\approx \bm{\rho}^\mathrm{stat,0}(x+a)}$.
}
The underlying reason is the translation invariance of the McRD equations if the boundaries are neglected.
Given a right eigenvector with (isolated, finite-dimensional) eigenvalue $0$ exists, also a left eigenvector with eigenvalue $0$, i.e., the zero vector of the adjoint operator exists if the operator $\bm{\mathcal{L}}$ is closed \cite{Kato1995}, which we assume to be the case.

With the zero mode of the adjoint operator $\mathbf{v}$, one can obtain the dynamics of $\delta\bm{\eta}$ independently of the profile perturbations $\delta\bm{\rho}$, which would have to be calculated separately.
To bypass this calculation, one projects Eq.~\eqref{eq:eta-stability-dyn} onto $\mathbf{v}$ by calculating the scalar product with $\mathbf{v}$.
Using the definition of the adjoint zero mode, the first term in Eq.~\eqref{eq:eta-stability-dyn} vanishes in the scalar product and one obtains the dynamics of the interfacial mass-redistribution potentials 
\begin{equation}\label{eq:eta-interface-dyn}
    \partial_t \, \delta\eta_\alpha \approx - \sum_{\alpha\beta}\delta\eta_\beta \int_{-b}^b\mathrm{d}x \, \frac{v_\alpha(x)}{\Delta v_\alpha} \,  \partial_{\eta_\beta}\Tilde{f}_\alpha \equiv - \sum_{\alpha\beta}\langle\partial_{\eta_\beta}\Tilde{f}_\alpha\rangle_v^\alpha \, \delta\eta_\beta.
\end{equation}
Here, we defined the average
\begin{equation}\label{eq:int-avg-v}
    \langle\bullet\rangle_v^\alpha
    = 
   \int_{-b}^b\mathrm{d}x \, \frac{v_\alpha(x)}{\Delta v_\alpha} \, \bullet
    \, .
\end{equation}
This average is localized to the interface region because $\mathbf{v}$ is localized at the interface. 
We discuss this localization in detail below in Sec.~\ref{sec:adjoint-zero-mode}.
The normalization factor is ${\Delta v_\alpha = \int_{-b}^b\mathrm{d}x\, v_\alpha}$.

Note that some species may have ${\Delta v_\alpha=0}$, and Eq.~\eqref{eq:eta-interface-dyn} cannot be divided by $\Delta v_\alpha$ for those. 
The dynamics of these species cannot be determined from Eq.~\eqref{eq:eta-interface-dyn}.
In particular, if one has ${\bm{\mathcal{L}}=\bm{\mathcal{L}}^\dagger}$ as discussed below in Sec.~\ref{sec:fully-solvable}, it holds that ${\mathbf{v}_\alpha = \partial_x\bm{\rho}^\mathrm{stat,0}}$ and ${\Delta v_\alpha = \Delta \rho_\alpha}$
with $\Delta \rho_\alpha$ denoting the jumps of the total densities across the interface from $\rho_\alpha^-$ at ${x\ll -\ell_\mathrm{int}/2}$ to $\rho_\alpha^+$ at ${x\gg \ell_\mathrm{int}/2}$ by
\begin{equation}
    \Delta\rho_\alpha
    \equiv
    \rho_\alpha^+ 
    -
    \rho_\alpha^-
    \, .
\end{equation}
In this case, the species not constrained by Eq.~\eqref{eq:eta-interface-dyn} fulfilling ${\Delta v_\alpha = 0}$ are those that do not show an interface in their density.

We assume that the analysis of the interface can be restricted to those species that strongly vary in their density across the interface and assume that those also have $\Delta v_\alpha \neq 0$.
Thus, restricting to the case where $\Delta v_\alpha \neq 0$ is fulfilled for all considered species, we define the reaction rates averaged over the interface region
\begin{equation}\label{eq:avg-rates-v}
    \big(
    \langle
    \partial_{\bm{\eta}}\Tilde{\mathbf{f}}
    \rangle_v
    \big)_{\alpha\beta} 
    = 
   \int_{-b}^b\mathrm{d}x \, 
   \frac{v_\alpha(x)}{\Delta v_\alpha} \, \partial_{\eta_\beta}\Tilde{f}_\alpha 
    \, .
\end{equation}
The dynamics of the mass-redistribution potentials, Eq.~\eqref{eq:eta-interface-dyn}, is stable if all the eigenvalues of the matrix $\langle\partial_{\bm{\eta}}\mathbf{f}\rangle_v$ are positive or have a positive real part.

In summary, the previous subsections give conditions that are necessarily fulfilled by stable stationary interface patterns.
In particular, in the interface region, the mass-redistribution potential must relax back to its stationary value.
These conditions restrict the type of reaction terms that sustain stable interface patterns.
Hence, we can use these properties to discuss the properties of the non-equilibrium Gibbs--Thomson relations derived below.

\subsection{General, stationary, weakly curved interfaces}
\label{sec:general-curved}

In this section, we determine the non-equilibrium Gibbs-Thomson relation at weakly curved interfaces.
The profile equation for the weakly curved, stationary interface reads [in the interface region, cf.~Eq.~\eqref{eq:profile-eq-weakly-bend}]
\begin{equation}\label{eq:profile-eq-weakly-curved-2species}
    0 \approx D_m^\alpha\left(\kappa \partial_x+\partial_x^2\right)\rho_\alpha^{\mathrm{stat},\kappa}(x) + \Tilde{f}_\alpha(\bm{\rho}^{\mathrm{stat},\kappa},\bm{\eta}^{\mathrm{stat},\kappa}) \, .
\end{equation}
Next, we expand the profile for the curved interface with respect to the flat interface ${\bm{\rho}^{\mathrm{stat},\kappa}(x)=\bm{\rho}^\mathrm{stat,0}(x)+\delta\bm{\rho}}(x)$ and similarly the mass-redistribution potential ${\bm{\eta}^{\mathrm{stat},\kappa}=\bm{\eta}^\mathrm{stat,0}+\delta\bm{\eta}}$.
By linearizing in the deviations $\delta\bm{\rho}$ and $\delta\bm{\eta}$, since they are expected to scale proportionally to the curvature $\kappa$, one finds to leading order in the interface curvature  (${\kappa \, \ell_\mathrm{int}\ll 1}$)
\begin{equation}\label{eq:profile-eq-weakly-curved-linearized}
    0 \approx D_m^\alpha \kappa \partial_x \rho_\alpha^{\mathrm{stat},0}(x) + \sum_\beta\left[ -\mathcal{L}_{\alpha\beta}\delta\rho_\beta + \delta\eta_\beta \partial_{\eta_\beta}\Tilde{f}_\alpha \right].
\end{equation}
Using that $\bm{\mathcal{L}}^\dagger \mathbf{v}=0$ [cf.\ Eq.~\eqref{eq:adjoint-zero-mode}], the projection on the adjoint zero mode $\mathbf{v}(x)$ yields (again employing ${\ell_\mathrm{int}\ll b\ll \kappa^{-1}}$ within the sharp-interface approximation)
\begin{equation}
    0 \approx \sum_{\alpha,\beta} \int_{-b}^b\mathrm{d}x\, v_\alpha\left[\delta_{\alpha\beta} D_m^\alpha \kappa\partial_x \rho^\mathrm{stat,0}_\alpha + \delta\eta_\beta \partial_{\eta_\beta} \Tilde{f}_\alpha|_{\bm{\rho}^\mathrm{stat,0}(x),\bm{\eta}^{\mathrm{stat},0}}\right].
\end{equation}
As above [cf.\ Eq.~\eqref{eq:eta-interface-dyn}], by projecting on the zero mode of the adjoint operator, the term dependent on the total-density profiles $\delta\bm\rho$ in Eq.~\eqref{eq:profile-eq-weakly-curved-linearized} vanishes.
Thus, one obtains the non-equilibrium Gibbs--Thomson relation [cf.\ Eq.~\eqref{eq:avg-rates-v}]
\begin{equation}\label{eq:deta-kappa-relation-gen}
    \Delta\mathbf{v} \cdot \langle\partial_{\bm{\eta}}\Tilde{\mathbf{f}}\rangle_v \, \delta\bm{\eta} = -\sigma\kappa
\end{equation}
relating $\delta\bm{\eta}$ (denoted $\delta\bm{\eta}_\mathrm{stat}$ in the main text) to the interface curvature $\kappa$ and the effective interfacial tension
\begin{equation}\label{eq:eta-interface-tension-v}
    \sigma =\sum_\alpha D_m^\alpha\int_{-b}^b\mathrm{d}x\, v_\alpha \partial_x \rho^\mathrm{stat,0}_\alpha.
\end{equation}
Assuming as for the stability at the straight interface that one may restrict the analysis to species with ${\Delta v_\alpha \neq 0}$, the matrix $\langle\partial_{\bm{\eta}}\Tilde{\mathbf{f}}\rangle_v$ must have eigenvalues with positive real parts (cf.\ Sec.~\ref{sec:interface-region-stability}).\footnote{
Note that the matrix $\langle\partial_{\bm{\eta}}\Tilde{\mathbf{f}}\rangle_v$ and the vector $\Delta\mathbf{v}$ are defined for the straight interface.
Therefore, within the sharp-interface limit, the dynamics cannot lead from a regime with positive eigenvalues to a regime with negative eigenvalues.}

Moreover, the integral in the effective interfacial tension, Eq.~\eqref{eq:eta-interface-tension-v}, is localized to the interface region because of the localization of both $\mathbf{v}$ and $\partial_x \bm\rho^{\mathrm{stat},0}$.
Because $\Delta\rho_\alpha$ is the density jump across the interface, in the interface region, one expects on dimensional grounds that gradients in the profile scale as ${\partial_x \rho^\mathrm{stat,0}_\alpha (x) \sim \Delta \rho_\alpha/\ell_\mathrm{int}^\alpha}$, where $\ell_\mathrm{int}^\alpha$ is a scale for the width of the interface profile $\rho^\mathrm{stat,0}_\alpha (x)$.
Thus, one estimates 
\begin{equation}
    \int_{-b}^b\mathrm{d}x\, v_\alpha \partial_x \rho^\mathrm{stat,0}_\alpha = \Delta v_\alpha \langle \partial_x \rho^\mathrm{stat,0}_\alpha \rangle_v^\alpha \sim \frac{\Delta v_\alpha \Delta \rho_\alpha}{\ell_\mathrm{int}^\alpha} 
    \, .
\end{equation}
This scaling results in
\begin{equation}
    \sigma \sim \sum_\alpha D_m^\alpha \frac{\Delta v_\alpha \Delta\rho_\alpha}{\ell_\mathrm{int}^\alpha} \sim \sum_\alpha \sqrt{D_m^\alpha f_\mathrm{r}^\alpha} \Delta v_\alpha \Delta\rho_\alpha\sim \sum_\alpha \ell_\mathrm{int}^\alpha f_\mathrm{r}^\alpha \Delta v_\alpha \Delta\rho_\alpha 
    \, ,
\end{equation}
where we used the diffusion length $\ell_\mathrm{int}^\alpha \sim \sqrt{D_m^\alpha/f_\mathrm{r}^\alpha}$ as an estimate for the interface width of species $\alpha$.
Here, $f_\mathrm{r}^\alpha = (\tau_\mathrm{r}^\alpha)^{-1}$ denotes an average rate of attachment and detachment processes (cf.\ Sec.~``Effective interfacial tension'' in the main text). 
The width estimate using the diffusion length scale is implied by the form of the profile equation Eq.~\eqref{eq:profile-eq-straight}.

Equation~\eqref{eq:deta-kappa-relation-gen} clearly resembles the thermodynamic Gibbs--Thomson relation Eq.~\eqref{eq:dMu-Gibbs-Thomson} and constitutes our generalization of it for multi-species McRD systems with two components for each species.
It states that the negative of a generalized Laplace pressure---given in terms of the nonequilibrium effective surface tension times the interface curvature---is balanced by the sum of terms proportional to the shifts in the mass-redistribution potential.
The shifts in the mass-redistribution potential take the place of the shifts in the chemical potentials.
Moreover, the density jumps $\Delta\rho_\alpha$ are replaced by the integrals $\Delta v_\alpha$, which are integrals over the adjoint zero mode.
In contrast, $\Delta\rho_\alpha$ are the integrals of the translation mode itself.
We discuss in the next sections, how the adjoint zero mode is related to the translation mode.

In the biologically relevant limit ${D_m^\alpha\ll D_c^\alpha}$, the mass-redistribution potentials are approximately given by the cytosolic concentrations, ${\eta_\alpha \approx c_\alpha}$ [cf.\ Eq.~\eqref{eq:mass-redistribution-potential}].
Thus, we expect ${(\langle\partial_{\bm{\eta}}\Tilde{\mathbf{f}}\rangle_v)_{\alpha\beta}\sim (\langle\partial_{\bm{c}}\Tilde{\mathbf{f}}\rangle_v)_{\alpha\beta}}$, i.e., the matrix contains the attachment rates of the different species on the diagonal.
If the cytosolic concentrations of the different species do not influence the attachment of each other, the matrix is approximately diagonal.
This assumption is fulfilled in conceptual models of intracellular protein patterns because the cytosolic concentrations are much lower than the membrane densities.
Thus, nonlinear reaction terms have to be considered for the membrane densities, but they are neglected for the cytosolic components.
Then, the matrix is diagonal, and the stability of the stationary interface (cf.\ Sec.~\ref{sec:interface-region-stability}) ensures that these averaged attachment rates are positive.
Thus, the attachment rates only scale the shifts $\delta\bm\eta$ in the non-equilibrium Gibbs--Thomson relation.
In particular, the terms ${\langle\partial_{\eta_\alpha}\Tilde{f}_\alpha\rangle_v \delta\eta_\alpha\sim \langle\partial_{c_\alpha}\Tilde{f}_\alpha\rangle_v \delta c_\alpha}$ with the (average) cytosolic shifts $\delta c_\alpha\approx\delta \eta_\alpha$ describe the additional attachment flow in the different species.
While in the thermodynamic Gibbs--Thomson relation Eq.~\eqref{eq:dMu-Gibbs-Thomson} the sum of the osmotic pressures of the different species balances the Laplace pressure, here, the sum of the additional attachment flows weighted by $\Delta v_\alpha$ balances the relative area change of the attachment and detachment zones quantified by the right-hand side $\sigma\kappa\sim\ell_\mathrm{int}\kappa$ of the non-equilibrium Gibbs--Thomson relation [cf.\ Sec.~\ref{sec:attachment-detachment-balance} and Main Text Sec.~``Effective interfacial tension''].
Below, we show for certain system classes explicitly that ${(\langle\partial_{\bm{\eta}}\Tilde{\mathbf{f}}\rangle_v)_{\alpha\beta}}$ indeed become the average attachment rates at the interface in the biologically relevant limit ${D_m^\alpha\ll D_c^\alpha}$.

Because the shifts in the mass-redistribution potential $\delta\bm{\eta}$ induce gradients in $\bm{\eta}$ and redistribute mass between differently curved interface regions within a QSS approximation, for an (approximately) diagonal, positive matrix $\langle\partial_{\eta_\alpha}\Tilde{f}_\alpha\rangle_v$, the interfaces will evolve analogously to a liquid interface with a positive interfacial tension if the integrals $\Delta v_\alpha$ over the adjoint zero mode have the same signs as the density jumps $\Delta\rho_\alpha$ [cf.\ Eq.~\eqref{eq:deta-kappa-relation-gen} with Eq.~\eqref{eq:dMu-Gibbs-Thomson}].
To analyze this analogy further, we have to analyze the signs of the components of $\Delta \mathbf{v}$ in more detail.
In the remainder of this section, we discuss different classes of systems where $\mathbf{v}(x)$ can be directly expressed in terms of $\bm{\rho}^\mathrm{stat,0}(x)$ such that $\Delta \mathbf{v}$ can be related to $\Delta \bm{\rho}$.
In the last subsection, we discuss that the adjoint zero mode $\mathbf{v}(x)$ is indeed localized to the interface region, a property we have used above.

\subsection{Symmetric second-order mutual-detachment systems}
\label{sec:fully-solvable}
As a minimal model for the PAR system, the two different protein species interact by a symmetric (reciprocal) second-order increase of the detachment terms \cite{Trong.etal2014}.
Generalizing this interaction to $N$ protein species, the reaction term for each protein species $\alpha$ takes the form (cf.\ Methods Sec.~A.1.a)
\begin{equation}\label{eq:potential-reaction-term}
    f_\alpha(\mathbf{m},\mathbf{c})
    = f_\alpha^\mathrm{self}(m_\alpha,c_\alpha)-\sum_{\beta\neq\alpha} k_{\alpha\beta} \, m_\beta^2 \, m_\alpha
    \, ,
\end{equation}
fulfilling reciprocity $k_{\alpha\beta} = k_{\beta\alpha}$.
While we restrict the structure of interaction terms between different species, reactions involving only components of the same species, that is, $f_\alpha^\mathrm{self}$ can be chosen freely.
Moreover, other than the reaction term, the diffusion coefficients can be chosen arbitrarily.

This system class is of interest because its mathematical form allows the stationary equations to be rewritten such that the linear operator $\bm{\mathcal{L}}$ is hermitian, which implies that its adjoint zero mode is equal to the translation mode, i.e., $\mathbf{v} = \partial_x \bm{\rho}^\mathrm{stat,0}$ and $\Delta\mathbf{v}=\Delta\bm{\rho}$.
For this relation, it is essential that the operator $\bm{\mathcal{L}}$ is hermitian, for which the exponent for the species $\beta$ must be one unit larger than that for $\alpha$ in the interaction terms $m_\beta^2 m_\alpha$.
However, we do not expect significant qualitative changes in the physics when we change the interaction order (the exponent of the species $\beta$) away from 2.

\subsubsection{Effective reaction potential}

In this section, we demonstrate that the term representing chemical reactions, Eq.~\eqref{eq:potential-reaction-term}, can be reformulated as the gradient of an effective potential.
This also enables us to recast the stationary pattern profile equations of mass-conserving reaction-diffusion processes in terms of a reaction functional.
However, this formulation \emph{does not} imply that the reaction--diffusion dynamics of this system are mapped onto gradient dynamics typically observed in thermodynamic systems.

Consider the profile equation Eq.~\eqref{eq:profile-eq-weakly-bend}.
It contains the modified reaction terms $\mathbf{f}_\alpha$ as a function of the total densities $\bm\rho$ and the mass-redistribution potentials $\bm\eta$, which, for the model class considered in this section, read
\begin{equation}
    \Tilde{f}_\alpha(\bm{\rho},\bm{\eta})
    = 
    (1-d_\alpha) 
    \bigg[
    f_\alpha^\mathrm{self} (\rho_\alpha,\eta_\alpha) 
    -
    \sum_{\beta\neq \alpha} k_{\alpha\beta} \left(\frac{\rho_\beta\,{-}\,\eta_\beta}{1\,{-}\,d_\beta}\right)^2\left(\frac{\rho_\alpha\,{-}\,\eta_\alpha}{1\,{-}\,d_\alpha}\right)
    \bigg].
\end{equation}
Here we have used the definition of the mass redistribution potential [Eq.~\eqref{eq:mass-redistribution-potential}], and we defined ${f_\alpha^\mathrm{self} (\rho_\alpha,\eta_\alpha) = f_\alpha^\mathrm{self}\big(m_\alpha(\rho_\alpha,\eta_\alpha),c_\alpha(\rho_\alpha,\eta_\alpha)\big)}$.
Interestingly, for the class of systems considered in this section, all reaction terms can be written as the gradients of a single potential $g(\bm\rho,\bm\eta)$,
\begin{equation}\label{eq:f-via-grad-g}
    \Tilde{f}_\alpha (\bm{\rho},\bm{\eta})
    =
    (1-d_\alpha)^2 \, \partial_{\rho_\alpha} g (\bm{\rho},\bm{\eta}),
\end{equation}
given by
\begin{align}
    g(\bm{\rho},\bm{\eta})  
    = \quad &\sum_\alpha \int\mathrm{d}\rho_\alpha \,  \frac{f_\alpha^\mathrm{self}(\rho_\alpha,\eta_\alpha)}{1-d_\alpha}\nonumber\\
    -&\sum_{\alpha<\beta} \frac{k_{\alpha\beta}}{2} \left(\frac{\rho_\beta-\eta_\beta}{1-d_\beta}\right)^2\left(\frac{\rho_\alpha-\eta_\alpha}{1-d_\alpha}\right)^2.
\end{align}
To arrive at this expression, it is necessary for the rates to be reciprocal and for the chosen non-linear form $m_\beta^2 m_\alpha$ to have an exponent for the species $\beta$ that is one unit larger than that for $\alpha$.\footnote{
For one-species, two-component mass-conserving reaction--diffusion systems, no Schwarz symmetry theorem for the derivatives of the reaction term must be fulfilled, and a reaction potential always exists as integral of the reaction term (see, for example, Ref.~\cite{Schlogl1972,Miller.etal2023}).}

As a side note, extending this potential density $g(\bm{\rho},\bm{\eta})$ by square-gradient terms proportional to $|\nabla \rho_\alpha|^2$, the profile equation Eq.~\eqref{eq:profile-eq} results from the extremal principle ${0 = \delta G/\delta\rho_\alpha}$ for a functional $G$ given by (specifying to two-dimensional systems)
\begin{equation}
    G [\bm{\rho},\bm{\eta}] 
    = \int\mathrm{d}^2x \left[\sum_\alpha  \, \frac{D_m^\alpha}{2(1-d_\alpha)^2} \, |\nabla \rho_\alpha|^2 -  g (\bm{\rho},\bm{\eta})\right].
\end{equation}
This is the standard form of many free energy functionals describing phase transitions, for example, for systems exhibiting phase separation, i.e., the Cahn-Hilliard model~\cite{Cahn.Hilliard1958}.

However, it's crucial to recognize a fundamental difference that distinguishes this functional $G$ from a free energy functional in thermodynamics.
In the present context of a reaction--diffusion system, the mass-redistribution potentials $\eta_\alpha$---analogues of the chemical potentials---are \textit{not} given as functional derivatives of the potential $G$ with respect to the densities $\rho_\alpha$.
Instead, $G$ is itself a function of these mass-redistribution potentials that obey a dynamic equation, Eq.~\eqref{eq:eta-dynamics}.
The dynamics of the mass-conserving reaction--diffusion (McRD) dynamics are
\begin{subequations}
\begin{align}
   \partial_t \rho_\alpha 
    &= D_c^\alpha \nabla^2 \eta_\alpha \, , \\
    \partial_t \eta_\alpha 
    &= 
    (D_m^\alpha+D_c^\alpha) \, \nabla^2 \eta_\alpha -  
    (1-d_\alpha)^2 \, \frac{\delta G}{\delta \rho_\alpha}
    \, ,
\end{align}
\end{subequations}
where the derivatives of $G$ contain the reactive dynamics of the mass-redistribution potentials.
This is different from the gradient dynamics of, for example, Cahn-Hilliard models of multicomponent systems \cite{Cahn.Hilliard1958,Eyre1993,Elliott.Luckhaus1991}, in which one has a closed equation for the densities $\rho_\alpha$ and the chemical potentials are given by gradients of the free energy functional.

\subsubsection{Curvature-induced shifts of the stationary mass-redistribution potentials---Non-equilibrium Gibbs--Thomson relation}

Using the reaction potential $g$, the profile equation for a weakly curved interface fulfilling ${\kappa \, \ell_\mathrm{int} \ll  1}$,  Eq.~\eqref{eq:profile-eq-weakly-bend}, can be written (to leading order) as
\begin{equation}\label{eq:profile-eq-weakly-bend-potential}
    0 = \frac{D_m^\alpha}{(1-d_\alpha)^2}\left(\kappa \partial_x+\partial_x^2\right)\rho_\alpha^{\mathrm{stat},\kappa}(x) + \partial_{\rho_\alpha} g|_{\bm{\rho}^{\mathrm{stat},\kappa},\bm{\eta}^{\mathrm{stat},\kappa}}
    \, ,
\end{equation}
with the coordinate $x$ normal to the interface line (see Fig.~\ref{fig:weakly-curved-int}).
We linearize this profile equation, Eq.~\eqref{eq:profile-eq-weakly-bend-potential}, in the deviations ${\delta\bm{\rho}=\bm{\rho}^{\mathrm{stat},\kappa}-\bm{\rho}^\mathrm{stat,0}}$ and ${\delta\bm{\eta}=\bm{\eta}^{\mathrm{stat},\kappa}-\bm{\eta}^\mathrm{stat,0}}$. 
One then projects on the translation mode $\partial_x\bm{\rho}^\mathrm{stat,0}$ using the scalar product Eq.~\eqref{eq:scalar-product}.
This gives to leading order in ${\kappa \, \ell_\mathrm{int}\ll 1}$
\begin{align}\label{eq:Melnikov-zero}
    0 &= \sum_{\alpha,\beta}\int_{-b}^b\mathrm{d}x \left(\partial_x \rho^\mathrm{stat,0}_\alpha\right)
    \left[ \frac{\delta_{\alpha\beta} D_m^\alpha \kappa}{(1-d_\alpha)^2} \, \partial_x \rho^\mathrm{stat,0}_\alpha 
    + \mathcal{L}_{\alpha\beta} \, \delta\rho_\beta 
    + 
    \delta\eta_\beta \, \partial_{\eta_\beta} \partial_{\rho_\alpha} g|_{\bm{\rho}^\mathrm{stat,0},\bm{\eta}^{\mathrm{stat},0}}\right]\nonumber\\
    &\approx M(\kappa, \delta\bm{\eta}) + \sum_{\alpha,\beta}\int_{-b}^b\mathrm{d}x\, \delta\rho_\beta \, \mathcal{L}^\dagger_{\beta\alpha}\left(\partial_x \rho^\mathrm{stat,0}_\alpha\right)
    \, .
\end{align}
Because ${\partial_{\rho_\alpha}\partial_{\rho_\beta}g = \partial_{\rho_\beta}\partial_{\rho_\alpha}g}$, the operator  $\bm{\mathcal{L}}$ is hermitian, ${\bm{\mathcal{L}}^\dagger = \bm{\mathcal{L}}}$, and the last term vanishes because ${0 = \bm{\mathcal{L}}\partial_x \bm{\rho}^\mathrm{stat,0} = \bm{\mathcal{L}}^\dagger \partial_x \bm{\rho}^\mathrm{stat,0}}$, i.e., one has $\mathbf{v} = \partial_x \bm{\rho}^\mathrm{stat,0}$.
Thus, the shift $\delta\bm{\eta}(\kappa)$ is determined by the condition ${0 = M}$ where we define
\begin{equation}\label{eq:Melnikov}
    M(\kappa, \delta\bm{\eta}) \equiv \sum_{\alpha,\beta}\int_{-b}^b \mathrm{d}x \left(\partial_x \rho^\mathrm{stat,0}_\alpha\right)\left[ \frac{\delta_{\alpha\beta} D_m^\alpha \kappa}{(1-d_\alpha)^2}\partial_x \rho^\mathrm{stat,0}_\alpha + \delta\eta_\beta \,  \partial_{\eta_\beta}\partial_{\rho_\alpha} g|_{\bm{\rho}^\mathrm{stat,0}(x),\bm{\eta}^{\mathrm{stat},0}}\right].
\end{equation}
As a side note, this term (within the sharp-interface approximation) corresponds to the Melnikov function in Hamiltonian mechanics \cite{Wiggins2003}.
Because, in the mechanical analogy (Secs.~\ref{sec:straight-int},\ref{sec:weakly-curved-int}), the profile equation corresponds to a dynamics which is Hamiltonian for ${\kappa=0}$, the Melnikov method is applicable to determine the change $\delta\bm{\eta}$ of the stationary mass-redistribution potential $\bm{\eta}^{\mathrm{stat},\kappa}$:
For the heteroclinic orbit to exist at a small, nonzero value of $\kappa$, i.e., for an interface solution to exist, the Melnikov function $M$ has to be zero.

One can write Eq.~\eqref{eq:Melnikov-zero} more concisely by using the interface average
\begin{equation}\label{eq:int-avg}
    \langle\bullet\rangle_\mathrm{int}^\alpha
    = 
   \int_{-b}^b\mathrm{d}x \, \frac{\partial_x \rho^\mathrm{stat,0}_\alpha}{\Delta\rho_\alpha} \, \bullet
    \, .
\end{equation}
Note that the derivatives $\partial_x \rho^\mathrm{stat,0}_\alpha$ are localized to the interface region.
One then has\footnote{
Although some species may have ${\Delta\rho_\alpha = 0}$, the product $\Delta\rho_\alpha \langle\partial_{\bm{\eta}}\Tilde{\mathbf{f}}\rangle_{\alpha\beta}$ is defined, which is all we need.}
\begin{equation}\label{eq:avg-rates}
    \big(\langle(1-\mathbf{d})^{-2}\partial_{\bm{\eta}}\tilde{\mathbf{f}}\rangle_\mathrm{int}\big)_{\alpha\beta} 
    = 
   \frac{1}{(1-d_\alpha)^2} \int_{-b}^b\mathrm{d}x \frac{\partial_x \rho^\mathrm{stat,0}_\alpha}{\Delta\rho_\alpha} \, \partial_{\eta_\beta}\Tilde{f}_\alpha 
    \, ,
\end{equation}
with $\mathbf{d}=\mathrm{diag}(D_m^1/D_c^1,\dots,D_m^N/D_c^N)$.
Additionally, we set
\begin{equation}\label{eq:mod-interfacial-tension-potential}
    \sigma =\sum_\alpha \frac{D_m^\alpha}{(1-d_\alpha)^2}\int_{-b}^b\mathrm{d}x \left(\partial_x \rho^\mathrm{stat,0}_\alpha\right)^2.
\end{equation}

Altogether, one can write the relation Eq.~\eqref{eq:Melnikov-zero} for the shifts of the stationary mass-redistribution potentials as the non-equilibrium Gibbs--Thomson relation
\begin{equation}
\label{eq:dEta-kappa-rel}
    \Delta\bm{\rho} \cdot 
    \big\langle (1-\mathbf{d})^{-2}\partial_{\bm{\eta}}\tilde{\mathbf{f}}
    \big\rangle_\mathrm{int} \,
    \delta\bm{\eta} = -\sigma\kappa
    \, .
\end{equation}
In the main text, we use the abbreviation ${\mathbf{F}_\mathrm{int} := \big\langle(1-\mathbf{d})^{-2}\partial_{\bm{\eta}}\tilde{\mathbf{f}}\big\rangle_\mathrm{int}}$.

The non-equilibrium Gibbs--Thomson relation~\eqref{eq:dEta-kappa-rel}, that applies if reactions can be written as derivatives of a reaction potential, is a special case of the general relation Eq.~\eqref{eq:deta-kappa-relation-gen}.
If a reaction potential exists, the adjoint zero mode is the translation mode of the interface itself (because $\bm{\mathcal{L}}$ is hermitian).
As a result, the obtained non-equilibrium Gibbs--Thomson relation resembles the thermodynamic Gibbs--Thomson relation more closely:
The left-hand side is given by products of the density jumps $\Delta\bm\rho$ and the mass-redistribution potentials $\delta\bm\eta$ while it reads $\Delta\bm\rho\cdot\delta\bm\mu$ in the thermodynamic relation Eq.~\eqref{eq:dMu-Gibbs-Thomson}.
In the following section, we discuss the reaction-rate matrix $\mathbf{F}_\mathrm{int}$ and show that it can be replaced by a diagonal attachment-rate matrix for $D_m^\alpha\ll D_c^\alpha$.
As a result, different average attachment rates of the species only rescale the shifts $\delta\bm\eta$.

\subsubsection{Reaction-rate matrix $\mathbf{F}_\mathrm{int}$}
\label{sec:comp-RDS}

In this section, we show that the reaction-rate matrix $\mathbf{F}_\mathrm{int}$ can be replaced by a matrix containing averaged rates of attachment in the interface region and which is diagonal with positive entries in the biologically relevant limit $D_m^\alpha\ll D_c^\alpha$.
From the definitions of the mass-redistribution potentials, Eq.~\eqref{eq:mass-redistribution-potential}, and $\tilde{f}_\alpha$, Eq.~\eqref{eq:mod-reac-term}, one has
\begin{equation}
    \partial_{\eta_\beta}\Tilde{f}_\alpha(\bm{\rho},\bm{\eta}) = (1-d_\alpha)\partial_{c_\beta}f_\alpha(\mathbf{m},\mathbf{c}) - \frac{1-d_\alpha}{1-d_\beta}\partial_{m_\beta}f_\alpha(\mathbf{m},\bm{\eta}-\mathbf{d}\,\mathbf{m})
\end{equation}
and $\partial_{m_\beta}f_\alpha(\mathbf{m},\bm{\eta}-\mathbf{d}\,\mathbf{m}) = \partial_{m_\alpha}f_\beta(\mathbf{m},\bm{\eta}-\mathbf{d}\,\mathbf{m})$.
Moreover, the chain rule gives
\begin{equation}
    \sum_\alpha\int_{-b}^b\mathrm{d}x \, 
    (\partial_x m_\alpha^\mathrm{stat,0}) \, 
    \partial_{m_\alpha}f_\beta(\mathbf{m},\bm{\eta}-\mathbf{d}\,\mathbf{m})|_{\mathbf{m}^{\mathrm{stat},0}(x),\bm\eta^{\mathrm{stat},0}} = f_\beta|_{b}-f_\beta|_{-b} \approx 0
    \, ,
\end{equation}
because the reaction term vanishes (approximately) in the plateaus, i.e., for $x\sim \pm b$.
As a result, one finds with ${\partial_x \rho_\alpha^{\mathrm{stat},0} 
= (1-d_\alpha)\partial_x m_\alpha^{\mathrm{stat},0}}$ for ${d_\alpha\ll 1}$
\begin{equation}
    \Delta\bm{\rho} \cdot (1-\mathbf{d})^{-2}\langle\partial_{\bm{\eta}}\tilde{\mathbf{f}}\rangle_\mathrm{int}
    \approx
    \Delta\bm{\rho} \langle\partial_{\mathbf{c}}\mathbf{f}\rangle_\mathrm{int}
    \, ,
\end{equation}
with
\begin{equation}
    \langle\partial_{\mathbf{c}}\mathbf{f}\rangle_\mathrm{int}
    =
   \mathrm{diag}\left(\langle\partial_{c_1} f_1^\mathrm{self}\rangle^1_\mathrm{int},\dots,\langle\partial_{c_N} f_N^\mathrm{self}\rangle^N_\mathrm{int}\right)
    \, .
\end{equation}
Thus, in the biologically relevant limit $d_\alpha\ll 1$, the non-equilibrium Gibbs--Thomson relation Eq.~\eqref{eq:dEta-kappa-rel} can be approximated as
\begin{equation}\label{eq:gen-Gibbs-Thomson-diag}
    \Delta\bm{\rho}\cdot \langle \partial_{\mathbf{c}}\mathbf{f}\rangle_\mathrm{int} \, 
    \delta\bm{\eta} 
    \approx 
    - \sigma\kappa
    \, .
\end{equation}
In comparison to phase separation in a thermodynamic system, the amplitude of the shifts $\delta\bm{\eta}$ are scaled by the reaction rates $\langle\partial_{c_\alpha} f_\alpha^\mathrm{self}\rangle^\alpha_\mathrm{int}$.
Thus, Eq.~\eqref{eq:gen-Gibbs-Thomson-diag} shows that the attachment rates $\langle\partial_{c_\alpha} f_\alpha^\mathrm{self}\rangle^\alpha_\mathrm{int}$, i.e., the changes of the reactive flux with the cytosolic concentrations, determine how strongly changes in the mass-redistribution potentials affect the reactive turnover in the interface region, balancing the relative area change of the attachment and detachment zones described by the right hand side of the non-equilibrium Gibbs--Thomson relation $- \sigma \kappa$.

If the averaged attachment rates $\langle\partial_{c_\alpha} f_\alpha^\mathrm{self}\rangle^\alpha_\mathrm{int}$ are comparable for the different species, one can introduce ${f_c\sim \langle\partial_{c_\alpha} f_\alpha^\mathrm{self}\rangle^\alpha_\mathrm{int}}$ and define the effective interfacial tension ${\sigma_f = \sigma/f_c}$. 
Then, Equation~\eqref{eq:gen-Gibbs-Thomson-diag} yields in analogy to multi-component phase separation [cf.~Eq.~\eqref{eq:dMu-Gibbs-Thomson}]
\begin{equation}\label{eq:dEta-kappa-rel-simplified}
    \Delta\bm{\rho}\cdot \delta\bm{\eta} \sim - \sigma_f\kappa.
\end{equation}
This result shows that the classical form of the thermodynamic Gibbs--Thomson relation for multi-component phase separation is recovered exactly for the mass-redistribution potentials in McRD systems in which the different species have similar attachment--detachment kinetics and interact via reciprocal second-order mutual detachment.
This simplest behavior applies for example in the symmetric model of the PAR protein system (cf.\ Methods Sec.~A.1.a)~\cite{Trong.etal2014}.

\subsubsection{Non-equilibrium Gibbs-Thomson relation for symmetric binary RD interfaces}

In this section, we shortly discuss how the non-equilibrium Gibbs--Thomson relation for multi-species systems connects to the simplified relation $\delta\eta \sim \sigma\kappa$ for one-species or symmetric binary systems ($\delta\mu \sim \sigma\kappa$ in incompressible binary liquid mixtures; cf.\ Main Text Sec.~``Effective interfacial tension'' and Methods Sec.~D).
We therefore focus on a binary interface where the domains of two different protein species 1 and 2 meet (A and B in the main text).
Then, Eq.~\eqref{eq:dEta-kappa-rel-simplified} simplifies to
\begin{equation}\label{eq:binary-gibbs-thomson-distinct-jumps}
    |\Delta\rho_1| \, \delta\eta_1 - |\Delta\rho_2| \, \delta\eta_2 \sim \sigma_f \kappa
    \, ,
\end{equation}
Here, we used that the density jumps $\Delta\rho_\alpha$ have opposite sign and chose the orientation of the interface such that ${\Delta\rho_1<0}$, i.e., the normal direction of the interface is chosen to point from species 1 to 2.
Thus, for ${\kappa > 0}$, the mass-redistribution potential of the outward-bend species 1 is increased compared to the mass-redistribution potential of the inward-bend species 2, which is decreased in comparison.

If the reaction terms and diffusion coefficients are chosen symmetrically for species 1 and 2, both species follow symmetric dynamics, leading to symmetric patterns fulfilling ${\Delta\rho_2 = - \Delta\rho_1 \equiv \Delta\rho}$.
The same occurs in thermodynamic, incompressible binary mixtures in which the volume fractions of both species together fill the whole available space such that the change in one density induces the same but opposite change in the other volume fraction.
The non-equilibrium Gibbs--Thomson relation thus reads
\begin{equation}\label{eq:Gibbs-Thomson-binary-symm}
    \delta\eta_1-\delta\eta_2 = \frac{\sigma_f}{\Delta\rho}\kappa.
\end{equation}
Moreover, because both species follow symmetric dynamics, the response to the interface curvature is symmetric.
As one species has an outward-bend interface as the other one has an inward-bend interface, one has ${\delta\eta_1 = - \delta\eta_2 \equiv \delta\eta}$, which gives 
\begin{equation}
    \delta\eta = \frac{\sigma_f}{2\Delta\rho}
    \, \kappa \, .
\end{equation}
The numerical verification of the relation Eq.~\eqref{eq:Gibbs-Thomson-binary-symm} for the symmetric PAR system is shown in the main text, Fig.~1g,h, as an example of the non-equilibrium Gibbs--Thomson relation.
In panel h, the membrane diffusion coefficient of one species is varied.
The induced difference in the diffusion coefficients breaks the symmetry between species A and B.
However, the numerical solution of the profile equation Eq.~\eqref{eq:profile-eq} shows that ${\Delta\rho_2 \approx - \Delta\rho_1}$ holds approximately also in this case if ${d_{1,2} \ll 1}$.
Thus, we again approximated relation Eq.~\eqref{eq:binary-gibbs-thomson-distinct-jumps} by the simplified equation Eq.~\eqref{eq:Gibbs-Thomson-binary-symm} which is shown in the main text figure.

Taken together, the non-equilibrium Gibbs--Thomson relation Eq.~\eqref{eq:gen-Gibbs-Thomson-diag} simplifies to the relation ${\delta\eta\sim\pm\kappa}$ for the individual mass-redistribution potentials of symmetric binary systems.

\subsubsection{Interface width in the effective interfacial tension}
\label{sec:interface-width-tension}
In the main text and SI Secs.~\ref{sec:multi-comp},~\ref{sec:general-curved}, we argue that the effective interfacial tension is proportional to the interface width.
We now demonstrate that the mathematical definition of the interfacial tension Eq.~\eqref{eq:mod-interfacial-tension-potential} indeed scales with the interface width.
To this end, we note that the distribution $\partial_x \rho^\mathrm{stat,0}_\alpha(x)/\Delta\rho_\alpha$ is localized to the interface region and (approximately) normalized on the interval $(-b,b)$.
The average value (height) of this distribution in the peak region is given by $\langle\partial_x \rho^\mathrm{stat,0}_\alpha/\Delta\rho_\alpha\rangle_\mathrm{int}^\alpha$.
As a result, the interface width 
$\ell_\mathrm{int}$ can be estimated by the quantity $\ell_\mathrm{int}^\alpha$ defined by $\ell_\mathrm{int}^\alpha\langle\partial_x \rho^\mathrm{stat,0}_\alpha/\Delta\rho_\alpha\rangle_\mathrm{int}^\alpha= 1$, i.e., by distributing the area under the curve $\partial_x \rho^\mathrm{stat,0}_\alpha/\Delta\rho_\alpha$ into a rectangle of height $\langle\partial_x \rho^\mathrm{stat,0}_\alpha/\Delta\rho_\alpha\rangle_\mathrm{int}^\alpha$ and width $\ell_\mathrm{int}^\alpha$ \cite{Weyer.etal2023}.
Inserting this definition into Eq.~\eqref{eq:mod-interfacial-tension-potential}, the effective interfacial tension $\sigma$ can be written as
\begin{align}
    \sigma &= \sum_\alpha \frac{D_m^\alpha}{(1-d_\alpha)^2} \int_{-\infty}^\infty\mathrm{d}x \left(\partial_x \rho^\mathrm{stat,0}_\alpha\right)^2\nonumber\\
    &= \sum_\alpha \frac{D_m^\alpha}{(1-d_\alpha)^2} \frac{\Delta\rho_\alpha^2}{\ell_\mathrm{int}^\alpha} \, .\label{eq:sigma-int-width-constr}
\end{align}
Next, we use that the interface width can also be approximated by the diffusive length scale set by membrane diffusion \cite{Brauns.etal2020}.
The interface width of species $\alpha$ will scale as ${\ell_\mathrm{int}^\alpha \sim\sqrt{D_m^\alpha \tau_\mathrm{r}^\alpha}}$ where $(\tau_\mathrm{r}^\alpha)^{-1} = f_\mathrm{r}^\alpha$ is an average rate of attachment and detachment of species $\alpha$ in the interface region (cf.\ Sec.~``Effective interfacial tension'' in the main text).
For the special case of two-component McRD systems, it can be shown that $\tau_\mathrm{r}^1=\partial_\eta\Tilde{f}$ with $\partial_\eta\Tilde{f}$ evaluated at the inflection point of the interface \cite{Brauns.etal2020}.
Consequently, inserting this estimate into Eq.~\eqref{eq:sigma-int-width-constr}, one obtains
\begin{equation}
     \sigma\sim \sum_\alpha \frac{\sqrt{D_m^\alpha f_c^\alpha} \Delta\rho_\alpha^2}{(1-d_\alpha)^2} \sim \sum_\alpha \frac{\ell_\mathrm{int}^\alpha f_c^\alpha \Delta\rho_\alpha^2}{(1-d_\alpha)^2} \, .
\end{equation}
Hence, the effective interfacial tension indeed scales with the interface widths $\ell_\mathrm{int}^\alpha$ of the different species.
If the reaction rates and density jumps are comparable, the largest interface width determines the interfacial tension.
This scaling is shown numerically in the main text, Fig.~1h.

Additionally to the interface width, the effective interfacial tension scales with the typical attachment--detachment rates $f_c^\alpha$.
However, the attachment rates also enter on the left-hand side of the non-equilibrium Gibbs--Thomson relation Eq.~\eqref{eq:gen-Gibbs-Thomson-diag}.
If the attachment rates are comparable, one can approximate the rate matrix to be proportional to the identity matrix, that is, one has $\mathbf{F}_\mathrm{int} \approx f_c \mathbf{id}$ [cf.\ Eq.~\eqref{eq:dEta-kappa-rel-simplified}].
In addition, one has $f_c\sim f_c^\alpha$ because both quantities estimate the attachment (and detachment) rates for species $\alpha$ in the interface region.
Inserting this approximation into Eq.~\eqref{eq:dEta-kappa-rel-simplified}, one obtains
\begin{equation}
    \sigma_f\sim \sum_\alpha \frac{\ell_\mathrm{int}^\alpha \Delta\rho_\alpha^2}{(1-d_\alpha)^2} \, .
\end{equation}
This result shows that in the case of comparable attachment rates for all species, that is, in the case that an exact analogy to the thermodynamic Gibbs--Thomson relation can be drawn [cf.\ Eq.~\eqref{eq:dEta-kappa-rel-simplified}], the effective interfacial tension solely scales with the interface width and the density jumps at the interface.
If the attachment rates are distinct, species with larger attachment rates contribute more strongly to the interfacial tension.

\subsection{Binary systems with similar ``self-reactions''}
\label{sec:binary-self-reac}
To construct the reaction functional in the previous section, we had to specify a particular form of the reactive interaction terms [cf.\ Eq.~\eqref{eq:potential-reaction-term}].
Here, we consider the opposite case of arbitrary interaction terms but symmetric ``self-reaction terms''.
The underlying idea we use to mathematically treat such systems is the following:
For two species $\alpha=1,2$, transposition of the linear operator $\bm{\mathcal{L}}$
is similar to an exchange of the two species, i.e., writing the equation for $\rho_2$ in the first and the equation for $\rho_1$ in the second line, or equivalently, labeling the density of the first species by $\rho_2$ and the density of the second species by $\rho_1$.
The only difference is that an exchange of the two species also exchanges the two diagonal terms, which is not the case if the operator is transposed.
If the diagonal terms are identical for both species, transposition can be formulated as the exchange of species.

Therefore, we consider systems that fulfill
\begin{equation}
\label{eq:assumption-selfInt}
    \frac{\partial_{\rho_1}\Tilde{f}_1}{D_m^1} 
    \approx 
    \frac{\partial_{\rho_2}\Tilde{f}_2}{D_m^2}
    \, .
\end{equation}
This applies, for example, to the inhibited-attachment system (see Methods Sec.~A.1.b) in the biologically relevant limit ${D_m^\alpha \ll D_c^\alpha}$ if the membrane-detachment rates and membrane diffusion coefficients are chosen as equal for the two species.
The reason is that only the linear detachment term depends on the membrane density of the own species and one has ${\partial_\rho c(\rho,\eta)\sim d \to 0}$ in the biologically relevant limit.
Thus, the condition Eq.~\eqref{eq:assumption-selfInt} can be easily checked to hold to lowest order in $d_\alpha$.

We introduce the modified linear operator
\begin{equation}
    \Tilde{\bm{\mathcal{L}}}
    \equiv 
    \mathbf{D}_m^{-1}\bm{\mathcal{L}} 
    \, .
\end{equation}
Here, we use the diffusion matrix ${\mathbf{D}_m = \mathrm{diag}(D_m^1, D_m^2)}$.
Then, the definition Eq.~\eqref{eq:eta-stability-dyn} of the linear operator $\bm{\mathcal{L}}$ and the assumption Eq.~\eqref{eq:assumption-selfInt} imply the relation
\begin{equation}
    \Tilde{\bm{\mathcal{L}}}^\dagger \approx \mathbf{S}^\pm \Tilde{\bm{\mathcal{L}}} \mathbf{S}^\pm\, ,
\end{equation}
with the matrices
\begin{equation}
    \mathbf{S}^\pm = \pm\begin{pmatrix}
        0 & 1\\
        1 & 0
    \end{pmatrix},
\end{equation}
which exchange the two species 1 and 2.
The defining equation for the adjoint zero mode ${0 = \bm{\mathcal{L}}^\dagger\mathbf{v}}$ can then be rewritten as
\begin{equation}
0 = \tilde{\bm{\mathcal{L}}}^\dagger \mathbf{D}_m\mathbf{v} = \mathbf{S}^\pm \mathbf{D}_m^{-1}\bm{\mathcal{L}} \mathbf{S}^\pm \mathbf{D}_m\mathbf{v}.
\end{equation}
Because the translation mode solves $\bm{\mathcal{L}}\partial_x\bm{\rho}^\mathrm{stat,0} = 0$, the adjoint zero mode $\mathbf{v}$ is obtained as
\begin{equation}\label{eq:v-constr}
\mathbf{v} = A_v \mathbf{D}_m^{-1}\mathbf{S}^\pm\partial_x \bm{\rho}^{\mathrm{stat},0}.
\end{equation}
We choose the symmetric term ${(D_m^1+D_m^2)/2}$ for the mode amplitude $A_v$ to obtain the same units for the adjoint zero mode as for the translation mode.
Moreover, because $\partial_x \rho_{1}^{\mathrm{stat},0}$ and $\partial_x \rho_{2}^{\mathrm{stat},0}$ have opposite signs, we choose $\mathbf{S}^-$ in the construction of $\mathbf{v}$ in order to preserve the sign of the components $v_\alpha$ with respect to the components of the translation mode $\partial_x \rho_\alpha^{\mathrm{stat},0}$.
Consequently, we have
\begin{equation}
    \mathbf{v} 
    = 
    \begin{pmatrix}
        -\frac{D_m^1+D_m^2}{2 D_m^1} \, \partial_x \rho_2^{\mathrm{stat},0}
        \\
        -\frac{D_m^1+D_m^2}{2 D_m^2} \, \partial_x \rho_1^{\mathrm{stat},0}
    \end{pmatrix} \, .
\end{equation}
Integration of the adjoint zero mode then yields
\begin{equation}
\Delta v_\alpha 
    = \int_{-b}^b\mathrm{d}x\, v_\alpha(x) =
    - \frac{D_m^1+D_m^2}{2 D_m^\alpha} \, 
    \Delta\rho_{3-\alpha}.
\end{equation}
Moreover, the averaged attachment rates defined in Eq.~\eqref{eq:avg-rates-v} read
\begin{equation}
    \left(\langle\partial_{\bm{\eta}}\Tilde{\mathbf{f}}\rangle_v\right)_{\alpha\beta} 
    = 
    \int_{-b}^{b}\mathrm{d}x \, 
    \frac{\partial_x \rho_{3-\alpha}^{\mathrm{stat},0}}{\Delta\rho_{3-\alpha}} \, 
    \partial_{\eta_\beta}\Tilde{f}_\alpha \, .
\end{equation}
Similarly as in Sec.~\ref{sec:comp-RDS}, one can show that these averaged reaction-rates can be replaced by the attachment rates $\partial_{c_\beta}f_\alpha$ in the non-equilibrium Gibbs--Thomson relation.
From the general form of the effective interfacial tension Eq. \eqref{eq:eta-interface-tension-v}, one obtains its expression
\begin{align}
    \sigma 
    = 
    -(D_m^1+D_m^2) \int_{-\infty}^{\infty}\mathrm{d}x\left(\partial_x\rho_1^{\mathrm{stat},0}\right)\left(\partial_x\rho_2^{\mathrm{stat},0}\right) \, .
\end{align}
Note that this effective interfacial tension is positive because $\partial_x\rho_1^{\mathrm{stat},0}$ and $\partial_x\rho_2^{\mathrm{stat},0}$ have opposite signs.
Moreover, the integral
\begin{equation}
    \frac{1}{\ell_\mathrm{int}^{12}} \equiv \frac{1}{\Delta\rho_1\Delta\rho_2} \int_{-\infty}^{\infty}\mathrm{d}x\left(\partial_x\rho_1^{\mathrm{stat},0}\right)\left(\partial_x\rho_2^{\mathrm{stat},0}\right)
\end{equation}
again measures the inverse of the interface width $\ell_\mathrm{int}$.
Taken together, the non-equilibrium Gibbs-Thomson relation Eq.~\eqref{eq:deta-kappa-relation-gen} reads if the reaction term of each species only depends on their own cytosolic concentration
\begin{equation}
    \left[ \frac{(1-d_2) \langle\partial_{c_2}\Tilde{f}_2\rangle_\mathrm{int}^2}{2 D_m^2|\Delta\rho_1|} \delta\eta_1
    -
    \frac{(1-d_1) \langle\partial_{c_1}\Tilde{f}_1\rangle_\mathrm{int}^1}{2 D_m^1|\Delta\rho_2|} \delta\eta_2\right] 
    \approx 
    \frac{\kappa}{\ell_\mathrm{int}^\mathrm{12}} \, .
\end{equation}
This gives an explicit form of the non-equilibrium Gibbs--Thomson relation for binary systems with arbitrary reactive interaction terms but symmetric ``self reactions''.
Importantly, all terms have the same sign as for liquid interfaces with a positive interfacial tension.
Interestingly, different from the systems with reaction potential, the shifts in the mass-redistribution potentials are not scaled by their own but by the average attachment rate of the other species.

\subsection{Interface localization of the adjoint zero mode}
\label{sec:adjoint-zero-mode}

Lastly, we argue that the adjoint zero mode $\mathbf{v}(x)$ is localized to the interface region and falls off exponentially in the plateau regions in general systems.

For our argument, we note that the defining relation $\bm{\mathcal{L}}^\dagger\mathbf{v}=0$ of the adjoint zero mode is the linear, ordinary differential equation [cf.\ Eq.~\eqref{eq:adjoint-L}]
\begin{equation}
    \partial_x^2 \mathbf{v} = - \mathbf{D}_m^{-1} (\partial_{\bm\rho}\tilde{\mathbf{f}})^\mathsf{T} \mathbf{v}\, .
\end{equation}
Here, we defined the diffusion matrix $\mathbf{D}_m =\mathrm{diag}(D_m^1,\dots, D_m^N)$.
Because the stationary pattern has approximately uniform densities in the plateau regions, the matrix $\partial_{\bm\rho}\tilde{\mathbf{f}}\approx\partial_{\bm\rho}\tilde{\mathbf{f}}|_{\bm\rho_p,\bm\eta_p}$ is approximately constant there, while it explicitly depends on $x$ in the interface region.

Consequently, in the plateau regions, the adjoint zero mode fulfills a linear ordinary differential equation with constant coefficients such that the adjoint zero mode can be written as the sum of exponential profiles $\sim \exp(\pm\sigma x)$ with (possibly complex) exponential rates $\pm\sigma$.
The squares of the rates $\sigma^2$ are the eigenvalues of $- \mathbf{D}_m^{-1} (\partial_{\bm\rho}\tilde{\mathbf{f}}|_{\bm\rho_p,\bm\eta_p})^\mathsf{T}$.
Finally, the adjoint zero mode is defined on the interval $[-b,b]$ with arbitrary $b\gg \ell_\mathrm{int}$.
Thus, the adjoint zero mode only remains finite as $b\to\infty$ if it is exponentially decaying in the plateaus.

Taken together, the adjoint zero mode $\mathbf{v}(x)$ is localized to the interface region, and decays exponentially with the distance from the interface in the plateau regions, analogously to the translation mode $\partial_x\bm\rho^{\mathrm{stat},0}(x)$.
Thus, the average $\langle\bullet\rangle_v^\alpha$ is an average over the interface region.

\section{Non-equilibrium Neumann law at triple interface junctions}
\label{sec:gen-Neumann-law}

In Sec.~``Non-equilibrium Neumann law'' in the main text, we argue that the balance of attachment and detachment for each species at the pattern interfaces prescribes the angles under which different interfaces meet at interface junctions.
In this section, we derive mathematically that this balance can be formulated analogously to the Neumann law for triple interface junctions in multi-component liquid mixtures undergoing phase separation.
This non-equilibrium Neumann law is given as Eq.~2 in the main text.
We derive this relation for three-species McRD systems (the three species that form the three domains meeting at the triple junction) in which each species is described by two components.
To derive the non-equilibrium Neumann law, we generalize the singular perturbation methods developed for multi-component Allen--Cahn and Cahn--Hilliard systems in Refs.~\cite{Bronsard.Reitich1993,Bronsard.etal1998}.

\subsection{Derivation of the non-equilibrium Neumann law}

We consider a two-dimensional multi-species system with a membrane and cytosolic component for each protein species [Eq.~\eqref{eq:multiSpecies-twoComps}].
Moreover, this system is chosen to include three species A, B, and C, assuming that the densities of all other species do not vary significantly throughout the junction of the AB, AC, and BC interfaces and can thus be neglected.

\begin{figure}[btp]
\centering
\includegraphics{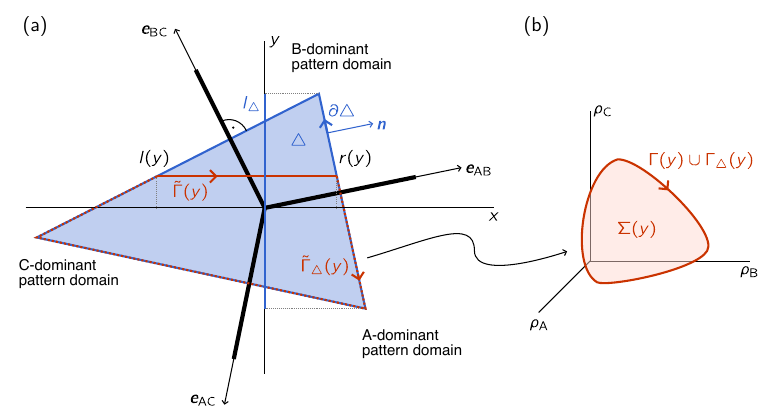}
\caption{
Derivation of the non-equilibrium Neumann law.
    (a) The triangle $\triangle$ (blue) is defined by choosing its sides perpendicular to each of the interfaces (thick, black lines) that meet in the junction
    and point into the direction $\mathbf{e}_\mathrm{AB,AC},$ and $\mathbf{e}_{\mathrm{BC}}$ along which the respective effective interfacial tension acts.
    The side length of the triangle is chosen large compared to the interface width $\ell_\mathrm{int}$ of all interfaces.
    The boundary of the triangle is $\partial\triangle$, and its outward-pointing normal vector is $\mathbf{n}$ (blue).
    The projection of the triangle $\triangle$ onto the $y$ axis is denoted by $I_\triangle$ (blue).
    We define the left and right sides of the triangle as the left and right paths ${\{(l(y),y)\}_{y\in I_\triangle}\subset \partial\triangle}$ and ${\{(r(y),y)\}_{y\in I_\triangle}\subset \partial\triangle}$ connecting the minimum and maximum of the triangle in $y$ direction.
    The axes can always be rotated such that one path contains two sides of the triangle (here it is the left one) and the other one the third side.
    The path $\Tilde{\Gamma}(y)$ connects the two points $(l(y),y)$ and $(r(y),y)$.
    This path is closed by the path $\Tilde{\Gamma}_\triangle(y) \subset \partial\triangle$.
    (b) The closed path $\Tilde{\Gamma}(y)\cup\Tilde{\Gamma}_\triangle(y)$ corresponds to a closed path $\Gamma(y)\cup\Gamma_\triangle(y)$ in the phase space of the total densities $(\rho_\mathrm{A},\rho_\mathrm{B},\rho_\mathrm{C})$.
    This path encloses the area $\Sigma(y)$ in phase space.
    }
    \label{fig:triple-junction}
\end{figure}

The stationary profiles at the triple junction have to fulfill the stationarity condition, Eq.~\eqref{eq:profile-eq}.
For a concise notation, we denote the stationary total-density patterns by ${\bm{\rho} = (\rho_\mathrm{A},\rho_\mathrm{B},\rho_\mathrm{C})}$ and the stationary mass-redistribution potentials by ${\bm{\eta} = (\eta_\mathrm{A},\eta_\mathrm{B},\eta_\mathrm{C})}$.
We also define the ${2 \times 3}$ dimensional matrix of density gradients ${\nabla\bm{\rho} = (\partial_i \rho_\alpha)}$, 
where ${\alpha \in \{A,B,C\}}$ denotes the species and ${i \in \{1,2\}}$  indicates the spatial directions $x_i$.

To find the non-equilibrium Neumann relation, we follow the calculation for Cahn--Hilliard systems given in Ref.~\cite{Bronsard.etal1998}.
In analogy to the derivation of the non-equilibrium Gibbs-Thomson relation at single interfaces [cf.~Eq.~\eqref{eq:Melnikov-zero}], one integrates the profile equation Eq.~\eqref{eq:profile-eq} multiplied by the derivatives $\partial_i \rho_\alpha$ over a triangular region that covers the interface junction, $\triangle$, as shown in Fig.~\ref{fig:triple-junction}a.
This region is chosen to be large compared to the junction core, that is, the junction core and interfaces are assumed to be narrow in comparison (sharp-interface approximation).
The edges of the triangle are aligned perpendicularly to the three interfaces that meet in the junction.
We assume that the interfaces are straight on the scale of the triangle.
From the profile equation Eq.~\eqref{eq:profile-eq} one then obtains
\begin{equation}\label{eq:reactive-turnover-balance-junction}
    0 = \sum_\alpha \int_\triangle\mathrm{d}^2x\left(\partial_i\rho_\alpha\right)\left[D_m^\alpha \nabla^2 \rho_\alpha + \Tilde{f}_\alpha(\bm{\rho},\bm{\eta})\right]
    \, .
\end{equation}
Noting the identity
\begin{equation}
    \sum_{j,\alpha}\left[\partial_j \big[(\partial_i \rho_\alpha)(\partial_j \rho_\alpha)\big] - \frac{1}{2} \partial_i (\partial_j\rho_\alpha)^2\right]
    =
    \sum_\alpha (\partial_i \rho_\alpha)(\nabla^2 \rho_\alpha)
\end{equation}
one can split Eq.~\eqref{eq:reactive-turnover-balance-junction} into a sum of three terms
\begin{align}\label{eq:reactive-turnover-balance-junction-2}
    0 &= \sum_\alpha\int_\triangle\mathrm{d}^2x\left(\partial_i\rho_\alpha\right)\Tilde{f}_\alpha(\bm{\rho},\bm{\eta})\nonumber\\
    &\quad - \frac{D_m^\alpha}{2} \sum_{j,\alpha}\int_\triangle\mathrm{d}^2x\,\partial_i ( \partial_j\rho_\alpha)^2\nonumber\\
    &\quad + D_m^\alpha \sum_{j,\alpha}\int_\triangle\mathrm{d}^2x\, \partial_j [ (\partial_i\rho_\alpha)(\partial_j\rho_\alpha) ] 
    \, .
\end{align}
The first term is an integral over the reaction term.
As we discuss in Sec.~\ref{sec:CH-core-turnover-vanishing} the integrand would be replaced by the local chemical potential for Cahn--Hilliard systems describing phase separation.
As the local chemical potential is derived as the gradient of the local free energy density in these thermodynamic systems, the integral can then be transformed into a boundary integral.
Here, in general, the reaction term does not derive from a reaction potential (cf.\ Sec.~\ref{sec:fully-solvable}) and, thus, cannot be mapped entirely onto the boundary of the triangle.
The second and third terms are integrals over total derivatives and can, therefore, be evaluated on the boundary.

In the following mathematical analysis, the different terms are mapped, as far as possible, onto the triangle boundaries.
Because the triangle boundary traverses only the isolated interfaces far from the interface junction, the boundary integrals depend only on properties of the isolated interfaces like their effective interfacial tensions.
Thus, by mapping the area integrals onto boundary integrals, the reactive turnover balance condition at the junction is mapped to the properties of the individual interfaces, resulting in the non-equilibrium Neumann relation.
We first focus on the total-derivative terms.

\noindent 
\textit{Second term.\;---}
Using Gauss' divergence theorem, and defining the coordinate unit vectors $\mathbf{e}_i$, $i=1,2$, pointing along the directions $x_i$, one has
\begin{align}
    - \frac{D_m^\alpha}{2}\sum_{j,\alpha}\int_\triangle\mathrm{d}^2x\,\partial_i ( \partial_j\rho_\alpha)^2 &= 
    - \frac{D_m^\alpha}{2}
    \int_\triangle\mathrm{d}^2x\,
    \nabla \cdot \left(
    \mathbf{e}_i ( \partial_j\rho_\alpha)^2
    \right)
    \nonumber\\
    & = 
    - \frac{1}{2} 
    \sum_{j,\alpha}
    \int_{\partial\triangle}\mathrm{d}s\,
    n_i D_m^\alpha (\partial_j\rho_\alpha)^2
    \nonumber\\ 
    & = 
    - \frac{1}{2} 
    \sum_{e=\mathrm{AB,AC,BC}} 
    \sum_{j,\alpha}
    \int_{\partial\triangle_e}\mathrm{d}s\,
    n_i D_m^\alpha (\partial_j\rho_\alpha)^2
     \, . \label{eq:second-term-constr-1}
\end{align}
Here, ${\mathbf{n}=(n_1,n_2)}$ denotes the outward-pointing normal of the triangle edges $\partial\triangle$ (parallel to the intersecting interface line, see Fig.~\ref{fig:triple-junction}), and $s$ is their arclength coordinate.
The edge of the triangle intersecting the interface $e$ is denoted by $\partial\triangle_e$.
Because we choose the triangle edges large compared to the interface width (sharp-interface approximation), the triangle edges intersect the interfaces far away from the junction, and each edge covers the whole interfacial region.
Thus, comparing with the effective interfacial tensions Eqs.~\eqref{eq:mod-interfacial-tension-potential},~\eqref{eq:eta-interface-tension-v} introduced through the non-equilibrium Gibbs--Thomson relation, we define the effective interfacial tension of interface $e$
\begin{equation}\label{eq:eff-int-tension-junction}
    \Tilde{\sigma}_e = \sum_{j,\alpha}\int_{\partial\triangle_e}\mathrm{d}s\,D_m^\alpha (\partial_j\rho_\alpha)^2
    \, ,
\end{equation}
which is reminiscent of, but not equal to, the definitions for the non-equilibrium Gibbs-Thomson relations:
The effective interfacial tension Eq.~\eqref{eq:eta-interface-tension-v} for general McRD systems with two components for each species, that we derived from the non-equilibrium Gibbs--Thomson relation, is given by the integral over the translation mode multiplied with the adjoint zero mode.
Here, we instead find, also in the general case, integrals over the square of the translation mode.

Defining the vectorial effective interfacial tension ${\Tilde{\bm\sigma}_e = \Tilde{\sigma}_e \mathbf{e}_e}$, where $\mathbf{e}_e$ is the unit vector parallel to the interface line $e$ pointing away from the junction, one has from Eq.~\eqref{eq:second-term-constr-1}
\begin{align}
    - \frac{D_m^\alpha}{2} \sum_{j,\alpha}\int_\triangle\mathrm{d}^2x\,\partial_i ( \partial_j\rho_\alpha)^2 
    = - \frac{1}{2} \sum_{e=\mathrm{AB,AC,BC}} \Tilde{\bm\sigma}_e\cdot \mathbf{e}_i
    \, .
\end{align}

\noindent 
\textit{Third term.\;---}
Analogously, applying Gauss' theorem to the third term, one finds
\begin{align}
    D_m^\alpha \sum_{j,\alpha}\int_\triangle\mathrm{d}^2x\, \partial_j [ (\partial_i\rho_\alpha)(\partial_j\rho_\alpha) ]
    & = \sum_{j,\alpha}\int_{\partial\triangle}\mathrm{d}s\,n_j D_m^\alpha (\partial_i\rho_\alpha)(\partial_j\rho_\alpha)\nonumber\\
    &= 0.
\end{align}
This term vanishes because $\sum_j n_j \partial_j\rho_\alpha$ is the directional derivative perpendicular to $\partial\triangle$.
Given our assumption that each edge of the triangle intersects a single interface perpendicularly and far away from the junction, this derivative is parallel to the (straight) interfaces and thus vanishes.\\

\noindent \textit{First term.\;---}
We denote the contribution from the reaction term as the modified core turnover
\begin{equation}\label{eq:mod-core-turnover}
    \Tilde{\mathbf{T}}_\mathrm{core} = \sum_\alpha \int_\triangle\mathrm{d}^2x\left(\nabla\rho_\alpha\right)\Tilde{f}_\alpha(\bm{\rho},\bm{\eta}).
\end{equation}
We will decompose this term further below.
This decomposition will show that the integral can be partly mapped onto the triangle boundary as well.
Before we construct this decomposition, we collect all terms.\\

Taking all terms together, Eq.~\eqref{eq:reactive-turnover-balance-junction-2} can be written as
\begin{equation}\label{eq:mod-Neumann-rel-projected}
    \sum_{e=\mathrm{AB,AC,BC}} \Tilde{\bm\sigma}_e\cdot \mathbf{e}_i 
    = 
    2 \,  \Tilde{\mathbf{T}}_\mathrm{core}\cdot \mathbf{e}_i.
\end{equation}
Because this equation has to hold for each spatial direction ${i}$, one obtains the vector identity
\begin{equation}
\label{eq:mod-Neumann-rel}
    \sum_{e=\mathrm{AB,AC,BC}} 
    \Tilde{\bm\sigma}_e = 
    2 \,  \Tilde{\mathbf{T}}_\mathrm{core}.
\end{equation}
This relation reformulates the balance of the total reactive turnovers of all three species Eq.~\eqref{eq:reactive-turnover-balance-junction} at the junction in terms of effective interfacial tensions of the individual interfaces.
After the given transformations, the vectorial sum of the effective interfacial tensions must be balanced by the integrated reactive turnover in the junction region.

In the following, we show that the modified core turnover $\tilde{\mathbf{T}}_\mathrm{core}$ decomposes into a part that can also be expressed in terms of the effective interfacial tensions and a part that only arises if the reaction term $\Tilde{\mathbf{f}}$ is not a gradient field.
To this end, we will express integrals in real space as integrals in the phase space of densities $(\rho_\mathrm{A},\rho_\mathrm{B},\rho_\mathrm{C})$ (cf.~Fig.~\ref{fig:triple-junction}). 
The contribution, arising only for non-gradient terms, will be related to integrals of closed paths in the density phase space.

We determine the decomposition for the component $ \tilde{T}_{\mathrm{core},1}$ of the modified core turnover along the spatial direction ${i=1}$ (``$x$'').
The complementary case ${i=2}$ (``$y$'') proceeds analogously.
First, the area integral in Eq.~\eqref{eq:mod-core-turnover} is rewritten as an integral over line integrals, which can then be expressed as line integrals in the density phase space.
To this end, we define the left and right sides of the triangle as the two paths ${\{(l(y),y)\}_{y\in I_\triangle}\subset \partial\triangle}$ and ${\{(r(y),y)\}_{y\in I_\triangle}\subset \partial\triangle}$ connecting the minimum and maximum of the triangle in $y$ direction (cf.\ Fig.~\ref{fig:triple-junction}a).
The axes can always be rotated such that one path contains two edges of the triangle and the other path is given by the third triangle edge.
Using $l(y)$ and $r(y)$, we first define
\begin{subequations}\label{eq:phi-def}
\begin{align}
    \phi(\mathbf{x}) 
    &\equiv \sum_\alpha\int_{l(y)}^{x}\mathrm{d}x' \left[\partial_{x'}\rho_\alpha((x',y))\right]\Tilde{f}_\alpha(\bm{\rho}((x',y)),\bm{\eta})
    \, , \\
    \phi(y) 
    &\equiv \phi(r(y), y)
    =
    \sum_\alpha \int_{l(y)}^{r(y)}\mathrm{d}x \left(\partial_x\rho_\alpha\right)\Tilde{f}_\alpha(\bm{\rho},\bm{\eta})\label{eq:phi-def-2}
    \, .
\end{align}
\end{subequations}
Here, the spatial variables ${\mathbf{x} = (x_1,x_2)\equiv (x,y)}$ are used.
With this, the modified core turnover reads
\begin{align}
   \tilde{T}_{\mathrm{core},1} &= \sum_\alpha \int_\triangle\mathrm{d}^2x\left(\partial_x\rho_\alpha\right)\Tilde{f}_\alpha(\bm{\rho},\bm{\eta}) \nonumber\\
    &= 
    \int_{I_\triangle}\mathrm{d}y\, \phi(y) \, ,
\label{eq:first-term-phi}
\end{align}
where the integration over $y$ extends over the whole projection $I_\triangle$ (an interval) of the triangle $\triangle$ onto the $y$ axis (cf.\ Fig.~\ref{fig:triple-junction}a).
The integral over $y$ can be transformed into an integral over the triangle boundary $\partial \Delta$ by noting that one has ${\phi((l(y),y))=0}$ on the left side of the triangle.
For the right side, the Jacobian relating the integral over $y$ to the integral over the triangle boundary parameterized by the arclength $s$ is the normal component $n_1$.
This yields
\begin{equation}
   \tilde{T}_{\mathrm{core},1} = \int_{I_\triangle}\mathrm{d}y\, \phi(y)
    = \int_{\partial\triangle}\mathrm{d}s\,n_1(s) \phi(y(s)) \, .
\end{equation}
Thus, we have now expressed the modified core turnover as an integral over the line integrals $\phi(y)$ in real space.

Next, we observe that the integrals $\phi(y)$, Eq.~\eqref{eq:phi-def-2}, can be rewritten as line integrals in the density phase space.
The path $\Tilde{\Gamma}(y)=(x,y)_{x\in[l(y),r(y)]}$ in real space corresponds to a path $\Gamma(y)=\bm{\rho}((x,y))_{x\in[l(y),r(y)]}$ in the phase space of the total densities $\bm\rho$ (red paths in Fig.~\ref{fig:triple-junction}).
Using the chain rule, the integral $\phi(y)$ corresponds to a line integral over the path $\Gamma(y)$ of the vector field $\Tilde{\mathbf{f}}$:
\begin{equation}
    \phi(y) = \int_{\Gamma(y)}\mathrm{d}\bm{\rho}\cdot\Tilde{\mathbf{f}}.
\end{equation}
Complementing the path $\Tilde{\Gamma}(y)$ with a path $\Tilde{\Gamma}_\triangle(y)\subset\partial\triangle$ over the triangle edges that connects the ends of $\Tilde{\Gamma}(y)$, shown as dashed red line in Fig.~\ref{fig:triple-junction}a, one obtains a closed path.
Concomitantly, the corresponding path $\Gamma_\triangle(y)$ in density phase space together with $\Gamma(y)$ encloses a surface $\Sigma(y)$ in phase space [see Fig.~\ref{fig:triple-junction}b].
Using Stokes' theorem, one has
\begin{equation}\label{eq:curl}
    \int_{\Gamma(y)}\mathrm{d}\bm{\rho}\cdot\Tilde{\mathbf{f}} + \int_{\Gamma_\triangle(y)}\mathrm{d}\bm{\rho}\cdot\Tilde{\mathbf{f}} = \int_{\Sigma(y)}\mathrm{d}\mathbf{S}_\rho \cdot \left(\nabla_\rho\times \Tilde{\mathbf{f}}\right)\equiv C_1(y)
    \, ,
\end{equation}
where the surface element $\mathrm{d}\mathbf{S}_\rho$ is a (three-dimensional) vector pointing in the normal direction of the surface $\Sigma(y)$.
The operator $\nabla_\rho = (\partial_{\rho_\mathrm{A}}, \partial_{\rho_\mathrm{B}}, \partial_{\rho_\mathrm{C}})$ denotes derivatives in the phase space and $\nabla_\rho\times \Tilde{\mathbf{f}}$ the rotation of the reaction term in the density phase space.

Using Eq.~\eqref{eq:curl}, one can decompose the modified core turnover Eq.~\eqref{eq:mod-core-turnover} into
\begin{equation}\label{eq:reactive-turnover-decomp}
    \tilde{T}_{\mathrm{core},1} = \int_{\partial\triangle}\mathrm{d}s\,n_1(s)\left[ C_1(y(s))-\int_{\Gamma_\triangle(y(s))}\mathrm{d}\bm{\rho}\cdot\Tilde{\mathbf{f}}\right].
\end{equation}
Importantly, the line integral $C_1(y)$ is an integral over the closed curve $\Gamma(y)\cup\Gamma_\triangle(y)$.
Thus, it is given by an area integral over the curl ${\nabla_\rho\times \Tilde{\mathbf{f}}}$.
Consequently, $C_1(y)$ vanishes if the reaction term is a gradient field in the density phase space.
Moreover, the integrals in the second term are restricted to the triangle boundaries.
Thus, these integrals only intersect single interfaces of the pattern far from the interface junction, and the intersections are perpendicular to the interface line.
This is the sought-for decomposition of the modified core turnover into a ``non-gradient'' term and terms evaluated on the triangle boundary.

Using the profile equation Eq.~\eqref{eq:profile-eq} and that, at each pattern interface, the arclength $s$ agrees with the coordinate normal to the interface, one thus finds
\begin{align}\label{eq:triangle-part-f}
    \int_{\Gamma_\triangle(y)}\mathrm{d}\bm{\rho}\cdot\Tilde{\mathbf{f}} 
    &=
    -\sum_\alpha D_m^\alpha \int_{\Tilde{\Gamma}_\triangle(y)}\mathrm{d}s\, (\partial_s\rho_\alpha) (\partial_s^2\rho_\alpha) \\
    &= \sum_\alpha\frac{D_m^\alpha}{2} \left[\big[\partial_s\rho_\alpha\big(\big(r(y(s)),y(s)\big)\big)\big]^2-\big[\partial_s\rho_\alpha\big(\big(l(y(s)),y(s)\big)\big)\big]^2\right]
    \nonumber \, .
\end{align}
We performed the integral by noting that $(\partial_s\rho_\alpha) (\partial_s^2\rho_\alpha) = \partial_s(\partial_s\rho_\alpha)^2/2$.
With this expression, one obtains as part of the modified core turnover Eq.~\eqref{eq:reactive-turnover-decomp} from the first term in Eq.~\eqref{eq:triangle-part-f}
\begin{align}
    \sum_\alpha\int_{\partial\triangle}\mathrm{d}s&\, n_1 \Big[\partial_s\rho_\alpha\Big(\big(l(x_I(s)),x_I(s)\big)\Big)\Big]^2
    \nonumber\\
    &\propto
    \sum_\alpha \int_{I_\triangle}\mathrm{d}y\,\big[\partial_s\rho_\alpha((l(y),y))\big]^2 - \sum_\alpha \int_{I_\triangle}\mathrm{d}y\,\big[\partial_s\rho_\alpha((l(y),y))\big]^2\nonumber\\
    & = 0
    \, .
\label{eq:triangle-part-f-2}
\end{align}
The projection factor $n_1$ translates the integral over the arclength $s$ of the triangle edges into an integral over $y$.
Because the integral over $\partial\triangle$ covers a closed path, the integral over $y$ appears twice with opposite signs, canceling each other.
In contrast, the first term in Eq.~\eqref{eq:triangle-part-f} gives under this integral
\begin{equation}
    \sum_\alpha \frac{D_m^\alpha}{2} \int_{\partial\triangle}\mathrm{d}s\, n_1 \Big[\partial_s\rho_\alpha\Big(\big(r(y(s)),y(s)\big)\Big)\Big]^2 = \frac{1}{2} \sum_{e=\mathrm{AB,AC,BC}} \Tilde{\bm\sigma}_e\cdot \mathbf{e}_1.\label{eq:triangle-part-f-1}
\end{equation}
Inserting Eqs.~\eqref{eq:triangle-part-f},~\eqref{eq:triangle-part-f-2},~\eqref{eq:triangle-part-f-1} into the modified core turnover Eq.~\eqref{eq:reactive-turnover-decomp}, one arrives at
\begin{align}
    \tilde{T}_{\mathrm{core},1}
    &=
    \int_{\partial\triangle}\mathrm{d}s\,n_1 \left[C_1(y(s)) - \int_{\Gamma_\triangle(y(s))}\mathrm{d}\bm{\rho}\cdot\Tilde{\mathbf{f}}\right]\nonumber\\
    &= 
    \int_{\partial\triangle}\mathrm{d}s\,n_1 C_1(y(s))
    -
    \frac{1}{2} \sum_{e=\mathrm{AB,AC,BC}} \Tilde{\bm\sigma}_e\cdot \mathbf{e}_1.\label{eq:derivation-core-turnover}
\end{align}
We thus define the core turnover
\begin{equation}\label{eq:core-turnover}
    \mathbf{T}_\mathrm{core} = \int_{\partial\triangle}\mathrm{d}s\begin{pmatrix}
        n_1 C_1(y(s))\\
        n_2 C_2(x(s))
    \end{pmatrix}.
\end{equation}
The $x$ component is the residual term depending on the curl contained in the $x$ component of the modified core turnover Eq.~\eqref{eq:derivation-core-turnover}.
Analogously, the $y$ component follows from the same decomposition of the $y$ component of the modified core turnover.
The quantity $C_2(x)$ is the closed line integral in the density phase space (integral over the curl $\nabla_\rho\times\Tilde{\mathbf{f}}$) associated with the path over the boundary of the triangle truncated in $x$ instead of in $y$ direction analogously to the path $\Tilde{\Gamma}(y)\cup\Tilde{\Gamma}_\triangle(y)$ shown in red in Fig.~\ref{fig:triple-junction}a.
Thus, the core turnover $\mathbf{T}_\mathrm{core}$ solely depends on the curl $\nabla_\rho\times\Tilde{\mathbf{f}}$ of the reaction term (line integral over a closed path).
We discuss in the next section in detail that the core turnover does not arise in Cahn--Hilliard systems because, in these systems, the reaction term is replaced by the local chemical potential (difference), which is the gradient (plus a constant) of the local free energy density in these thermodynamic systems.

With this definition of the core turnover and Eq.~\eqref{eq:derivation-core-turnover}, the relation Eq.~\eqref{eq:mod-Neumann-rel-projected} is transformed into
\begin{equation}
    \sum_{e=\mathrm{AB,AC,BC}} \Tilde{\bm\sigma}_e\cdot \mathbf{e}_i = \mathbf{T}_\mathrm{core}\cdot \mathbf{e}_i,
\end{equation}
with $i=1$ (``$x$'').
Because this relation also holds for $i=2$ (``$y$''), one arrives at the non-equilibrium Neumann law [cf.~Eq.~(2) in the main text]
\begin{equation}
    \sum_{e=\mathrm{AB,AC,BC}} \Tilde{\bm\sigma}_e = \mathbf{T}_\mathrm{core}.
\end{equation}
To calculate the core turnover numerically to compare the non-equilibrium Neumann law with simulations in Fig.~2d,e, we use that Eq.~\eqref{eq:mod-Neumann-rel} implies that the core turnover can be expressed as $\mathbf{T}_\mathrm{core} = 2 \Tilde{\mathbf{T}}_\mathrm{core}$.
The numerical analysis of the junctions is described in detail in Sec.~\ref{sec:junction-analysis}.

Moreover, note that $C_1(y)$ [and $C_2(x)$ analogously] is (approximately) zero if the integration path $\Tilde{\Gamma}(y)\cup \Tilde{\Gamma}_\triangle(y)$ does not include the core region because the area $\Sigma(y)$ in the phase space vanishes.
The reason is that, within the sharp-interface approximation, the closed path then cuts an isolated interface(s) twice in opposite directions.
Thus, the area $\Sigma(y)$ collapses (approximately) to a line in the density phase space.
Because it only contributes if the core region is included, we interpret the term $\mathbf{T}_\mathrm{core}$ as a turnover contribution due to the junction core.

In summary, we rewrote the reactive turnover balance Eq.~\eqref{eq:reactive-turnover-balance-junction} within a triangular region around the junction, with its edges aligned perpendicularly to the three interfaces, in terms of integrals over individual interfaces.
These integrals give the vectorial sum of the effective interfacial tensions.
If the reaction term is not a gradient field in the total-density phase space, the reactive turnover balance in the junction region contains an additional contribution that cannot be expressed in terms of the individual interfaces.
This contribution to the turnover is associated with the junction core where all three pattern domains meet.
We show explicitly that this core turnover is given by line integrals over closed paths in the density phase space, or equivalently, integrals over the curl ${\nabla_\rho\times\Tilde{\mathbf{f}}}$ of the reaction term.

\subsection{Vanishing of the core turnover}

The finite core turnover only arises in the non-equilibrium Neumann law.
It does not occur in phase-separating liquid mixtures approaching thermodynamic equilibrium.
In this section, we show the difference explicitly for multi-species Cahn--Hilliard systems that describe phase separation \cite{Bronsard.etal1998}.
Moreover, we show that the core turnover can be eliminated by a reformulation of the non-equilibrium Neumann law if the reaction term can be derived from a reaction potential [cf.\ Eq.~\eqref{eq:f-via-grad-g}].
Finally, we give a heuristic argument as to why the core turnover also vanishes in the non-equilibrium Neumann law at the analogue of the transition to full wetting.
As a result, the transition to full wetting in McRD systems is the same as for liquid mixtures (see Sec.~``Non-equilibrium Neumann law'' and Fig.~2e in the main text).

\subsubsection{Multi-species Cahn--Hilliard dynamics}
\label{sec:CH-core-turnover-vanishing}

The dynamics of a $N+1$-species phase-separating system is described by a $N$-component Cahn--Hilliard system for $N$ volume fractions $\bm\phi$ while the species ${N+1}$ is the solvent occupying the rest of the volume (cf.\ Sec.~\ref{sec:Gibbs-Thomson-phase-sep}).
The $N$-component Cahn--Hillard system reads \cite{Eyre1993,Elliott.Luckhaus1991}
\begin{align}
    \partial_t \bm\phi 
    &= 
    \nabla( \bm{\Gamma}\nabla \bm\mu)
    \, , \\
    \bm\mu 
    &= 
    -\mathbf{K} \nabla^2 \bm\phi + \nabla_\phi g(\bm\phi)
    \, ,
\end{align}
where $g(\bm\phi)$ denotes the locel free-energy density.
The mobility matrix $\bm\Gamma$ and the rigidity matrix $\mathbf{K}$ are assumed to be constant and diagonal and, in the simplest case of symmetric mixtures, both are (proportional to) the identity matrix.
The amplitude of the rigidity matrix sets the width of the interfaces.

A stationary state is reached in a system with no-flux or periodic boundary conditions if the chemical potentials $\bm\mu_\mathrm{stat}$ are spatially uniform.
Consequently, the stationary profiles of the volume fractions $\bm\phi_\mathrm{stat}$ are determined by
\begin{equation}\label{eq:profile-eq-CH}
    0 = \mathbf{K} \nabla^2 \bm\phi_\mathrm{stat} + \bm\mu_\mathrm{stat}-\nabla_\phi g(\bm\phi_\mathrm{stat}) \, .
\end{equation}
The mathematical form of this equation agrees with the profile equation Eq.~\eqref{eq:profile-eq} for McRD systems with two components for each species if the total densities $\bm\rho$ are identified with the volume fractions $\bm\phi$, the membrane diffusion coefficients $\mathbf{D}_m$ with $\mathbf{K}$, and the reaction term $\Tilde{\mathbf{f}}$ with ${\bm\mu_\mathrm{stat}-\nabla_\phi g(\bm\phi)}$ (cf.~\cite{Weyer.etal2023}).
Hence, our derivation of the non-equilibrium Neumann law also holds for liquid mixtures.
Thus, the core turnover can be calculated using Eqs.~\eqref{eq:curl},~\eqref{eq:core-turnover} by integrating the curl ${\nabla_\phi\times [\bm\mu_\mathrm{stat}-\nabla_\phi g(\bm\phi_\mathrm{stat})]}$.
In general, the curl of a gradient field vanishes such that it holds ${\nabla_\phi\times [\bm\mu_\mathrm{stat}-\nabla_\phi g(\bm\phi_\mathrm{stat})] = 0}$ and the core turnover ${\mathbf{T}_\mathrm{core}}$ vanishes.
The reason is the gradient structure due to the underlying free energy.
As a result, the interface junctions in multi-species Cahn--Hilliard systems fulfill the (classical) Neumann law, as derived in Ref.~\cite{Bronsard.etal1998}.

For McRD systems, the derivation of the (generalized) Neumann law links the reactive turnover balance at the junction to the balance of the vectorial effective interfacial tensions (and the core turnover).
Applying it to liquid mixtures, it derives the force balance of the interfacial tensions from the osmotic pressure balance at the junction:
This becomes clear from the integral of the ``reaction term'' in the initial balance equation for the triangle Eq.~\eqref{eq:reactive-turnover-balance-junction}.
Consider, for instance, the integral in $x$ direction in the $x$ component
\begin{equation}
    \sum_\alpha\int_{l(y)}^{r(y)} \mathrm{d}x\,  \partial_x \phi_\alpha [\mu_\alpha - \partial_{\phi_\alpha}g] = [\bm\phi(r)-\bm\phi(l)]\cdot \bm\mu - [g(\bm\phi(r))-g(\bm\phi(l))].
\end{equation}
If the integral boundaries $l(y),r(y)$ lie outside the interface regions, $\bm\phi(l),\bm\phi(r)$ are the densities of the different phases.
Because the chemical potentials are equal in all phases in thermodynamic equilibrium and fulfill $\bm\mu = \nabla_{\bm\phi} g(\bm\phi(l))= \nabla_{\bm\phi} g(\bm\phi(r))$, the integral gives the difference
\begin{equation}
    [g(\bm\phi(l))-\bm\phi(l)\cdot\nabla_{\bm\phi} g(\bm\phi(l))]-[g(\bm\phi(r))-\bm\phi(r)\cdot\nabla_{\bm\phi} g(\bm\phi(r))] = p(\bm\phi(l))-p(\bm\phi(l)).
\end{equation}
The two terms are the osmotic pressures of the phases at the two integral boundaries.
Thus, for liquid mixtures, the balance condition Eq.~\eqref{eq:reactive-turnover-balance-junction} calculates the balance of the osmotic pressures acting along the triangle boundaries.

Integrating the stationary profile equation Eq.~\eqref{eq:profile-eq-CH} across a single straight interface multiplied by $\partial_x \bm\phi_\mathrm{stat}$ instead of a junction, i.e., the analogue of the balance condition Eq.~\eqref{eq:reactive-turnover-balance-junction} for a single interface, one obtains that this osmotic-pressure difference must be zero.
Hence, for a single interface, the integral gives the condition that the osmotic pressures are the same in both phases, leading to the common tangent construction.
Integrating over the junction, this balance condition becomes the force balance of the interfacial tensions instead.

\subsubsection{McRD systems with a reaction potential}
In Sec.~\ref{sec:fully-solvable}, we discuss for the non-equilibrium Gibbs--Thomson relation that, in certain cases, the reaction term can be derived from a reaction potential by some integrating factor [$(1-d_\alpha)^2$ in Eq.~\eqref{eq:f-via-grad-g}].
Because the reaction term then becomes a gradient field [multiplied by the matrix of integrating factors; cf.\ Eq.~\eqref{eq:f-via-grad-g}], these systems closely reproduce the behavior of liquid mixtures.

In particular, one can divide the individual profile equations Eq.~\eqref{eq:profile-eq} by the integrating factors, and then calculate the reactive turnover balance in the junction region Eq.~\eqref{eq:reactive-turnover-balance-junction} for these modified profile equations.
For a three-species second-order mutual detachment system, as discussed in Sec.~\ref{sec:fully-solvable}, this yields the reactive turnover balance in the junction region
\begin{equation}
    0 = \sum_\alpha \int_\triangle\mathrm{d}^2x\left(\partial_i\rho_\alpha\right)\left[\frac{D_m^\alpha}{(1-d_\alpha)^2} \nabla^2 \rho_\alpha + \partial_{\rho_\alpha} g\right]
    \, .
\end{equation}
Consequently, the reaction term is a gradient field.
At the same time, the membrane-diffusion coefficients are modified by the integrating factor.

From this reformulation, one obtains a modified version of the non-equilibrium Neumann law:
As the reaction term is transformed into a gradient field, the core turnover vanishes, although it has a finite value if one does not divide by the integrating factor.
Moreover, the modified diffusion coefficients alter the effective interfacial tensions $\tilde{\sigma}_e$, $e =\mathrm{AB,AC,BC}$, into [cf.\ Eq.~\eqref{eq:eff-int-tension-junction}]
\begin{equation}
    \Tilde{\sigma}_e = \sum_{j,\alpha}\frac{D_m^\alpha}{(1-d_\alpha)^2}\int_{\partial\triangle_e}\mathrm{d}s\, (\partial_j\rho_\alpha)^2.
\end{equation}
Importantly, as in thermodynamic liquid mixtures, these agree with the effective interfacial tensions $\sigma_\mathrm{AB,AC,BC}$ Eq.~\eqref{eq:mod-interfacial-tension-potential} that one derives from the non-equilibrium Gibbs--Thomson relation.

As a result, if the reaction term can be derived from a reaction potential the non-equilibrium Neumann law can be brought in the form of the classical Neumann law without the core turnover and with effective interfacial tensions that are the same as in the non-equilibrium Gibbs--Thomson relation.

\subsubsection{Transition to ``full wetting''}

In the main text, Fig.~2e, we observe: If one effective interfacial tension, say $\Tilde{\sigma}_\mathrm{BC}$, is larger than the sum of the other two $\Tilde{\sigma}_\mathrm{AB}+\Tilde{\sigma}_\mathrm{AC}$, this interface (BC) vanishes, and it is replaced by the other two interfaces as species A covers B and C completely (analogue of full wetting in phase separation).
Consequently, the condition for ``full wetting'' is observed to be the same as for phase-separating systems.
Numerically, one finds that the core turnover $\mathbf{T}_\mathrm{core}$ vanishes at the threshold [cf.\ Fig.~2e].
We shortly discuss why we expect this to be generic for the McRD systems with two components for each species.

If the pattern depends continuously on the variation of the parameter that is tuned to cross the threshold to ``full wetting'' in McRD systems, ($k_\mathrm{BC}$ in Main Text Fig.~2e), the profile of the BC interface continuously approaches the profile of an AB interface next to an AC interface.
Beyond the threshold, the junction has vanished, and only separated interfaces exist.
Now consider the definition of the (modified) core turnover, Eq.~\eqref{eq:mod-core-turnover},
\begin{equation}\label{eq:turnover-integral-singleInterface}
    \Tilde{\mathbf{T}}_\mathrm{core} = \frac{\mathbf{T}_\mathrm{core}}{2} = \int_\triangle\mathrm{d}^2x\left(\nabla\rho_\alpha\right)\Tilde{f}_\alpha(\bm{\rho},\bm{\eta})
    \, .
\end{equation}
How does the core turnover behave at the analogue of the threshold to full wetting for McRD systems?
If this turnover integral would only cover a single interface, we can evaluate the integral by introducing the new coordinates $(s,r)$ with $s$ perpendicular and $r$ parallel to the interface.
Then, integrating over a domain whose edges cut the interface perpendicularly (as $\triangle$ does), one has
\begin{equation}
    \int\mathrm{d}^2x\left(\partial_i\rho_\alpha\right)\Tilde{f}_\alpha(\bm{\rho},\bm{\eta})\propto \int\mathrm{d}s\left(\partial_s\rho_\alpha\right)\Tilde{f}_\alpha(\bm{\rho},\bm{\eta}) 
    \, ,
\end{equation}
because all derivatives parallel to the interface vanish ($\partial_r\rho_\alpha = 0$).
The resulting integral vanishes due to the reactive turnover balance Eq.~\eqref{eq:reactive-turnover-balance-2c} of single interfaces if the integration domain extends over the whole interface width.

As a result, the turnover integral Eq.~\eqref{eq:turnover-integral-singleInterface} vanishes where the integral only covers isolated interfaces and its boundaries are perpendicular to the interfaces.
As the threshold to ``full wetting'' is approached, the gradual decomposition of the BC into an AB and an AC interface implies that the core-turnover integral ${\Tilde{\mathbf{T}}_\mathrm{core}=\mathbf{T}_\mathrm{core}/2}$ covers increasingly separated, single interfaces.
Thus, assuming that the area of the junction core does not diverge, the magnitude of the core turnover decreases because the junction core decomposes into increasingly isolated interfaces.
At the threshold, the emerging AB and AC interfaces become perfectly isolated and the core turnover vanishes.

\section{Distribution of random vertex angles}
\label{sec:random-angles}

\begin{figure}[btp]
    \centering
    \includegraphics[width=\textwidth]{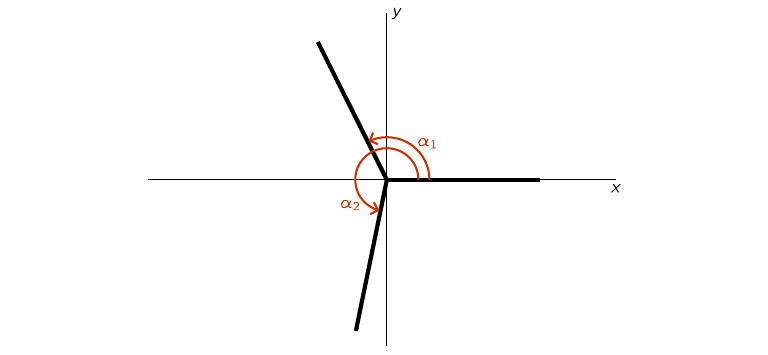}
    \caption{Construction of the random vertex angle distribution.
    One interface of the triple vertex is fixed and the relative directions of the other two interfaces are described by the angles $\alpha_{1,2}$.
    }
    \label{fig:random-angles}
\end{figure}

In the main text, Fig.~4d, we compare the histogram of experimentally measured vertex angles with the distribution of random vertex angles.
To calculate the distribution of random vertex angles $\theta$, we consider triple vertices at which three interfaces meet.
Without loss of generality, the vertex is positioned such that one interface points along the positive $x$-axis (Fig.~\ref{fig:random-angles}).
The two angles $\alpha_{1,2}\in [0,2\pi)$ describe the directions of the other two interfaces.
Let us consider the case ${\alpha_1 < \alpha_2}$.
The opposite case ${\alpha_1 \geq \alpha_2}$ is obtained by exchanging the indices.

Given this ordering, a vertex angle $\theta$ can occur either between $0$ and $\alpha_1$, $\alpha_1$ and $\alpha_2$, or between $\alpha_2$ and $2\pi$. 
The first possibility occurs if one has ${\alpha_1 = \theta}$ and ${\alpha_2 > \alpha_1}$.
In the second case, one has ${\alpha_1 < 2\pi-\theta}$ and ${\alpha_2 = \alpha_1+\theta}$.
In the last case, the angle $\theta$ arises if ${\alpha_2 = 2\pi-\theta}$ and ${\alpha_2 > \alpha_1}$.
Adding the probabilities for the three possibilities, the probability density $P(\theta)$ for the vertex angle $\theta$ is given by
\begin{align}\label{eq:random-vertex-angles-deriv}
    P(\theta) 
    = 
    \frac{2}{3} 
    \big[
    &P(\alpha_1=\theta) P(\alpha_2>\theta) \nonumber\\
    &+ P(\alpha_1<2\pi-\theta) P(\alpha_2=\alpha_1+\theta) \nonumber\\
    &+ P(\alpha_2=2\pi-\theta) P(\alpha_1<2\pi-\theta)
    \big] \, ,
\end{align}
where ${P(\alpha_i<\phi) = \int_0^\phi\mathrm{d}\alpha\, P(\alpha_i=\alpha)}$ is the cumulative probability.
The prefactor $2$ describes that all three configurations also occur with interchanged indices $1$ and $2$, i.e., for ${\alpha_1 \geq \alpha_2}$.
In addition, the different configurations allow $\theta$ to occur as first, second, or third vertex angle such that one has to divide by three to normalize the probability density to one.

Random interface orientations imply uniform probability densities on the unit circle, that is, one has ${P(\alpha_i=\phi) = (2\pi)^{-1}}$.
As a result, one obtains from Eq.~\eqref{eq:random-vertex-angles-deriv}
\begin{equation}
    P(\theta) = 2 \, \frac{2\pi-\theta}{(2\pi)^2} = \frac{1}{\pi}\left(1-\frac{\theta}{2\pi}\right).
\end{equation}
Measuring the angles in degrees, one has
\begin{equation}
    P_\mathrm{d}(\theta) = \frac{1}{\SI{180}{\degree}}\left(1-\frac{\theta}{\SI{360}{\degree}}\right).
\end{equation}
To obtain the histogram of random vertex angles in Fig.~4d, the probability density is multiplied by the total number of vertex angles times the bin width.

\section{Junction analysis}
\label{sec:junction-analysis}

To analyze the triple-junction angles and positions in numerical simulations, the junctions are fitted in a two-step procedure that we describe in this section. In short, rough junction positions are determined first.
In the second step, the exact junction position and the interface directions are determined by fitting a junction model parameterized by the position and the three interface directions.
As described below, tailored fitting procedures are used to determine the parameter dependence of a single junction (Fig.~2d,e in the main text) and the junctions in the coarsening simulations (Fig.~3) in the main text.
Moreover, we describe below how the non-equilibrium Neumann law is tested in the numerical simulations.
The analysis is implemented in Mathematica 13.1 and the notebooks are available at \url{https://github.com/henrikweyer/Turing-foams}.

For the single junctions analyzed in Fig.~2d,e in the main text, the junction is positioned in a circular domain with radius ${R = 15}$.
In steady state, the junction connects interfaces which are either straight or circular arcs (in the sharp-interface limit).
To determine the interface lines, straight lines or circular arcs are fitted to
the squared gradients of the membrane densities of species B and C $|\nabla m_\mathrm{B,C}|^2$ which are peaked at the interfaces.
Importantly, as one junction angle approaches zero, the junction moves towards the edge of the domain.
To resolve the junction close to the transition to full wetting, the domain is enlarged to ${R = 30}$ for the three smallest angles measured in Fig.~2e in the main text.
However, due to the finite system size, the exact parameter value of the transition to full wetting cannot be determined (red-shaded area in Fig.~2e in the main text).
To compare these measurements of the junction angles with the non-equilibrium Neumann relation, the effective interfacial tensions are measured at the interfaces far from the junction (in the same simulation).
In addition, the core turnover is determined as integral over the junction core (see Sec.~\ref{sec:gen-Neumann-law}).
Importantly, one has to correct for the interface curvature, which shifts the turnover balance and results in a non-zero contribution compared to straight interfaces.
To this end, the curvature-induced contribution is determined by integrating the core-turnover expression at the interfaces far away from the junction.
This contribution is subtracted from the integral over the junction core, assuming that the interfaces are circular arcs up to their meeting point.

In contrast to the simulations of single junctions, in the coarsening simulations Fig.~3 in the main text, the junctions are close to each other, and the junction angles have to be fitted using only short parts of the interfaces around the junctions.
Therefore, we fit a parametrized, approximate junction profile to ${I(\mathbf{x}) = \sum_{\alpha=A,B,C}D_\mathrm{m}^\alpha |\nabla \rho_\alpha|^2}$.
To this end, the junction positions are roughly determined by finding the branch points of $I(\mathbf{x})$, which is only nonzero close to the interfaces, with Mathematica's morphological algorithms.
Then, the junction is fitted by minimizing a loss function $L$ explained below using the data within a radius ${R = \min(1,r)}$ around the branch point, where $r$ is the distance to the closest junction (``junction data'').
If a junction is close to a neighboring junction (we choose the threshold ${r < 0.5}$), this particular junction is excluded from the set of junctions that is fitted at the given time point because the direction of the short branch cannot be determined reliably.

To construct the parameterized, approximate junction model that will be fitted to $I(\mathbf{x})$, the profiles of $I(\mathbf{x})$ across isolated interfaces $s_\mathrm{AB,AC,BC}(x)$ are determined far away from the junction core in the simulations of single junctions at the same parameter values (Main Text Fig.~2d,e).
Here, $x$ denotes the normal coordinate of the interface (cf.\ Sec.~\ref{sec:multi-comp}).
Then, given a junction position $\mathbf{p}$, interface directions $\mathbf{e}_i$, and interface normal vectors $\mathbf{n}_i$, ${i \in \{ \mathrm{AB,AC,BC} \}}$ and ${|\mathbf{e}_i| = |\mathbf{n}_i| = 1}$, the junction model is constructed as
\begin{equation}
    j(\mathbf{x}) = s_{i(\mathbf{x})}\left((\mathbf{x}-\mathbf{p})\cdot\mathbf{n}_{i(\mathbf{x})}\right),
\end{equation}
where ${i(\mathbf{x})}$ denotes the closest interface $i$, i.e., the interface for which ${|(\mathbf{x}-\mathbf{p})\cdot\mathbf{n}_{i}|}$ is smallest and one has ${(\mathbf{x}-\mathbf{p})\cdot\mathbf{e}_{i}>0}$.

This junction model is fitted by minimizing a loss function $L(\mathbf{p},\{\gamma_i\}_{i=\mathrm{AB,AC,BC}})$.
It depends on the junction position $\mathbf{p}$ and the interface angles $\gamma_i$ defined by ${\mathbf{e}_i = [\cos(\gamma_i),\sin(\gamma_i)]}$.
The loss function reads
\begin{equation}\label{eq:loss-junctions}
    L(\mathbf{p},\{\gamma_i\}) = \frac{1}{\sum_k w(\mathbf{x}_k)}\sum_k w(\mathbf{x}_k) \left[I(\mathbf{x}_k)-j(\mathbf{x}_k)\right]^2,
\end{equation}
where the index $k$ denotes the individual data points of the junction data (``pixels'' of the density-field simulations).
The contributions from the different data points $k$ are weighted by the weight functions $w(\mathbf{x}_k)$.
Because the interface profiles $s_i(x)$ in the junction model $j$ are determined far from the junction core, the weights are chosen such that the contributions from data points in the junction core are suppressed.
We choose ${w(\mathbf{x}) = \frac{1}{2}[1+\tanh(|(\mathbf{x}-\mathbf{p})\cdot\mathbf{e}_{i(\mathbf{x})}|-0.3)]}$.
Before the loss function is minimized, intermediate steps based on finding the maxima of the angular histogram of the squared density gradients and the fitting of the single density domains ensure adequate initial conditions for the minimization.

Lastly, the direction of movement of the junctions (cf.\ Fig.~3 in the main text) is determined by comparing their position between consecutive time points $t_l$ on the logarithmic scale ${\log_{10} t_l = 2,2.1,2.2,\dots,5}$.
A junction at $t_l$ is related to that junction at $t_{l+1}$ which is closest.
The histograms are constructed from the collection of measurements for the pairs of consecutive time points and the different junctions.

\section{Turing-foam analysis}
\label{sec:foam-analysis}
The vertices of the Turing foam (Figs.~4,~5 in the main text) are analyzed using similar ideas as for the junctions (see Sec.~\ref{sec:junction-analysis}).
The analysis is again implemented in Mathematica 13.1 and the notebooks are available at \url{https://github.com/henrikweyer/Turing-foams}.

In the simulations, the Turing-foam structure (steady-state pattern) is recorded at time ${t = 10^{10}\si{s}}$.
The skeleton network of the MinE membrane density ${m_\mathrm{e}^\mathrm{tot} = m_\mathrm{de}+m_\mathrm{e}}$ (shown in the main text in Fig.~4h and Fig.~5a,g in cyan) is determined using Mathematica's morphological algorithms.
For further analysis, the maximum of $m_\mathrm{e}^\mathrm{tot}$ is normalized to one.
The branch points of the skeleton network give the vertices, which are then fitted similarly to the interface junctions, fitting branches instead of single interfaces; see Sec.~\ref{sec:junction-analysis}.
The fit data is chosen within the radius $R=\min(\SI{5}{\micro m},r)$ around each branch point, where $r$ is the distance to the closest neighboring branch point.
The branch cross-section is approximated by a Gaussian profile, i.e., analogously to the junction model in Sec.~\ref{sec:junction-analysis} we use the vertex model ${j(\mathbf{x}) = \exp\left[-4 \log(2) ((\mathbf{x}-\mathbf{p})\cdot\mathbf{n}_{i(\mathbf{x})})^2/\delta^2\right]}$.
Here, $\delta$ is the full width at half maximum of the branches, which is approximated by the number of pixels of $m_\mathrm{e}^\mathrm{tot}$ with more than half the maximum intensity divided by the number of pixels contained in the skeleton network (multiplied by the pixel side length).
The loss function minimized during fitting is chosen as [cf.~Eq.~\eqref{eq:loss-junctions}]
\begin{equation}
    L(\mathbf{p},\{\gamma_i\}) = \sum_k \left[m_\mathrm{e}^\mathrm{tot}(\mathbf{x}_k)-j(\mathbf{x}_k)\right]^2.
\end{equation}
A varying weight $w(\mathbf{x}_k)$ for the different data points depending on their distance from the vertex core was not necessary to obtain reliable fits.
The angle histograms are created by collecting the three vertex angles of all vertices.
Because the three angles sum to $\SI{360}{\degree}$ exactly, the average of the angle histograms is exactly $\SI{120}{\degree}$ [cf.\ Fig.~4d,i in the main text].

The domain areas, as well as internal MinE branches, are determined by measuring the single domains enclosed by the skeleton network using Mathematica's morphological algorithms.
To analyze the dynamics of single domains over time, the domains are tracked by identifying those that are closest to each other after one time step.
The time points follow a logarithmic scale.

The experimental mesh patterns are analyzed similarly.
Again, the skeleton network and its branch points are determined.
The vertices are fitted assuming Gaussian branch cross sections.
To determine the mesh branches reliably, we combine the fluorescence measurements $f_\mathrm{D}(\mathbf{x})$ and $f_\mathrm{E}(\mathbf{x})$ of labeled MinD and MinE, respectively.
To this end, the set of pixel values $\{(f_\mathrm{D}, f_\mathrm{E})(\mathbf{x})\}$ is clustered into two clusters (describing domains and branches) using the k-means method and the cluster centroids are determined.
We use the orthogonal projections $f(\mathbf{x})$ of the pixel values $(f_\mathrm{D},f_\mathrm{E})(\mathbf{x})$ onto the axis connecting the two centroids as values distinguishing between the (MinE) branches and (MinD) domains.
The width $\delta$ is determined by the same method as in the simulations but using $f(\mathbf{x})$.
The maximum branch intensity of $f(\mathbf{x})$ is estimated from the image histogram and normalized to one (as are all higher values).
Vertices are identified as triple vertices if the distance to the closest vertex fulfills ${r \gtrsim  \SI{2.13}{\micro m}}$ (corresponding to 10 pixels while the branch widths range between 2 to 6 pixels).
If the distance to the next vertex is less, the vertex is classified as a four-fold vertex.
The vertices are fitted to the data within the radius ${R \approx \min(\SI{4.25}{\micro m},r)}$ from the branch point.

The domain areas are determined as for the simulated Turing foam.
To analyze the domain evolution in the time-series data (Fig.~8 in the main text), consecutive images are averaged to reduce noise.
The domain areas are again determined as for the simulated Turing foam.

\section{Movie captions}
The following captions describe the Supplementary Movies 1 to 6.
\medskip

\textbf{Movie 1: Coarsening in the three-species mutual detachment system.}
Movie 1 displays two simulations of the three-species mutual detachment system given in Eq.~A4 in Methods Sec.~A.1.a.
The movie is recorded on a logarithmic timescale and shows the coarsening process of the pattern domains starting out from random initial conditions.
The membrane densities of the three species A, B, and C are shown in blue, red, and yellow, respectively.
The simulations are performed on a quadratic domain with edge length $L=30$ and no-flux boundary conditions.
The reaction rates and cytosolic diffusion coefficients are specified in Tab.~I in the Methods.
In the left simulation, the membrane diffusion coefficients of all species are chosen equally as $D_\mathrm{m}^{\mathrm{A,B,C}} = 0.01$.
In the right simulation, the membrane diffusion coefficient $D_\mathrm{m}^{\mathrm{A}} = 0.2$ is increased ($D_\mathrm{m}^{\mathrm{B,C}} = 0.01$).
Fig.~3a,e in the main text shows snapshots from the left and right simulation, respectively.
The tracked junctions are marked in black.
\medskip

\textbf{Movie 2: Emergence of 2D Turing foam from random initial conditions in simulations of the Min system.}
The movie shows six simulations of the model of the Min system including the MinE switch and MinE persistent membrane binding, as described in Methods Sec.~A.2.
During pattern growth, 4-fold vertices split into two 3-fold vertices before the pattern reaches a steady state.
The total MinE membrane density $m_\mathrm{de}+m_\mathrm{e}$ is shown in cyan on a black background.
The simulation parameters are given in Tab.~II in the Methods.
The simulation is performed on a quadratic domain with edge length ${L=\SI{100}{\micro m}}$ and no-flux boundary conditions.
As initial condition, the total MinD and MinE mass is distributed uniformly in the cytosolic components $c_\mathrm{DD}$ and $c_\mathrm{E,l}$, respectively.
The uniform densities are perturbed by noise uniformly distributed between $-1\%$ and $+1\%$ of the uniform densities (six independent runs shown).
The movie is recorded on a logarithmic timescale with the time in seconds shown in the movie.
\medskip

\textbf{Movie 3: Interrupted coarsening in simulations of the Min system.}
The movie shows the transient coarsening of the Min Turing foam in six independent simulations of the Min system (same setup and color code as in Movie 2).
The initial condition is chosen as the final pattern from the simulations shown in Movie 2.
Interrupted coarsening is induced by increasing all diffusion coefficients by a factor $2$, which corresponds to a rescaling of the edge length of the simulation domain by a factor $1/\sqrt{2}$.
As a result, the foam-domain areas (in the rescaled units) are reduced by a factor of $2$, and the foam domains transiently grow via a coarsening process induced by the collapse of domains with 5 edges before the coarsening process arrests.
Fig.~5a in the main text shows snapshots from the simulation shown in the top left.
\medskip

\textbf{Movie 4: Domain splitting in simulations of the Min system.}
Movie 4 shows the splitting of foam domains in the Min Turing foam in six independent simulations of the Min system (same setup and color code as in Movie 2).
Again, the initial condition is chosen as the final pattern from the simulations shown in Movie 2.
Domain splitting is induced by decreasing all diffusion coefficients by a factor of $2$.
This corresponds to a rescaling of all lengths by a factor of $\sqrt{2}$, such that the areas of the foam domains correspond to areas increased by a factor $2$.
The domains split until the domain areas (in rescaled units) become comparable to their original size (cf.\ Fig.~4h in the main text).
Fig.~5g in the main text shows snapshots from the simulation shown in the top left.

\textbf{Movie 5: Evolution of the Turing foam in \textit{in vitro} experiments with the Min system.}
The movie shows the time evolution (imaged every 10 seconds) of a mesh pattern formed by MinD (magenta) and MinE-His (cyan) on a supported lipid bilayer.
The experiment was performed as part of Ref.~\cite{Glock.etal2019}.
In this experiment, the total concentrations of MinD and MinE in the buffer solution are $\SI{2}{\micro M}$ and $\SI{8}{\micro M}$, respectively.
We highlighted four- and five-sided domains that collapse during the course of the movie by white squares and pentagons.
Examples of newly growing MinE branches are labeled by white lines.
The scale bar is $\SI{10}{\micro\meter}$ long (white, lower left corner).
\medskip

\textbf{Movie 6: Evolution of the Turing foam in simulations of the ferrocyanide--iodate--sulfite (FIS) reaction--diffusion system.}
The three panels display the Turing foam in simulations of the ferrocyanide--iodate--sulfite (FIS) reaction--diffusion system, recorded on a logarithmic timescale.
The simulation uses the four-species G\'asp\'ar--Showalter model with the parameters employed in Ref.~\cite{Lee.Swinney1995}, Fig.~17, and described in Methods Sec.~A.3.
The density of species $y$ ($\mathrm{H}^+$) is shown in cyan on a black background.
The simulation parameters are given in Tab.~III in the Methods.
The simulation was performed in rescaled length units (by rescaling the diffusion coefficients), which correspond to a quadratic domain with edge length $L = \sqrt{2}\si{\milli\meter}$ with no-flux boundary conditions.
The first panel (initial growth) shows the growth from initial perturbations.
The species $x$, $z$, and $a$ were initialized with uniform concentrations of $10^{-8}\si{M}$.
The species $y$ was initialized with a uniform concentration of $10^{-4}\si{M}$.
Because large initial perturbations were required to initialize pattern formation, we added uniformly distributed noise with a maximal amplitude of $10\%$ and additionally a sinusoidal perturbation of $30\%$ amplitude to the uniform base concentration of $y$.
The interrupted coarsening (Von Neumann evolution) and domain splitting (Domain splitting) are observed by using the final pattern from the simulation of the initial growth and scaling all diffusion coefficients by a factor $2$ and $1/2$, respectively.
The time in seconds is shown in the movie.

\clearpage

\input{SI_Turing_Foam.bbl}

%% file: Turing_Foam.bbl
%